\documentclass[12pt]{article}
\usepackage{amsfonts}
\usepackage[dvips]{graphicx}
\usepackage{latexsym,amsmath,amssymb,graphics,stmaryrd}
\usepackage{cite}
\usepackage{array}
\usepackage{subfigure}
\def\showlabelfont{\tiny\tt}
\def\SL@margintext#1{{\showlabelfont\SL@prlabelname{#1}}}
\usepackage{multirow}
\usepackage{epsfig}
\usepackage[dvips]{graphicx}
\usepackage[dvips]{color}
\usepackage{pstricks}
\usepackage{pst-node}
\usepackage{pst-plot}
\usepackage{dsfont}
\usepackage{amsthm}
\usepackage{graphics}
\usepackage{subfigure}
\usepackage{fancyhdr}
\usepackage[dvips]{color}

\usepackage{rotating}
\newlength{\arlength}

\usepackage{feynmp}
\usepackage{Dalgebra}

\fancypagestyle{plain}{
\fancyhf{}
\newcommand{\fpage}{\iffloatpage{}{\thepage}}
\fancyfoot[C]{\fpage}

}

\newcommand{\col}{~,}
\newcommand{\pnt}{~.}

\newcommand{\CS}{\mathrm{CS}}
\newcommand{\FP}{\mathrm{FP}}
\newcommand{\gf}{\mathrm{gf}}

\newcommand{\la}{\lambda}

\newcommand{\s}{\sigma}
\newcommand{\bla}{\bar\lambda}
\newcommand{\veps}{\varepsilon}

\newcommand{\be}{\begin{equation}}
\newcommand{\ee}{\end{equation}}
\newcommand{\NN}{{\mathcal N}}

\newcommand{\unitmatrix}{\mathds{1}}
\newcommand{\comm}[2]{\left[#1\smash[b]{\mathbin{,}}#2\right]}
\newcommand{\acomm}[2]{\left\{#1\smash[b]{\mathbin{,}}#2\right\}}

\newcommand{\de}{\operatorname{d}\!}

\newcommand{\e}{\operatorname{e}}

\newcommand{\pfour}[4]{{}\{#1,#2,#3,#4\}{}}
\newcommand{\pthree}[3]{{}\{#1,#2,#3\}{}}
\newcommand{\ptwo}[2]{{}\{#1,#2\}{}}
\newcommand{\pone}[1]{{}\{#1\}{}}

\newlength{\neglength}
\newlength{\diameter}
    
\newcommand{\nvml}[3][1]{%
\fmfcmd{%
begingroup;
save a, vp, tvp, nvp, tv, nv, ip, ts, tt, is, it, n, m, scale, t, r, s, ttpr, tnpr, ep, mm;
path lcirc;
pair vp[][], tvp[][], tv[][], nvp[][], nv[][], ip[][], ts[], is[], tt[], it[], ep[], mid;
n := #2;
m:=3;
for i=1 upto n:
for j=1 upto m:
a[i][j] := arctime ((j-1)/(m-1)*arclength pm[i]) of pm[i];
vp[i][j] := point a[i][j] of pm[i];
tvp[i][j] := unitvector direction a[i][j] of pm[i];
nvp[i][j] := tvp[i][j] rotated -90;
endfor;
endfor;
if(vp[1][1]=vp[n][m]):
vp[n+1][1] := vp[1][1];
tvp[0][m] := tvp[n][m];
nvp[0][m] := nvp[n][m];
tvp[n+1][1] :=tvp[1][1];
nvp[n+1][1] :=nvp[1][1];
else:
vp[n+1][1] := vp[n][m];
tvp[0][m] := (0,0);
nvp[0][m] := (0,0);
tvp[n+1][1] :=tvp[n][m];
nvp[n+1][1] :=nvp[n][m];
fi;
s := 1;
for i=1 upto n:
for j=1 upto m:
if (j=1):
tv[i][1] := (tvp[i-1][m]+tvp[i][1]);
nv[i][1] := (nvp[i-1][m]+nvp[i][1]);
if (not(tv[i][1]=(0,0))):
tv[i][1] := unitvector tv[i][1];
fi;
if (not(nv[i][1]=(0,0))):
nv[i][1] := unitvector nv[i][1];
fi;
ttpr := tvp[i][1] dotprod tvp[i-1][m];
tnpr := tvp[i][1] dotprod nvp[i-1][m];
elseif (j=m):
tv[i][m] := (tvp[i][m]+tvp[i+1][1]);
nv[i][m] := (nvp[i][m]+nvp[i+1][1]);
if (not(tv[i][m]=(0,0))):
tv[i][m] := unitvector tv[i][m];
fi;
if (not(nv[i][m]=(0,0))):
nv[i][m] := unitvector nv[i][m];
fi;
ttpr := tvp[i][m] dotprod tvp[i+1][1];
tnpr := -tvp[i][m] dotprod nvp[i+1][1];
else:
nv[i][j] :=nvp[i][j];
tv[i][j] :=tvp[i][j];
fi;
scale := 25;
if ((j=1) or (j=m)):
 if ((tnpr<=0) and not((tv[i][j]=(0,0)) or (nv[i][j]=(0,0)))):
  ip[i][j] := vp[i][j] shifted(0.15*scale*nvp[i][j]);
  ts[s] := tvp[i][j];
  is[s] := ip[i][j];
  s:=s+1;
 else:
  if ((j=1) and (ttpr>0)):
  fi;
 fi;
else:
 ip[i][j] := vp[i][j] shifted(0.15*scale*nv[i][j]);
 ts[s] := tv[i][j];
 is[s] := ip[i][j];
 s:=s+1;
fi;
endfor;
endfor;
if(vp[1][1]=vp[n][m]):
ts[s] := ts[1];
is[s] := is[1];
else:
tv[n+1][1] := unitvector (tvp[n][m]+tvp[n+1][1]);
nv[n+1][1] := unitvector (nvp[n][m]+nvp[n+1][1]);
ip[n+1][1] := vp[n+1][1] shifted(0.15*scale*nv[n+1][1]);
ts[s] := tv[n+1][1];
is[s] := ip[n+1][1];
fi;
t=#1;
lcirc:=is[1];
for k=2 upto s:
lcirc := lcirc{ts[k-1]}..tension t..{ts[k]}is[k];
endfor;
mm := arctime (0.5* arclength lcirc) of lcirc;
if(vp[1][1]=vp[n][m]):
ep1 := point arctime (0* arclength lcirc) of lcirc of lcirc;
ep2 := point mm of lcirc;
mid := 1/2[ep1,ep2];
else:
ep1 := point mm of lcirc;
ep2 :=unitvector direction mm of lcirc rotated -90;
mid:= ep1 shifted(0.2*scale*ep2);
fi;
draw(lcirc) withpen pencircle scaled 0.25;
drawarrow(subpath(mm*0.8,mm*1.1) of lcirc) withpen pencircle scaled 0.25;
endgroup;
}
\fmfiv{label=#3,l.dist=0}{mid}
}

\newcommand{\svertex}[2]{%
\fmfiequ{#1}{point length(#2)/2 of #2}
}
\newcommand{\dvertex}[3]{%
\fmfiequ{#1}{point length(#3)/3 of #3}
\fmfiequ{#2}{point 2length(#3)/3 of #3}
}
\newcommand{\vvertex}[3]{%
\fmfipath{px}
\fmfiequ{px}{(0,ypart(#2))..(100,ypart(#2))}
\fmfiequ{#1}{point xpart(#3 intersectiontimes px) of #3}
}

\newcommand{\vacpol}[3][0.5]{%
\fmfcmd{
begingroup;
save t,v,tv,do,di,ppol,pstr,dia;
path ppol,pstr;
pair v[],tv[],do[],di[];
ppol=vpath(__#2,__#3);
t1=arctime (1/3*arclength ppol) of ppol;
t2=arctime (2/3*arclength ppol) of ppol;
v1=point t1 of ppol;
v2=point t2 of ppol;
t3=arctime (0.5*arclength ppol) of ppol;
v3=point t3 of ppol;
dia=#1*0.5 arclength ppol; 
fill(fullcircle scaled dia shifted v3) withcolor 0.2black;
endgroup;
}
}

\newcommand{\vacpolp}[2][0.5]{%
\fmfcmd{
begingroup;
save t,v,tv,do,di,ppol,pstr,dia;
path ppol,pstr;
pair v[],tv[],do[],di[];
ppol=#2;
t3=arctime (0.5*arclength ppol) of ppol;
v3=point t3 of ppol;
dia=#1 arclength ppol; 
fill(fullcircle scaled dia shifted v3) withcolor 0.2black;
endgroup;
}
}

\newcommand{\plainwrap}[4]{%
\fmfipath{pi[]}
\fmfiset{pi1}{vloc(__#1) ..controls (-0.175w,ypart(vloc(__#1))) and (-0.175w,-0.15w) .. (xpart(vloc(__#2)),-0.15w)}
\fmfiset{pi2}{(xpart(vloc(__#2)),-0.15w) ..(xpart(vloc(__#3)),-0.15w)}

\fmfiset{pi3}{(xpart(vloc(__#3)),-0.15w) ..controls (1.175w,-0.15w) and (1.175w,ypart(vloc(__#4))) .. vloc(__#4)}
\fmfi{plain}{pi1 ..pi2 ..pi3}
}

\newcommand{\wigglywrap}[4]{%
\fmfipath{pi[]}
\fmfiset{pi1}{#1 ..controls (-0.175w,ypart(#1)) and (-0.175w,-0.15w) .. (xpart(vloc(__#2)),-0.15w)}
\fmfiset{pi2}{(xpart(vloc(__#2)),-0.15w) ..(xpart(vloc(__#3)),-0.15w)}

\fmfiset{pi3}{(xpart(vloc(__#3)),-0.15w) ..controls (1.175w,-0.15w) and (1.175w,ypart(#4)) .. #4}
\fmfi{photon}{pi1}
\fmfi{photon}{pi2}
\fmfi{photon}{pi3}
}

\newcommand{\Ifourtwotwoq}[8]{%
\settoheight{\eqoff}{$\times$}%
\setlength{\eqoff}{0.5\eqoff}%
\addtolength{\eqoff}{-10\unitlength}%
\raisebox{\eqoff}{%
\fmfframe(0,0)(0,0){%
\begin{fmfchar*}(20,20)
  \fmfleft{vl}
  \fmfright{vr}
  \fmftop{vt}
  \fmfbottom{vb}
  \fmf{phantom,tension=1}{vl,v1}
  \fmf{phantom,tension=1}{vr,v3}
  \fmf{phantom,tension=1}{vt,v4}
  \fmf{plain,tension=1}{vb,v2}
\fmffixed{(0.9w,0)}{v1,v3}
\fmfpoly{phantom}{v1,v2,v3,v4}
  \fmf{#4,right=0.25}{v1,v2}
  \fmf{#3,right=0.25}{v2,v3}
  \fmf{#2,right=0.25}{v3,v4}
  \fmf{#1,right=0.25}{v4,v1}
\fmffreeze
  \fmf{#8}{v0,v1}
  \fmf{#7}{v0,v2}
  \fmf{#6}{v0,v3}
  \fmf{#5}{v0,v4}
\end{fmfchar*}}}
}

\newcommand{\Ifourtwotwob}[8]{%
\settoheight{\eqoff}{$\times$}%
\setlength{\eqoff}{0.5\eqoff}%
\addtolength{\eqoff}{-10\unitlength}%
\raisebox{\eqoff}{%
\fmfframe(0,0)(0,0){%
\begin{fmfchar*}(20,20)
  \fmfleft{vl}
  \fmfright{vr}
  \fmftop{vt}
  \fmfbottom{vb}
  \fmf{phantom,tension=1}{vl,v1}
  \fmf{phantom,tension=1}{vr,v3}
  \fmf{phantom,tension=1}{vt,v4}
  \fmf{plain,tension=1}{vb,v2}
\fmffixed{(0.9w,0)}{v1,v3}
\fmfpoly{phantom}{v1,v2,v3,v4}
  \fmf{#4,right=0.25}{v1,v2}
  \fmf{#3,right=0.25}{v2,v3}
  \fmf{#2,right=0.25}{v3,v4}
  \fmf{#1,right=0.25}{v4,v1}
\fmffreeze
  \fmf{#8}{v0,v1}
  \fmf{phantom}{v0,v2}
  \fmf{#7}{v0,v3}
  \fmf{phantom}{v0,v4}
\fmffreeze
  \fmf{#6,right=0.25}{v0,v4}
  \fmf{#5,left=0.25}{v0,v4}
\end{fmfchar*}}}
}

\newcommand{\Ifourtwobbbtwo}[6]{%
\settoheight{\eqoff}{$\times$}%
\setlength{\eqoff}{0.5\eqoff}%
\addtolength{\eqoff}{-10\unitlength}%
\raisebox{\eqoff}{%
\fmfframe(1,0)(1,0){%
\begin{fmfchar*}(10,20)
  \fmfleft{vl}
  \fmfright{vr}
  \fmftop{vt}
  \fmfbottom{vb}
  \fmf{phantom,tension=1}{vt,v1}
  \fmf{phantom,tension=1}{vl,v2}
  \fmf{phantom,tension=1}{vb,v3}
  \fmf{plain,tension=1}{vr,v4}
\fmffixed{(0,0.9h)}{v3,v1}
\fmfpoly{phantom}{v1,v2,v4}
\fmfpoly{phantom}{v3,v4,v2}
  \fmf{#1,left=0.25}{v2,v1}
  \fmf{#4,right=0.25}{v2,v3}
  \fmf{#5,left=0.25}{v2,v3}
  \fmf{plain,right=0.25}{v4,v1}
  \fmf{#6,left=0.25}{v4,v3}
  \fmf{#2,right=0.25}{v2,v1}
  \fmf{#3,left=0.25}{v4,v1}
\end{fmfchar*}}}
}

\newcommand{\Ifourtwobbbthree}[6]{%
\settoheight{\eqoff}{$\times$}%
\setlength{\eqoff}{0.5\eqoff}%
\addtolength{\eqoff}{-10\unitlength}%
\raisebox{\eqoff}{%
\fmfframe(1,0)(1,0){%
\begin{fmfchar*}(10,20)
  \fmfleft{vl}
  \fmfright{vr}
  \fmftop{vt}
  \fmfbottom{vb}
  \fmf{phantom,tension=1}{vt,v1}
  \fmf{phantom,tension=1}{vl,v2}
  \fmf{phantom,tension=1}{vb,v3}
  \fmf{plain,tension=1}{vr,v4}
\fmffixed{(0,0.9h)}{v3,v1}
\fmfpoly{phantom}{v1,v2,v4}
\fmfpoly{phantom}{v3,v4,v2}
  \fmf{#1,left=0.25}{v2,v1}
  \fmf{#2}{v2,v1}
  \fmf{plain,right=0.25}{v2,v1}
  \fmf{#3,right=0.25}{v4,v1}
  \fmf{#4,left=0}{v2,v4}
  \fmf{#5,right=0.25}{v2,v3}
  \fmf{#6,right=0.25}{v3,v4}
\end{fmfchar*}}}
}

\newcommand{\Ifourtwoe}[7]{%
\settoheight{\eqoff}{$\times$}%
\setlength{\eqoff}{0.5\eqoff}%
\addtolength{\eqoff}{-7.5\unitlength}%
\raisebox{\eqoff}{%
\fmfframe(0,0)(0,0){%
\begin{fmfchar*}(15,15)
  \fmftop{vtl,vtr}
  \fmfbottom{vb}
  \fmf{phantom,tension=1}{vb,v3}
  \fmf{phantom,tension=1}{vtl,v1}
  \fmf{phantom,tension=1}{vtr,v12}
\fmffixed{(0.9h,0)}{v1,v2}
\fmfpoly{phantom}{v3,v2,v1}
  \fmf{#4,right=0.25}{v1,v3}
  \fmf{#1,left=0.25}{v1,v2}
  \fmf{#7,right=0.25}{v3,v2}
  \fmf{phantom}{v0,v3}
  \fmf{#3}{v0,v2}
  \fmf{#2}{v1,v0}
\fmffreeze
  \fmf{#6,left=0.25}{v0,v3}
  \fmf{#5,right=0.25}{v0,v3}
\end{fmfchar*}}}
}

\newcommand{\chioneca}[6]{%
\fmf{plain,tension=0,left=0.25}{#1,vc}
\fmf{plain,tension=1}{vc,#2}
\fmf{plain,tension=0,right=0.25}{#3,vc}
\fmf{plain,tension=0,right=0.25}{va,#4}
\fmf{plain,tension=1}{#5,va}
\fmf{plain,tension=0,left=0.25}{va,#6}
\fmf{plain,tension=1}{vc,va}
\fmfposition
\fmfipath{p[],pca}
\fmfipair{vm[],vo[],vi[],vmm,vom,vim}
\fmfiset{p1}{vpath(__#1,__vc)}
\fmfiset{p2}{vpath(__vc,__#2)}
\fmfiset{p3}{vpath(__#3,__vc)}
\fmfiset{pca}{vpath(__vc,__va)}
\fmfiset{p4}{vpath(__va,__#4)}
\fmfiset{p5}{vpath(__#5,__va)}
\fmfiset{p6}{vpath(__va,__#6)}
\svertex{vm1}{p1}
\dvertex{vo1}{vi1}{p1}
\svertex{vm2}{p2}
\dvertex{vi2}{vo2}{p2}
\svertex{vm3}{p3}
\dvertex{vo3}{vi3}{p3}
\svertex{vmm}{pca}
\dvertex{vom}{vim}{pca}
\svertex{vm4}{p4}
\dvertex{vi4}{vo4}{p4}
\svertex{vm5}{p5}
\dvertex{vo5}{vi5}{p5}
\svertex{vm6}{p6}
\dvertex{vi6}{vo6}{p6}
}

\newcommand{\chionerangefourl}{%
\fmftop{v3}
\fmfbottom{v4}
\fmfforce{(0.125w,h)}{v3}
\fmfforce{(0.125w,0)}{v4}
\fmffixed{(0.25w,0)}{v1,veu1}
\fmffixed{(0.25w,0)}{v2,v1}
\fmffixed{(0.25w,0)}{v3,v2}
\fmffixed{(0.25w,0)}{v4,v5}
\fmffixed{(0.25w,0)}{v5,v6}
\fmffixed{(0.25w,0)}{v6,ved1}
\fmf{plain}{ved1,veu1}
\chioneca{v1}{v2}{v3}{v4}{v5}{v6}
\fmfipair{ve,veo[],vem[],vei[],veom,vemm,veim}
\fmfipath{pe}
\fmfiset{pe}{vpath(__ved1,__veu1)}
\vvertex{veo1}{vo1}{pe}
\vvertex{vem1}{vm1}{pe}
\vvertex{vei1}{vi1}{pe}
\vvertex{veom}{vom}{pe}
\vvertex{vemm}{vmm}{pe}
\vvertex{veim}{vim}{pe}
\vvertex{veo6}{vo6}{pe}
\vvertex{vem6}{vm6}{pe}
\vvertex{vei6}{vi6}{pe}
\svertex{ve}{pe}
}

\newcommand{\chionetwog}{%
\fmftop{v1}
\fmfbottom{v5}
\fmfforce{(0.125w,h)}{v1}
\fmfforce{(0.125w,0)}{v5}
\fmffixed{(0.25w,0)}{v1,v2}
\fmffixed{(0.25w,0)}{v2,v3}
\fmffixed{(0.25w,0)}{v3,v4}
\fmffixed{(0.25w,0)}{v5,v6}
\fmffixed{(0.25w,0)}{v6,v7}
\fmffixed{(0.25w,0)}{v7,v8}
\fmf{plain,right=0.25}{v1,v4c}
\fmf{plain,left=0.25}{v4c,v2}
\fmf{plain,left=0.25}{v3,v4c}
\fmf{plain,right=0.25}{v4c,v4}
\fmf{plain,right=0.25}{v4a,v5}
\fmf{plain,left=0.25}{v6,v4a}
\fmf{plain,left=0.25}{v4a,v7}
\fmf{plain,right=0.25}{v8,v4a}
\fmf{phantom,tension=4}{v4c,v4a}
\fmffreeze
\fmfposition
\fmfipath{p[]}
\fmfipair{vgm[],vgd[],vgu[]}
\fmfiset{p1}{vpath(__v1,__v4c)}
\fmfiset{p2}{vpath(__v4c,__v4)}
\fmfiset{p3}{vpath(__v4a,__v5)}
\fmfiset{p4}{vpath(__v8,__v4a)}
\svertex{vgm1}{p1}
\svertex{vgm2}{p2}
\dvertex{vgu3}{vgd3}{p3}
\svertex{vgm3}{p3}
\dvertex{vgd4}{vgu4}{p4}
\svertex{vgm4}{p4}
}

\DeclareMathOperator{\tr}{tr}

\DeclareMathOperator{\diag}{diag}

\DeclareMathOperator{\perm}{P}
\DeclareMathOperator{\Kop}{K}

\DeclareMathOperator{\D}{D}
\DeclareMathOperator{\barD}{\vphantom{\D}\smash[t]{\bar{\mathrm{D}}}}
\DeclareMathOperator{\Ld}{L}

\numberwithin{equation}{section}
\newlength{\eqoff}

\newlength{\unit}
\psset{xunit=\unit,yunit=\unit,runit=\unit}
\newlength{\linew}
\psset{linewidth=\linew}
\begin{document}
\begin{fmffile}{CSgraphs}
\fmfcmd{%
input Dalgebra
}

\fmfcmd{%
marksize=2mm;
def draw_mark(expr p,a) =
  begingroup
    save t,tip,dma,dmb; pair tip,dma,dmb;
    t=arctime a of p;
    tip =marksize*unitvector direction t of p;
    dma =marksize*unitvector direction t of p rotated -45;
    dmb =marksize*unitvector direction t of p rotated 45;
    linejoin:=beveled;
    draw (-.5dma.. .5tip-- -.5dmb) shifted point t of p;
  endgroup
enddef;
style_def derplain expr p =
    save amid;
    amid=.5*arclength p;
    draw_mark(p, amid);
    draw p;
enddef;
def draw_marks(expr p,a) =
  begingroup
    save t,tip,dma,dmb,dmo; pair tip,dma,dmb,dmo;
    t=arctime a of p;
    tip =marksize*unitvector direction t of p;
    dma =marksize*unitvector direction t of p rotated -45;
    dmb =marksize*unitvector direction t of p rotated 45;
    dmo =marksize*unitvector direction t of p rotated 90;
    linejoin:=beveled;
    draw (-.5dma.. .5tip-- -.5dmb) shifted point t of p withcolor 0white;
    draw (-.5dmo.. .5dmo) shifted point t of p;
  endgroup
enddef;
style_def derplains expr p =
    save amid;
    amid=.5*arclength p;
    draw_marks(p, amid);
    draw p;
enddef;
def draw_markss(expr p,a) =
  begingroup
    save t,tip,dma,dmb,dmo; pair tip,dma,dmb,dmo;
    t=arctime a of p;
    tip =marksize*unitvector direction t of p;
    dma =marksize*unitvector direction t of p rotated -45;
    dmb =marksize*unitvector direction t of p rotated 45;
    dmo =marksize*unitvector direction t of p rotated 90;
    linejoin:=beveled;
    draw (-.5dma.. .5tip-- -.5dmb) shifted point t of p withcolor 0white;
    draw (-.5dmo.. .5dmo) shifted point arctime a+0.25 mm of p of p;
    draw (-.5dmo.. .5dmo) shifted point arctime a-0.25 mm of p of p;
  endgroup
enddef;
style_def derplainss expr p =
    save amid;
    amid=.5*arclength p;
    draw_markss(p, amid);
    draw p;
enddef;
style_def dblderplain expr p =
    save amidm;
    save amidp;
    amidm=.5*arclength p-0.75mm;
    amidp=.5*arclength p+0.75mm;
    draw_mark(p, amidm);
    draw_mark(p, amidp);
    draw p;
enddef;
style_def dblderplains expr p =
    save amidm;
    save amidp;
    amidm=.5*arclength p-0.75mm;
    amidp=.5*arclength p+0.75mm;
    draw_mark(p, amidm);
    draw_marks(p, amidp);
    draw p;
enddef;
style_def dblderplainss expr p =
    save amidm;
    save amidp;
    amidm=.5*arclength p-0.75mm;
    amidp=.5*arclength p+0.75mm;
    draw_mark(p, amidm);
    draw_markss(p, amidp);
    draw p;
enddef;
style_def dblderplainsss expr p =
    save amidm;
    save amidp;
    amidm=.5*arclength p-0.75mm;
    amidp=.5*arclength p+0.75mm;
    draw_marks(p, amidm);
    draw_markss(p, amidp);
    draw p;
enddef;
}

\begin{titlepage}
\begin{flushright}\footnotesize
\texttt{UUITP-34/10} \\
\texttt{IFUM-965-FT}\\
\texttt{HU-MATH-2010-16}
\end{flushright}
\Large
\begin {center}
{\bf
Superspace calculation of the 
four-loop spectrum 
in
$\NN=6$ supersymmetric Chern-Simons theories
}
\end {center}

\renewcommand{\thefootnote}{\fnsymbol{footnote}}

\large
\vspace{1cm}
\centerline{
M.\ Leoni ${}^{a,b}$,
A.\ Mauri ${}^{a,b}$, 
J.\ A.\ Minahan ${}^c$,
O.\ Ohlsson Sax ${}^c$,
}
\centerline{
A.\ Santambrogio ${}^b$,
C.\ Sieg ${}^{d,e}$,
G.\ Tartaglino-Mazzucchelli ${}^c$
\footnote[1]{\noindent \tt
matias.leoni@mi.infn.it, \\
\hspace*{6.3mm}andrea.mauri@mi.infn.it, \\
\hspace*{6.3mm}joseph.minahan@fysast.uu.se, \\
\hspace*{6.3mm}olof.ohlsson-sax@physics.uu.se, \\
\hspace*{6.3mm}alberto.santambrogio@mi.infn.it, \\
\hspace*{6.3mm}csieg@nbi.dk, csieg@math.hu-berlin.de \\
\hspace*{6.3mm}gabriele.tartaglino-mazzucchelli@fysast.uu.se}}
\vspace{4ex}
\normalsize
\begin{center}
\emph{$^a$  Dipartimento di Fisica, Universit\`a degli Studi di Milano \\
Via Celoria 16, 20133 Milano, Italy}\\
\vspace{0.2cm}
\emph{$^b$ INFN-Sezione di Milano\\
Via Celoria 16, 20133 Milano, Italy}\\
\vspace{0.2cm}
\emph{$^c$  Department of Physics and Astronomy, Uppsala University\\
SE-751 08 Uppsala, Sweden}\\
\vspace{0.2cm}
\emph{$^d$ The Niels Bohr International Academy\\ The Niels Bohr Institute\\
Blegdamsvej 17, 2100 Copenhagen, Denmark}\\
\vspace{0.2cm}
\emph{$^e$ Institute f\"ur Mathematik/Physik, 
Humboldt-Universit\"at zu Berlin\\
Rudower Chaussee 25, 12489 Berlin, Germany}
\end{center}
\vspace{0.5cm}
\rm
\abstract
\normalsize
Using $\mathcal{N}=2$ superspace techniques we compute the
four-loop spectrum of single trace operators in the $SU(2)\times SU(2)$ sector of ABJM and ABJ supersymmetric Chern-Simons theories. 
Our computation
yields a four-loop contribution to the function $h^2(\lambda)$ (and its ABJ generalization) in
the magnon dispersion relation which has fixed maximum
transcendentality and coincides with 
the findings 
in components 
given in the
revised versions of
arXiv:0908.2463 and arXiv:0912.3460. 
We also discuss 
possible scenarios for an all-loop 
function $h^2(\lambda)$ 
that interpolates between weak and strong couplings.
\\[0.3cm]

\vfill
\end{titlepage}

\fmfcmd{%
thin := 1pt; 
thick := 2thin;
arrow_len := 4mm;
arrow_ang := 15;
curly_len := 3mm;
dash_len := 1.5mm; 
dot_len := 1mm; 
wiggly_len := 2mm; 
wiggly_slope := 60;
zigzag_len := 2mm;
zigzag_width := 2thick;
decor_size := 5mm;
dot_size := 2thick;
}

\newcommand{\threelthreeg}{
\fmfipair{v[],ve[]}
\fmfipath{ls[]}
\fmfipair{a[]}
\fmftop{vt}
\fmfbottom{vb}
\fmffixed{(0,0.1h)}{vo,vt1}
\fmf{phantom}{vt1,vt}
\fmf{phantom}{vb,vb1}
\fmffixed{(0,0.1h)}{vb1,vi}
\fmffixed{(0,0.75h)}{vi,vo}
\fmf{phantom,right=0.5}{vi,vo}
\fmf{phantom,right=0.5}{vo,vi}
\fmf{phantom}{vi,v0}
\fmf{phantom}{v0,vo}
\fmffreeze
\fmfposition
\fmfiset{ls1}{vpath(__vo,__vi)}
\fmfiset{ls2}{vpath(__vi,__vo)}
\fmfiequ{v3}{point length(ls2)/2 of ls2}
\fmfiset{ls3}{v3--vloc(__v0)}
\fmfiset{ls4}{vloc(__v0)--(vloc(__v0) shifted (100 unitvector direction 1 of ls3 rotated -60))}
\fmfiset{ls5}{vloc(__v0)--(vloc(__v0) shifted (100 unitvector direction 1 of ls3 rotated 60))}
\fmfiequ{a1}{ls1 intersectiontimes ls4}
\fmfiequ{a2}{ls1 intersectiontimes ls5}
\fmfiequ{v1}{point xpart(a1) of ls1}
\fmfiequ{v2}{point xpart(a2) of ls1}
\fmfiequ{v3}{point length(ls2)/2 of ls2}
\fmfiequ{ve1}{v1+(0,0.1h)}
\fmfiequ{ve2}{v2-(0,0.1h)}
\fmfiequ{ve3}{v3+(0.1h,0)}
\fmffreeze
}

%
\newcommand{\vacpolD}[9]{%
\settoheight{\eqoff}{$\times$}%
\setlength{\eqoff}{0.5\eqoff}%
\addtolength{\eqoff}{-10\unitlength}%
\raisebox{\eqoff}{%
\fmfframe(1,0)(1,0){%
\begin{fmfchar*}(30,20)
\fmfleft{v1}
\fmfright{v2}
\fmffixed{(0.4w,0)}{vc1,vc2}
\fmf{#1}{v1,vc1}
\fmf{#3}{vc2,v2}
\fmf{#2,tension=0.5,left=1}{vc1,vc2}
\fmf{#2,tension=0.5,left=1}{vc2,vc1}
\fmffreeze
\fmfposition
\fmfipath{p[]}
\fmfiset{p1}{vpath(__v1,__vc1)}
\fmfiset{p2}{vpath(__vc1,__vc2)}
\fmfiset{p3}{reverse vpath(__vc2,__vc1)}
\fmfiset{p4}{vpath(__vc2,__v2)}
\fmfis{phantom,ptext.clen=6,ptext.hout=3,ptext.oout=12,ptext.out=#4,ptext.sep=;}{p1}
\fmfis{phantom,ptext.clen=6,ptext.hin=3,ptext.hout=3,ptext.oin=10,ptext.oout=10,ptext.in=#5,ptext.out=#6,ptext.sep=;}{p2}
\fmfis{phantom,label.side=right,ptext.clen=6,ptext.hin=-4,ptext.hout=-4,ptext.oin=10,ptext.oout=10,ptext.in=#7,ptext.out=#8,ptext.sep=;}{p3}
\fmfis{phantom,ptext.clen=6,ptext.hin=3,ptext.oin=12,ptext.in=#9,ptext.sep=;}{p4}
\end{fmfchar*}}}}

\newcommand{\swftwoone}[3]{%
\settoheight{\eqoff}{$\times$}%
\setlength{\eqoff}{0.5\eqoff}%
\addtolength{\eqoff}{-7.5\unitlength}%
\raisebox{\eqoff}{%
\fmfframe(1,0)(1,0){%
\begin{fmfchar*}(20,15)
\fmfleft{v1}
\fmfright{v2}
\fmffixed{(0.66w,0)}{vc1,vc2}
\fmf{plain}{v1,vc1}
\fmf{plain}{vc2,v2}
\fmf{plain,tension=0.5}{vc1,vc2}
\fmf{#1,tension=0.5,left=#2}{vc1,vc2}
\fmf{#1,tension=0.5,right=#3}{vc2,vc1}
\fmffreeze
\fmfposition
\fmfipath{p[]}
\fmfiset{p1}{vpath(__v1,__vc1)}
\fmfiset{p2}{vpath(__vc1,__vc2)}
\fmfiset{p3}{reverse vpath(__vc2,__vc1)}
\fmfiset{p4}{vpath(__vc2,__v2)}
\end{fmfchar*}}}}

\newcommand{\swftwotwo}[2]{%
\settoheight{\eqoff}{$\times$}%
\setlength{\eqoff}{0.5\eqoff}%
\addtolength{\eqoff}{-7.5\unitlength}%
\raisebox{\eqoff}{%
\fmfframe(1,0)(1,0){%
\begin{fmfchar*}(20,15)
\fmfleft{v1}
\fmfright{v2}
\fmffixed{(0.33w,0)}{vc1,vc2}
\fmffixed{(0.33w,0)}{vc2,vc3}
\fmf{plain}{v1,vc1}
\fmf{plain}{vc1,vc2}
\fmf{plain}{vc2,vc3}
\fmf{plain}{vc3,v2}
\fmf{photon,tension=0.5,left=#2}{vc1,vc2}
\fmf{photon,tension=0.5,left=#1}{vc1,vc3}
\fmffreeze
\fmfposition
\fmfipath{p[]}
\end{fmfchar*}}}}

\newcommand{\swftwothree}[2]{%
\settoheight{\eqoff}{$\times$}%
\setlength{\eqoff}{0.5\eqoff}%
\addtolength{\eqoff}{-7.5\unitlength}%
\raisebox{\eqoff}{%
\fmfframe(1,0)(1,0){%
\begin{fmfchar*}(20,15)
\fmfleft{v1}
\fmfright{v2}
\fmffixed{(0.33w,0)}{vc1,vc2}
\fmffixed{(0.33w,0)}{vc2,vc3}
\fmf{plain}{v1,vc1}
\fmf{plain}{vc1,vc2}
\fmf{plain}{vc2,vc3}
\fmf{plain}{vc3,v2}
\fmf{photon,tension=0.5,left=#2}{vc2,vc3}
\fmf{photon,tension=0.5,left=#1}{vc1,vc3}
\fmffreeze
\fmfposition
\fmfipath{p[]}
\end{fmfchar*}}}}

\newcommand{\swftwofour}[2]{%
\settoheight{\eqoff}{$\times$}%
\setlength{\eqoff}{0.5\eqoff}%
\addtolength{\eqoff}{-7.5\unitlength}%
\raisebox{\eqoff}{%
\fmfframe(1,0)(1,0){%
\begin{fmfchar*}(20,15)
\fmfleft{v1}
\fmfright{v2}
\fmffixed{(0.33w,0)}{vc1,vc2}
\fmffixed{(0.33w,0)}{vc2,vc3}
\fmf{plain}{v1,vc1}
\fmf{plain}{vc1,vc2}
\fmf{plain}{vc2,vc3}
\fmf{plain}{vc3,v2}
\fmf{photon,tension=0.5,left=#1}{vc1,vc2}
\fmf{photon,tension=0.5,left=#2}{vc2,vc3}
\fmffreeze
\fmfposition
\fmfipath{p[]}
\end{fmfchar*}}}}

\newcommand{\swftwofive}[1]{%
\settoheight{\eqoff}{$\times$}%
\setlength{\eqoff}{0.5\eqoff}%
\addtolength{\eqoff}{-7.5\unitlength}%
\raisebox{\eqoff}{%
\fmfframe(1,0)(1,0){%
\begin{fmfchar*}(20,15)
\fmfleft{v1}
\fmfright{v2}
\fmffixed{(0.22w,0)}{vc1,vc2}
\fmffixed{(0.22w,0)}{vc2,vc3}
\fmffixed{(0.22w,0)}{vc3,vc4}
\fmf{plain}{v1,vc1}
\fmf{plain}{vc1,vc2}
\fmf{plain}{vc2,vc3}
\fmf{plain}{vc3,vc4}
\fmf{plain}{vc4,v2}
\fmf{photon,tension=0.5,left=#1}{vc1,vc3}
\fmf{photon,tension=0.5,left=-#1}{vc2,vc4}
\fmffreeze
\fmf{photon,tension=0.5}{vc2,vc}
\fmffreeze
\fmfposition
\fmfipath{p[]}
\fmfiset{p1}{vpath(__v1,__vc1)}
\fmfiset{p2}{vpath(__vc1,__vc2)}
\fmfiset{p3}{vpath(__vc2,__vc3)}
\fmfiset{p4}{vpath(__vc3,__vc4)}
\fmfiset{p5}{vpath(__vc4,__v2)}
\end{fmfchar*}}}}

\newcommand{\swftwosix}{%
\settoheight{\eqoff}{$\times$}%
\setlength{\eqoff}{0.5\eqoff}%
\addtolength{\eqoff}{-7.5\unitlength}%
\raisebox{\eqoff}{%
\fmfframe(1,0)(1,0){%
\begin{fmfchar*}(20,15)
\fmfleft{v1}
\fmfright{v2}
\fmffixed{(0.33w,0)}{vc1,vc2}
\fmffixed{(0.33w,0)}{vc2,vc3}
\fmffixed{(0,0.33w)}{vc2,vc}
\fmf{plain}{v1,vc1}
\fmf{plain}{vc1,vc2}
\fmf{plain}{vc2,vc3}
\fmf{plain}{vc3,v2}
\fmf{photon,tension=0.5,left=0.5}{vc1,vc}
\fmf{photon,tension=0.5,left=0.5}{vc,vc3}
\fmffreeze
\fmf{photon,tension=0.5}{vc2,vc}
\fmffreeze
\fmfposition
\fmfipath{p[]}
\fmfiset{p1}{vpath(__v1,__vc1)}
\fmfiset{p2}{vpath(__vc1,__vc2)}
\fmfiset{p3}{vpath(__vc2,__vc3)}
\fmfiset{p4}{vpath(__vc3,__v2)}
\fmfiset{p5}{vpath(__vc1,__vc)}
\fmfiset{p6}{vpath(__vc2,__vc)}
\fmfiset{p7}{vpath(__vc,__vc3)}
\end{fmfchar*}}}}

\newcommand{\swftwoseven}[1]{%
\settoheight{\eqoff}{$\times$}%
\setlength{\eqoff}{0.5\eqoff}%
\addtolength{\eqoff}{-7.5\unitlength}%
\raisebox{\eqoff}{%
\fmfframe(1,0)(1,0){%
\begin{fmfchar*}(20,15)
\fmfleft{v1}
\fmfright{v2}
\fmffixed{(0.66w,0)}{vc1,vc2}
\fmf{plain}{v1,vc1}
\fmf{plain}{vc2,v2}
\fmf{plain,tension=0.5}{vc1,vc2}
\fmf{photon,tension=0.5,left=#1}{vc1,vc2}
\fmffreeze
\fmfposition
\fmfipath{p[]}
\fmfiset{p1}{vpath(__v1,__vc1)}
\fmfiset{p2}{vpath(__vc1,__vc2)}
\fmfiset{p3}{reverse vpath(__vc2,__vc1)}
\fmfiset{p4}{vpath(__vc2,__v2)}
\vacpol{vc1}{vc2}
\end{fmfchar*}}}}

\newcommand{\swftwoeight}[1]{%
\settoheight{\eqoff}{$\times$}%
\setlength{\eqoff}{0.5\eqoff}%
\addtolength{\eqoff}{-7.5\unitlength}%
\raisebox{\eqoff}{%
\fmfframe(1,0)(1,0){%
\begin{fmfchar*}(20,15)
\fmfleft{v1}
\fmfright{v2}
\fmf{plain}{v1,vc1}
\fmf{plain}{vc1,v2}
\fmf{photon,tension=1,right=#1}{vc1,vc1}
\fmffreeze
\fmfposition
\fmfipath{p[]}
\fmfiset{p1}{vpath(__v1,__vc1)}
\fmfiset{p2}{vpath(__vc1,__v2)}
\fmfiset{p3}{reverse vpath(__vc1,__vc1)}
\vacpol{vc1}{vc1}
\end{fmfchar*}}}}

\newcommand{\swftwosixD}[1]{%
\settoheight{\eqoff}{$\times$}%
\setlength{\eqoff}{0.5\eqoff}%
\addtolength{\eqoff}{-7.5\unitlength}%
\raisebox{\eqoff}{%
\fmfframe(1,0)(1,0){%
\begin{fmfchar*}(20,15)
\fmfleft{v1}
\fmfright{v2}
\fmffixed{(0.33w,0)}{vc1,vc2}
\fmffixed{(0.33w,0)}{vc2,vc3}
\fmffixed{(0,0.33w)}{vc2,vc}
\fmf{plain}{v1,vc1}
\fmf{plain}{vc1,vc2}
\fmf{plain}{vc2,vc3}
\fmf{plain}{vc3,v2}
\fmf{photon,tension=0.5,left=0.5}{vc1,vc}
\fmf{photon,tension=0.5,left=0.5}{vc,vc3}
\fmffreeze
\fmf{photon,tension=0.5}{vc2,vc}
\fmffreeze
\fmfposition
\fmfipath{p[]}
\fmfiset{p1}{vpath(__v1,__vc1)}
\fmfiset{p2}{vpath(__vc1,__vc2)}
\fmfiset{p3}{vpath(__vc2,__vc3)}
\fmfiset{p4}{vpath(__vc3,__v2)}
\fmfiset{p5}{vpath(__vc1,__vc)}
\fmfiset{p6}{vpath(__vc2,__vc)}
\fmfiset{p7}{vpath(__vc,__vc3)}
\fmfcmd{fill fullcircle scaled 15 shifted vloc(__vc) withcolor black ;}
\fmfiv{plain,label=$\textcolor{white}{#1}$,l.dist=0}{vloc(__vc)}
\end{fmfchar*}}}}

\newcommand{\superpot}[5][plain]{%
\fmf{#1,tension=1}{#2,vc}
\fmf{#1,tension=1}{vc,#3}
\fmf{#1,tension=1}{#4,vc}
\fmf{#1,tension=1}{vc,#5}
\fmffreeze
\fmfposition
\fmfipath{p[],pca}
\fmfipair{vm[],vo[],vi[]}
\fmfiset{p1}{vpath(__#2,__vc)}
\fmfiset{p2}{vpath(__vc,__#3)}
\fmfiset{p3}{vpath(__#4,__vc)}
\fmfiset{p4}{vpath(__vc,__#5)}
\svertex{vm1}{p1}
\dvertex{vo1}{vi1}{p1}
\svertex{vm2}{p2}
\dvertex{vi2}{vo2}{p2}
\svertex{vm3}{p3}
\dvertex{vo3}{vi3}{p3}
\svertex{vm4}{p4}
\dvertex{vi4}{vo4}{p4}
}


\newcommand{\cvert}[6]{%
\settoheight{\eqoff}{$\times$}%
\setlength{\eqoff}{0.5\eqoff}%
\addtolength{\eqoff}{-7\unitlength}%
\raisebox{\eqoff}{%
\fmfframe(1,1)(1,1){%
\begin{fmfchar*}(12,12)
\fmfright{v2,v3}
\fmfleft{v1}
\fmf{#1}{v1,vc1}
\fmf{#2}{vc1,v2}
\fmf{#3}{vc1,v3}
\fmffreeze
\fmfposition
\fmfipath{p[]}
\fmfiset{p1}{vpath(__v1,__vc1)}
\fmfiset{p2}{vpath(__vc1,__v2)}
\fmfiset{p3}{vpath(__vc1,__v3)}
\fmfis{phantom,ptext.clen=6,ptext.hout=3,ptext.oout=12,ptext.out=#4,ptext.sep=;}{p1}
\fmfis{phantom,label.side=right,ptext.clen=6,ptext.hin=-3,ptext.oin=10,ptext.in=#5,ptext.sep=;}{p2}
\fmfis{phantom,ptext.clen=6,ptext.hin=3,ptext.oin=10,ptext.in=#6,ptext.sep=;}{p3}
\end{fmfchar*}}}
}

\newcommand{\qvert}[8]{%
\settoheight{\eqoff}{$\times$}%
\setlength{\eqoff}{0.5\eqoff}%
\addtolength{\eqoff}{-7\unitlength}%
\raisebox{\eqoff}{%
\fmfframe(1,1)(1,1){%
\begin{fmfchar*}(12,12)
\fmfleft{v2,v1}
\fmfright{v3,v4}
\fmfforce{(0,h)}{v1}
\fmfforce{(0,0)}{v2}
\fmfforce{(w,0)}{v3}
\fmfforce{(w,h)}{v4}
\fmf{#1}{v1,vc1}
\fmf{#2}{v2,vc1}
\fmf{#3}{vc1,v3}
\fmf{#4}{vc1,v4}
\fmffreeze
\fmfposition
\fmfipath{p[]}
\fmfiset{p1}{vpath(__v1,__vc1)}
\fmfiset{p2}{vpath(__v2,__vc1)}
\fmfiset{p3}{vpath(__vc1,__v3)}
\fmfiset{p4}{vpath(__vc1,__v4)}
\fmfis{phantom,ptext.clen=6,ptext.hout=3,ptext.oout=12,ptext.out=#5,ptext.sep=;}{p1}
\fmfis{phantom,ptext.clen=6,ptext.hout=3,ptext.oout=12,ptext.out=#6,ptext.sep=;}{p2}
\fmfis{phantom,label.side=right,ptext.clen=6,ptext.hin=-3,ptext.oin=12,ptext.in=#7,ptext.sep=;}{p3}
\fmfis{phantom,ptext.clen=6,ptext.hin=3,ptext.oin=12,ptext.in=#8,ptext.sep=;}{p4}
\end{fmfchar*}}}
}


\newpage
\setcounter{page}{1}
\renewcommand{\thefootnote}{\arabic{footnote}}
\setcounter{footnote}{0}


\section{Introduction}

The ABJM model is an $\NN=6$ supersymmetric $U(N)\times U(N)$ Chern-Simons theory with opposite levels coupled to matter \cite{Aharony:2008ug}.  Like its cousin $\NN=4$ super Yang-Mills in four
dimensions, its two point functions of single trace operators map to an integrable system in the planar limit \cite{Minahan:2008hf,Gaiotto:2008cg,Bak:2008cp,Gromov:2008qe}.   For $\NN=4$ SYM, the integrability has been used as a powerful tool to  interpolate between strong and weak coupling, where one can see the perturbative behavior of the gauge theory morph into the stringy behavior expected from the AdS/CFT conjecture \cite{Gromov:2009zb,Frolov:2010wt}.

The ABJM model has two extra features that give it a richer structure than $\NN=4$ SYM, at least as far as the integrability of the two point functions is concerned.  The first is that the Bethe equations and the dispersion relations contain an undetermined function $h^2(\la)$ of the 't Hooft coupling, $\la=N/k$, where $k$ is the Chern-Simons level \cite{Gromov:2008qe}.  The second is that the theory can be deformed into a $U(M)\times U(N)$ gauge theory while still maintaining the $\NN=6$ supersymmetry \cite{Aharony:2008gk}.  In this ABJ case there are now two 't~Hooft parameters, 
\begin{equation}
\lambda=\frac{M}{k}\col\qquad\hat\lambda=\frac{N}{k}\col
\end{equation}
and, if integrability is maintained, a single function $h^2(\bar\la,\s)$, where
\begin{equation}\label{barlambdasigmadef}
\bar\lambda=\sqrt{\lambda\hat\lambda}\col\qquad
\sigma=\frac{\lambda-\hat\lambda}{\bar\lambda}\pnt
\end{equation}

The spin-chain that appears in the 
ABJ(M)
models has $OSp(6|4)$ symmetry and 
is of alternating type, with the spins on the odd sites in the 
singleton
representation of the supergroup and the spins on the even sites in the 
anti-singleton
representation \cite{Minahan:2008hf,Gaiotto:2008cg,Bak:2008cp,Zwiebel:2009vb,Minahan:2009te}.  In order to find $h^2(\bla,\s)$ it is only necessary to consider the compact subgroup $SU(2)\times SU(2)$ of $OSp(6|4)$, with the spins on the odd sites transforming in the $(\mathbf{2},\mathbf{1})$ representation and the spins on the even sites transforming in the $(\mathbf{1},\mathbf{2})$ representation.  
The ground state has all spins aligned and the excitations (or magnons)  are  flipped spins that live on either odd or even sites.   
The dispersion relations for these two types of magnons are given by

\begin{equation}
\begin{aligned}\label{Eoddeven}
E_\text{odd}(p)=\sqrt{Q^2+4h^2(\bar\lambda,\sigma)\sin^2\tfrac{p}{2}}-Q
\col\qquad
E_\text{even}(p)=E_\text{odd}(p)\big|_{\sigma\to-\sigma}\,,
\end{aligned}
\end{equation}
where $Q=1/2$ for fundamental magnons while larger values of $Q$ correspond to magnon bound states.

At weak coupling the function $h^2(\bla,\s)$ can be computed perturbatively.  The leading contribution appears at two-loop order and is relatively easy to compute, both for ABJM \cite{Minahan:2008hf,Gaiotto:2008cg,Bak:2008cp}, and ABJ \cite{Bak:2008vd,Minahan:2009te}, where one finds
\be
h^2(\bla,\s)=\bla^2+\mathcal{O}(\bla^4)\,.
\ee
However, at strong coupling on the ABJM slice where $\s=0$,  one readily finds from the string sigma model 
\cite{Nishioka:2008gz,Gaiotto:2008cg,Grignani:2008is}.
\be
h^2(\bla,0)=\frac{1}{2}\, \bla +\mathcal{O}(1)\,.
\ee
Hence, $h^2(\bar{\la},\s)$
is an interpolating function and can be expected to have corrections at every even order of perturbation theory, with a general structure
\begin{equation}\label{h4expansion}
\begin{aligned}
h^2(\bar\lambda,\sigma)=\bar\lambda^2+\sum_{n=2}^\infty\bar\lambda^{2n}h_{2n}(\sigma)\pnt
\end{aligned}
\end{equation}

The 
four-loop
term in (\ref{h4expansion})  was computed in \cite{Minahan:2009aq,Minahan:2009wg}, where it was found that\footnote{\label{meaculpa}A different result for $h_4(\s)$ in (\ref{h4res}) was given in earlier versions of \cite{Minahan:2009aq,Minahan:2009wg}.  After it became clear that those results were in conflict with the results presented in this paper, an overall sign error was discovered for three of the Feynman graphs.}
\be\label{h4res}
h_4(\s)=-(4+\s^2)\zeta(2)\,.
\ee
This calculation was done using the explicit component action and involved the computation of dozens of Feynman diagrams.   A straightforward extension of the methods in  \cite{Minahan:2009aq,Minahan:2009wg}  to higher loops would lead to a mind boggling number of diagrams.  Moreover, one would like to verify (or disprove) that the ABJ theory is integrable, even at the 
four-loop order.   The $SU(2)\times SU(2)$ sector is trivially integrable at four loops, so it would be necessary to go beyond this sector to find a nontrivial check of integrability at this order.  But even this seemingly modest task is extremely daunting in component language.

In this paper we compute $h_4(\s)$ in (\ref{h4res}) using the superspace formalism.  Superspace techniques have proven to be very effective in computing 
the dilatation operator \cite{Sieg:2010tz} and in evaluating
wrapping corrections 
\cite{Serban:2004jf,Sieg:2005kd}
in $\NN=4$ SYM \cite{Fiamberti:2008sh,Fiamberti:2009jw} 
and  in 
its $\beta$-deformation
\cite{Mauri:2005pa,Fiamberti:2008sm,Fiamberti:2008sn}.
Naturally, one would also like  to apply them to the 
 ABJ(M) models.
 Their main virtue is that they drastically reduce the number of Feynman diagrams that one must compute.  
 We will later summarize several
 restrictions on the allowed diagrams \cite{Sieg:2010tz}
that greatly limit the number that can contribute to  $h_4(\s)$.
As we will see in this paper, at the two-loop order there is only one diagram in superspace that contributes to $h^2(\bla,\s)$.  At the 
four-loop
 order there are 15 (plus reflections of some of the diagrams).  Contrast this to the component calculation in \cite{Minahan:2009aq,Minahan:2009wg}, where one has many times more diagrams.  Not only does this demonstrate the formalism's power, but it is also crucial in verifying that (\ref{h4res}) is actually correct (see footnote \ref{meaculpa}).

One can also see from (\ref{h4res}) that $h_4(\s)$ has uniform 
transcendentality two.
From the component point of view this seems almost miraculous since many diagrams have rational coefficients (that is, they have  
transcendentality zero),
others have 
transcendentality two,
and some are mixed.  When everything is combined one finds that the rational coefficients cancel.  In superspace, while there are still diagrams with rational coefficients, their cancellation appears more natural 
due to correlations between the single and double poles.

We will also present two possible scenarios for an all-loop function for $h^2(\la)$, including one that might work.  It reproduces the first two orders of perturbation theory as well as the leading sigma-model contribution at strong-coupling. The one-loop sigma-model contribution to $h^2(\lambda)$ depends on how a sum is carried out over an infinite number of modes.  Our proposal disagrees with the more conventional prescription in \cite{McLoughlin:2008he}, but agrees with the prescription in \cite{Gromov:2008fy}.
  The other proposal looks for a connection with matrix models on a Lens space.  These arise in the study of supersymmetric Wilson loops in 
  ABJ(M)
  models \cite{Kapustin:2009kz,Drukker:2009hy,Marino:2009jd,Drukker:2010nc}.   In particular, we consider the free energy of the matrix model which is a function of $\la$.  
  We will see  that $h^2(\lambda)$ has a structure similar to the derivative of  the  matrix model free energy, both at small and large $\lambda$.  But the coefficients in their respective expansions do not quite line up.

In order to complete the four-loop analysis in the $SU(2)\times SU(2)$ subsector, we will apply the superspace formalism to compute the 
leading
 wrapping corrections for a length 
 four
 operator in the {\bf (1,1)} representation of $SU(2)\times SU(2)$.
Here we find that the wrapping corrections {\it per se} differ from those computed in component language.  However, other range five
interactions must be subtracted and this subtracted piece also differs from the corresponding term in the component calculation.  The two effects combine to give the same 
four-loop
anomalous dimension for this operator as was found using components. 

The rest of the paper is organized as follows:  In section \ref{sec:ABJMsuper} we review the 
ABJ(M)
models in $\NN=2$ superspace.  In section \ref{Dilatation-OP} we discuss the relation of the 
dilatation operator to $h^2(\bla,\s)$.  In section \ref{FComputation} we  enumerate and compute all Feynman diagrams that contribute to 
the four-loop term $h_4(\s)$.
   In section \ref{sec:hallorder} we discuss our investigation into possible all-loop functions for $h^2(\la)$.  In section \ref{sec:wrapping} we apply the superspace formalism to the wrapping corrections for operators of length 
   four.
     In section \ref{sec:concl} we present our conclusions, which includes suggestions for further work.  Many further details, 
      including the four-loop decoupling of odd and even site magnons and the consistency of double poles due to UV subdivergences
  can be found in the several appendices.

\section{ABJ(M) models in $\mathcal{N}=2$ superspace}
\label{sec:ABJMsuper}

In this section we review the $\NN=2$ superspace formulation for $\mathcal{N}=6$ superconformal 
Chern-Simons theory.   This 
was first given in \cite{Benna:2008zy}, but 
in this paper we follow the notations used in \cite{Leoni:2010az}
which are adapted from the ones of \cite{Gates:1983nr}. 
For the first papers on the $\NN=2$ superspace formulation of Chern-Simons 
 theory coupled to matter see
  \cite{Zupnik:1988en,Ivanov:1991fn,Gates:1991qn,Nishino:1991sr}.
Appendix
 \ref{app:conventions} collects our notation and conventions. 

The $U(M)\times U(N)$ supersymmetric Chern-Simons 
theory
 has 
two $\NN=2$ vector supermultiplets, $V$ and $\hat V$, 
with $V$ transforming  in the adjoint of $U(M)$ and $\hat V$ in the adjoint of $U(N)$. 
In order to extend the supersymmetry to $\NN=6$, the ABJ(M) action also contains two sets of chiral matter superfields, 
$Z^A$ and $W_A$ with $A=1,2$. 
$Z^A$ and $W_A$ transform respectively in the 
$({\bf 2}, {\bf 1})$ and $({\bf 1}, {\bf 2})$
of the global $SU(2)\times SU(2)$ 
flavour subgroup
described in the introduction. 
Moreover, they transform in the bifundamental representations $({\bf M}, {\bf \bar{N}})$ and
$({\bf N}, {\bf \bar{M}})$ of the  $U(M)\times U(N)$ gauge group.

Each gauge group in the gauge fixed ${\cal N}=2$ superspace action 
has associated with it a pair of 
chiral ghost superfields, $c,c'$ for $U(M)$ and
$\hat{c},\hat{c}'$ for $U(N)$ 
\cite{Avdeev:1991za,Avdeev:1992jt,Bianchi:2009ja}. 
Including all of these ingredients, the gauge fixed ABJ(M) action in 
${\cal N}=2$
 superspace reads
\begin{equation}
\begin{aligned}
S_\CS+S_\gf
&=\frac{k}{4\pi}\Big[\int\de^3x\de^4\theta\int_0^1\de t\tr V\Big(\barD^\alpha\e^{-tV}\D_\alpha\e^{tV}+\frac{1}{2}\Big(\frac{1}{\alpha}\D^2+\frac{1}{\bar\alpha}\barD^2\Big)V\Big)\\
&\phantom{{}={}\frac{k}{4\pi}\Big[}
-\int\de^3x\de^4\theta\int_0^1\de t\tr\hat V\Big(\barD^\alpha\e^{-t\hat V}\D_\alpha\e^{t\hat V}+\frac{1}{2}\Big(\frac{1}{\hat\alpha}\D^2+\frac{1}{\hat{\bar\alpha}}\barD^2\Big)\hat V\Big)\Big]
\col\\
S_{\FP}&=\frac{k}{4\pi}\Big[\int\de^3x\de^4\theta\tr(c'+\bar c')\Ld_{\frac{1}{2}V}(c+\bar c+\coth\Ld_{\frac{1}{2}V}(c-\bar c))\\
&\phantom{{}={}\frac{k}{4\pi}\Big[}
-\int\de^3x\de^4\theta\tr(\hat c'+\hat{\bar c}')\Ld_{\frac{1}{2}\hat V}(\hat c+\hat{\bar c}+\coth\Ld_{\frac{1}{2}\hat V}(\hat c-\hat{\bar c}))
\Big]\nonumber
\end{aligned}
\end{equation}
\begin{equation}
\begin{aligned}
S_{\text{mat}}&=\frac{k}{4\pi}\int\de^3x\de^4\theta\tr\big(\bar Z_A\e^VZ^A\e^{-\hat V}+\bar W^B\e^{\hat V}W_B\e^{-V}\big)
\col\\
S_{\text{pot}}&=\frac{k}{4\pi}\frac{i}{2}\Big[
\int\de^3x\de^2\theta
\epsilon_{AC}\epsilon^{BD}\tr Z^AW_BZ^CW_D\\
&\phantom{{}={}\frac{k}{4\pi}\frac{1}{2}\Big[}
+\int\de^3x\de^2\bar\theta
\epsilon^{AC}\epsilon_{BD}
\tr\bar Z_A\bar W^B\bar Z_C\bar W^D\Big]
\col
\end{aligned}
\end{equation}
where $\Ld_VX=\comm{V}{X}$ and $\alpha$ and $\hat\alpha$ are gauge fixing parameters.  

Many of the terms in this action have an infinite expansion, but for our purposes it is only necessary to retain the first few orders of any expansion. 
The first term in the Chern-Simons Lagrangian expands to
 \begin{equation}
\begin{aligned}
\int_0^1\de t\tr V\barD^\alpha\e^{-tV}\D_\alpha\e^{tV}
=
\frac{1}{2}\tr V\barD^\alpha\D_\alpha V
-\frac{1}{6}\tr V\barD^\alpha\comm{V}{\D_\alpha V}
+\dots
\pnt
\end{aligned}
\end{equation}
The quadratic piece in this expression, 
together with the $\alpha$- and $\hat{\alpha}$-dependent gauge fixing terms,
determines the gauge 
superfield propagators.
In order to simplify the $\D$-algebra manipulations we will choose the Landau 
gauge where $\alpha=\hat\alpha=0$.
The leading expansion for the Fadeev-Popov action is
\begin{equation}
\begin{aligned}
S_{\FP}
&=
\frac{k}{4\pi}\int\de^3x\de^4\theta\tr\Big(
\bar c'c-c'\bar c
+\frac{1}{2}(c'+\bar c')\comm{V}{c+\bar c}
\Big)
+\dots
\col\\
\end{aligned}
\end{equation}
while 
the leading expansion for the matter action $D$-terms is
\begin{equation}
\begin{aligned}
S_{\rm mat}&=\frac{k}{4\pi}
\int\de^3x\de^4\theta\Big[\tr\bar Z_A\Big(Z^A+VZ^A-Z^A\hat V+\frac{1}{2}(V^2Z^A+Z^A\hat V^2)-VZ^A\hat V\Big)\
+\dots\\
&\phantom{{}={}\frac{k}{4\pi}\int\de^3x\de^4\theta\Big[}
+\tr\bar W^A\Big(W_A+\hat VW_A-W_A V+\frac{1}{2}(\hat V^2W_A+W_A V^2)-\hat VW_A V\Big)\\
&\phantom{{}={}\frac{k}{4\pi}\int\de^3x\de^4\theta\Big[}
+\dots\Big]
\,.\\
\end{aligned}
\end{equation}

We have collected the Feynman rules which follow from the action and the above expansions  in 
appendix \ref{app:Feynmanrules}.
The supergraphs are then constructed from the Feynman rules
and are reducible   to ordinary 
 integrals using standard D-algebra techniques \cite{Gates:1983nr}. 
 
 The advantage of using superspace as opposed to the component approach is that the number of diagrams is significantly smaller.  Furthermore, one can often find cancellation patterns between different supergraphs or demonstrate finiteness theorems for classes of diagrams 
\cite{Fiamberti:2008sh,Sieg:2010tz}.
Such generalized finiteness conditions \cite{Sieg:2010tz}
that follow from power counting arguments and some of their 
implications are summarized in section \ref{sec:finitenesscond}. They
predict the finiteness of many diagrams and will be of great use to us in 
our calculations.

\section{The dilatation operator and $h^2({\bar \lambda},\sigma)$}
\label{Dilatation-OP}

The dilatation operator ${\cal D}$ is the natural tool to study the anomalous dimensions of 
composite operators in field theory.
It can be defined as the operator that by acting on composite operators 
${\cal O}_a$ 
provides the  matrix of 
scaling
dimensions
\begin{equation}
{\cal D}{\cal O}_a=\Delta_a{}^b({\cal O}){\cal O}_b
\pnt
\end{equation}
Note that $\Delta_a{}^b$ leads in general to the mixing between operators.
As known, the matrix of 
dimensions, and therefore the dilatation operator, can be 
extracted from the perturbative renormalization of the composite operators
$\mathcal{O}_a$
\begin{equation}\label{opren}
\mathcal{O}_{a,\text{ren}}=\mathcal{Z}_{a}{}^b\mathcal{O}_{b,\text{bare}}
\col\qquad
\mathcal{Z}=\unitmatrix+\bar\lambda^2\mathcal{Z}_2+\bar\lambda^4\mathcal{Z}_4+\dots
\pnt
\end{equation}
The matrix $\mathcal{Z}$ is such that $\mathcal{O}_{a,\text{ren}}$
is free from perturbative quantum divergences and can be computed in perturbation theory
by means of standard methods.
In this paper we use dimensional reduction with the space-time dimension $D$ given by
\begin{equation}\label{Ddef}
D=3-2\varepsilon\col
\end{equation}
in order to regularize quantum divergences that show up
as inverse powers of $\varepsilon$ in the limit  $\varepsilon\to0$. 
By introducing the 't Hooft mass $\mu$ and the 
dimensionful
combination
$\bar\lambda\mu^{2\varepsilon}$ 
the dilatation operator is then extracted from $\mathcal{Z}$ as
\begin{equation}
\label{DinZ2}
\mathcal{D}=
\mathcal{D}_{{\rm classical}}
+\mu\frac{\de}{\de\mu}\ln\mathcal{Z}(\bar\lambda\mu^{2\varepsilon},\varepsilon)
=\mathcal{D}_{{\rm classical}}+\lim_{\varepsilon\rightarrow0}\left[2\varepsilon\bar\lambda
\frac{\de}{\de\bar\lambda}\ln\mathcal{Z}(\bar\lambda,\varepsilon)\right]\pnt
\end{equation}
In a loop expansion of the dilatation operator, the $l$th loop
order is then simply given by the $\bar{\lambda}^{2l}$
 coefficient of the 
$1/\varepsilon$ pole of $\ln\mathcal{Z}$ multiplied by $2l$.
The higher order poles must be absent in  $\ln\mathcal{Z}$; this will be later
used as a consistency check for our result.

As discussed in the introduction, in the ABJ(M) models
the dilatation operator can be 
mapped to the long range Hamiltonian of a spin-chain system for the whole 
$OSp(6|4)$ symmetry group \cite{Minahan:2008hf,Minahan:2009te}.
We focus on the $SU(2)\times SU(2)$ subsector where
the magnons propagating along the spin chain form two sectors: the ones living  on the odd sites belong to the first $SU(2)$, while those on the even sites are associated with
 the other $SU(2)$.
As demonstrated in appendix \ref{app:Z4mixed},
in our four-loop analysis the two different types of magnons can 
be regarded as non-interacting, 
since the contributions to the dilatation
operator of the respective diagrams that could lead to these interactions 
cancel. The all-loop Bethe Ansatz \cite{Gromov:2008qe} predicts that 
such interactions start at eight loops.
In analogy with the ${\cal N}=4$ case,
the spin-chain is interpreted as a quantum mechanical system in 
which the ground state 
of length $2L$ can be chosen to be
\begin{equation}
\Omega=\tr{(W_1Z^1)^L}
\pnt
\end{equation}
With a single excitation $W_2$ of an odd site
the momentum eigenstate is defined as
\begin{equation}\label{onemagnonstate}
\psi_p=\sum_{k=0}^{L-1}
e^{ipk}
(W_1Z^1)^kW_2Z^1(W_1Z^1)^{L-k-1}
\end{equation}
This describes a single magnon excitation with momentum $p$.
The main difference between the ${\cal N}=6$ CS and the ${\cal N}=4$ SYM case
is the existence in the former of two different $SU(2)$ 
excitations corresponding to the sectors mentioned above.

Up to four loops, the dilatation operator for a chain of lenght $2L$ 
then  expands as
\begin{equation}\label{Ddecomp}
{\cal D}=L
+\bar\lambda^2({\cal D}_{2,\text{odd}}
+{\cal D}_{2,\text{even}})
+\bar\lambda^4({\cal D}_{4,\text{odd}}(\sigma)
+{\cal D}_{4,\text{even}}(\sigma))
+\mathcal{O}(\bar\lambda^6)
\col
\end{equation}
where the individual parts act non-trivially on odd
and even sites only.

In the $\mathcal{N}=4$ SYM case chiral functions have been introduced in 
\cite{Fiamberti:2008sh} as a very convenient basis for the dilatation operator
of the $SU(2)$ subsector. 
The chiral functions directly
capture the structure of the chiral superfields in the
Feynman diagrams. As in the $\mathcal{N}=4$ SYM case, also in
the $\mathcal{N}=6$ CS case the elementary building block for the 
chiral function of the $SU(2)\times SU(2)$ subsector 
is constructed from the superpotential by contracting one chiral and 
one anti-chiral vertex with a single chiral propagator. The resulting 
flavour structure then yields the simplest non-trivial chiral function.

The chiral functions that are relevant 
to two loops in $\mathcal{N}=4$ SYM and to 
four loops in $\mathcal{N}=6$ CS theory turn out to have identical form 
in terms of the respective permutation structures and read
\begin{equation}\label{chifunc}
\begin{aligned}
\chi(a,b)&=\ptwo{a}{b}-\pone{a}-\pone{b}+\pone{}
\col\\
\chi(a)&=\pone{a}-\pone{}
\col\\
\chi()&=\pone{}
\pnt
\end{aligned}
\end{equation}
However, the permutation structures in both theories slightly differ. 
In the $\mathcal{N}=6$ CS case they are given by
\cite{Minahan:2009wg}
\begin{equation}\label{permstruc}
\begin{aligned}
\pfour{a_1}{a_2}{\dots}{a_m}=
\sum_{i=0}^{L-1}
\perm_{2i+a_1\,2i+a_1+2}\perm_{2i+a_2\,2i+a_2+2}\dots\perm_{2i+a_m\,2i+a_m+2}\col
\end{aligned}
\end{equation}
where we identify $L+i\equiv i$,
such that the product of permutations, in which $\perm_{a\,a+2}$ permutes 
the flavours at sites $a$ and $a+2$,
is inserted at every second site
of the cyclic spin chain of length 
$2L$.\footnote{Note that the permutation structures 
obey
\begin{equation}
\begin{aligned}\label{permrules}
\pfour{\dots}{a}{b}{\dots}&=\pfour{\dots}{b}{a}{\dots}\col\qquad
|a-b|\neq 2\col \\
\pthree{a}{\dots}{b}&=\pthree{a+2n}{\dots}{b+2n}\pnt
\end{aligned}
\end{equation}}
The insertion at each second site thereby allows for the decomposition
of the dilatation operator into two separate pieces acting 
only on odd or even sites as in \eqref{Ddecomp}.
The decomposition of the dilatation operator to four loops
\cite{Minahan:2009wg} in terms of chiral functions 
then reads
\begin{equation}\label{D4}
\begin{aligned}
{\cal D}_{2,\text{odd}}&=-\chi(1)\col\\
{\cal D}_{2,\text{even}}&=-\chi(2)\col\\
{\cal D}_{4,\text{odd}}(\sigma)
&=-\chi(1,3)-\chi(3,1)+(2-h_4(\sigma))\chi(1)
\col\\
{\cal D}_{4,\text{even}}(\sigma)
&=-\chi(2,4)-\chi(4,2)+(2-h_4(-\sigma))\chi(2)
\pnt
\end{aligned}
\end{equation}
The coefficients are thereby fixed by the magnon dispersion relation
\eqref{Eoddeven} in terms of the four-loop contribution
$h_4(\sigma)$ of the a priori undetermined function 
$h^2(\bar\lambda,\sigma)$ in \eqref{h4expansion}.
As explained in \cite{Minahan:2009wg} to obtain the above result, one 
just has to compare the expansion of the magnon dispersion relation
to the momentum dependence when the individual terms are applied to the 
single magnon momentum eigenstate \eqref{onemagnonstate}.

The function $h_4(\sigma)$ can be computed in the weak 
coupling limit from a direct perturbative calculation. 
This has been done by using component fields techniques 
in \cite{Minahan:2009wg}. 
Here we present its calculation 
by using ${\cal N}=2$ supergraphs. 
As in the component calculation \cite{Minahan:2009wg}, also here it 
suffices to only consider the odd part of the dilatation operator, i.e.\
the contributions with chiral functions that have odd integers as 
arguments.\footnote{As we mentioned before, odd and even site magnons are decoupled here, there is therefore no contribution with chiral functions with both odd and even integer arguments. We explicitly demonstrate their absence at four loops in appendix \ref{app:Z4mixed}.} 
The supergraphs computation of the full
${\cal D}_{4,{\rm odd}}$, and in particular of $h_4(\sigma)$,
is the main result of our paper.

\section{Feynman diagram calculation}
\label{FComputation}

Before starting with the explicit evaluation of Feynman diagrams we 
will summarize 
the previously mentioned finiteness conditions 
which allow us to disregard entire classes of diagrams.

\subsection{Finiteness conditions}
\label{sec:finitenesscond}

Based on power counting and structural properties of the Feynman rules, 
in \cite{Sieg:2010tz} finiteness conditions for Feynman diagrams of 
$\mathcal{N}=4$ SYM theory in terms of $\mathcal{N}=1$ superfields and for
$\mathcal{N}=6$ CS theory in terms of $\mathcal{N}=2$ superfields were 
derived. They hold for each diagram that contributes to the renormalization of 
chiral operators in the respective $SU(2)$ or 
$SU(2)\times SU(2)$ subsectors.
In Landau
 gauge, such a diagram with interaction range $R\ge2$ has no 
overall UV divergence,
if at least one of the following criteria is matched:\footnote{$R\ge 2$ means, the composite operator is 1PI connected with the rest of the diagram, not including the non-interacting fields of the operator.}
\begin{enumerate}
\item All of its chiral vertices are part of any loop.
\item One of its spinor derivative $\D_\alpha$ is brought outside the loops.
\item The number of its spinor derivatives $\barD_\alpha$ brought outside 
loops becomes equal or bigger than twice the number of chiral vertices 
that are not part of any loop.
\end{enumerate}
In the flavour $SU(2)\times SU(2)$ subsector, a chiral vertex that 
is not part of any loop always generates flavour permutations and therefore a
non-trivial chiral structure of the diagram.
Analogously to the $\mathcal{N}=4$ SYM case,
the above finiteness conditions hence imply the following rule:
\begin{itemize}
\item 
All diagrams with interaction range $R\ge2$ and trivial chiral structure
$\chi()$ are finite.
\end{itemize}
Together with the conformal invariance on the quantum level, i.e.\ 
the finiteness of the chiral self energy, this implies that 
any diagram which does not manipulate the flavour, i.e.\ it has
trivial chiral structure $\chi()$ defined in 
\eqref{chifunc}, has no overall UV divergence.

Since the propagators of the vector fields in  Landau
 gauge carry $\D\barD$,
the finiteness conditions imply the following statement:
\begin{itemize}
\item
A diagram with interaction range $R\ge2$ has no overall UV divergence,
if it contains at least one cubic gauge-matter interaction with a  
chiral field line which is not part of any loop. In particular,
if in the diagram exactly one of the chiral vertices appears outside
the loops, then it also has no overall UV divergence 
if the anti-chiral field of at least one 
cubic gauge-matter interaction is not part of any loop.
\end{itemize}
According to this statement, there are no contributions to the dilatation 
operator that come from diagrams in which the chiral line of a 
cubic gauge-matter vertex is an external line.
In appendix \ref{IR_cancellation} we will, however, evaluate such diagrams
with IR divergences explicitly to show that indeed all IR divergences
cancel out in the renormalization constant $\mathcal{Z}$ in \eqref{opren}.

\subsection{Two loops}

Before attacking the more involved four-loop case, let us see how the two-loop result 
is obtained
from supergraphs.
There is only one non-vanishing logarithmically divergent diagram contributing.
It evaluates to
\begin{equation}
\begin{aligned}
\settoheight{\eqoff}{$\times$}%
\setlength{\eqoff}{0.5\eqoff}%
\addtolength{\eqoff}{-8.5\unitlength}%
\raisebox{\eqoff}{%
\fmfframe(0,1)(-5,1){%
\begin{fmfchar*}(20,15)
\fmftop{v3}
\fmfbottom{v4}
\fmfforce{(0.125w,h)}{v3}
\fmfforce{(0.125w,0)}{v4}
\fmffixed{(0.25w,0)}{v2,v1}
\fmffixed{(0.25w,0)}{v3,v2}
\fmffixed{(0.25w,0)}{v4,v5}
\fmffixed{(0.25w,0)}{v5,v6}
\chioneca{v1}{v2}{v3}{v4}{v5}{v6}
\fmf{plain,tension=1,left=0,width=1mm}{v4,v6}
\end{fmfchar*}}}
&\to\frac{(4\pi)^2}{k^2}MN\, I_2\, \chi(1)
=\frac{\lambda\hat\lambda}{4}\frac{1}{\varepsilon}\chi(1)
\col
\label{two-loop}
\end{aligned}
\end{equation}
where the two-loop integral $I_2$ is given in (\ref{I2}).
As already discussed, to obtain  the contribution to the dilatation operator
one has to take the coefficient of the pole $1/\varepsilon$ and multiply it by
$-2l$, in this case equal  to $-4$.
Once a factor 
$\bar\lambda^2=\lambda\hat\lambda$ is removed one gets 
\begin{equation}
\begin{aligned}
\mathcal{D}_2=-\chi(1)
\pnt
\end{aligned}
\end{equation}
This coincides with the results found in 
\cite{Minahan:2008hf,Bak:2008cp,Bak:2008vd} in components.

\subsection{Four loops}

Now, let us move to the four-loop contributions to the dilatation operator. 
We will separate them according to the range of the 
interactions.
We will explicitly present
only the diagrams surviving the finiteness 
conditions
of \cite{Sieg:2010tz} that are summarized in section 
\ref{sec:finitenesscond}.
It is important to note that, according to these arguments, 
an 
overall
 UV divergence can be present  
in superficially logarithmically divergent 
diagrams if at least one purely chiral vertex remains outside the loops.
This implies that the minimum range of interaction at any loop is three.
This is consistent with the fact that the minimal structure that 
can appear in the dilatation operator is $\chi(1)$.
The range varies between three and the maximum one which at 
four loops is five.

Note that
together with the $1/\varepsilon$ poles
we will also keep the 
higher order poles 
that display the presence of subdivergences.
Here, to four-loop order the only appearing higher order poles are double 
poles.  In 
appendix
 \ref{DoublePolesCancellation} their cancellation in
 $\ln{\cal Z}$ will be explicitly demonstrated as an important consistency 
check of our calculation.

We note that, for the convenience of the reader, all the integrals appearing in the 
following are collected in the appendix \ref{Integrals}.

\subsubsection{Range five interactions}

At four loops there is only one supergraph that involves the maximum number 
of five neighbouring fields in the interaction. It is given by
\begin{equation}
\begin{aligned}
S_{\text{r}5}=
\settoheight{\eqoff}{$\times$}%
\setlength{\eqoff}{0.5\eqoff}%
\addtolength{\eqoff}{-8.5\unitlength}%
\raisebox{\eqoff}{%
\fmfframe(-1,1)(4,1){%
\begin{fmfchar*}(20,15)
\fmftop{v1}
\fmfbottom{v6}
\fmfforce{(0.125w,h)}{v1}
\fmfforce{(0.125w,0)}{v6}
\fmffixed{(0.25w,0)}{v1,v2}
\fmffixed{(0.25w,0)}{v2,v3}
\fmffixed{(0.25w,0)}{v3,v4}
\fmffixed{(0.25w,0)}{v4,v5}
\fmffixed{(0.25w,0)}{v6,v7}
\fmffixed{(0.25w,0)}{v7,v8}
\fmffixed{(0.25w,0)}{v8,v9}
\fmffixed{(0.25w,0)}{v9,v10}
\fmffixed{(0,0)}{v10,vr}
\fmffixed{(0.25w,0)}{v6,vl}
\fmf{plain}{v7,v4la}
\fmf{plain,tension=0,right=0.25}{v4la,v6}
\fmf{plain}{v4la,v4lc}
\fmf{plain,tension=0,left=0.25}{v4lc,v1}
\fmf{plain,left=0}{v4lc,v2}
\fmf{plain,tension=0,right=0.25}{v4lc,v3}
\fmf{plain,tension=0,left=0.25}{v8,v4ra}
\fmf{plain}{v9,v4ra}
\fmf{plain,tension=0,right=0.25}{v10,v4ra}
\fmf{plain}{v4ra,v4rc}
\fmf{plain,left=0}{v4rc,v4}
\fmf{plain,tension=0,left=0.25}{v5,v4rc}
\fmffreeze
\fmfposition
\fmf{plain,tension=1,left=0,width=1mm}{v6,v10}
\fmfi{plain}{vloc(__v4la){dir 0}..{dir 0}vloc(__v4rc)}
\end{fmfchar*}}}
&\to
\frac{(4\pi)^4}{k^4}(MN)^2I_4\chi(1,3)
=\frac{(\lambda\hat\lambda)^2}{16}\Big(-\frac{1}{2\varepsilon^2}+\frac{2}{\varepsilon}\Big)\chi(1,3)
\pnt
\end{aligned}
\end{equation}
By taking into account the reflected diagram,
the maximum range contribution to the renormalization constant is\footnote{By ${\cal R}$
we indicate the reflection of a supergraph at the vertical axis. As in 
\cite{Minahan:2009wg}, the operation preserves the type of chiral 
function, i.e.\ if it belongs to the odd or even sector.
In case of an even number of neighbours interacting with 
each other the operation therefore 
involves a shift of the interaction by one site along the 
composite operator. Effectively, ${\cal R}$ therefore exchanges
$\lambda$ with $\hat{\lambda}$ and $\chi(a,b)$ with $\chi(b,a)$.}
\begin{equation}
\begin{aligned}\label{Zr5}
\mathcal{Z}_{\text{r}5,
\text{odd}
}
=-(1+{\cal{R}})S_{\text{r}5}
=\frac{(\lambda\hat\lambda)^2}{16}\Big(\frac{1}{2\varepsilon^2}-\frac{2}{\varepsilon}\Big)(\chi(1,3)+\chi(3,1))
\pnt
\end{aligned}
\end{equation}

\subsubsection{Range four interactions}

There are four diagrams which
have range four interactions and contribute to 
the structure $\chi(1)$ in the dilatation operator. 
According to section \ref{sec:finitenesscond}, for an overall
UV divergence to be present,
at least one purely chiral vertex has to remain outside the loops,
and a single gauge propagator can not end up on an external leg.
Therefore, the only relevant contributions turn out to be
\begin{equation}
\begin{aligned}
S_{\text{r}4}=
\settoheight{\eqoff}{$\times$}%
\setlength{\eqoff}{0.5\eqoff}%
\addtolength{\eqoff}{-8.5\unitlength}%
\raisebox{\eqoff}{%
\fmfframe(0,1)(0,1){%
\begin{fmfchar*}(20,15)
\fmftop{v1}
\fmfbottom{v5}
\fmfforce{(0.125w,h)}{v1}
\fmfforce{(0.125w,0)}{v5}
\fmffixed{(0.25w,0)}{v1,v2}
\fmffixed{(0.25w,0)}{v2,v3}
\fmffixed{(0.25w,0)}{v3,v4}
\fmffixed{(0.25w,0)}{v5,v6}
\fmffixed{(0.25w,0)}{v6,v7}
\fmffixed{(0.25w,0)}{v7,v8}
\fmffixed{(0,whatever)}{v2,v4lc}
\fmffixed{(0,whatever)}{v4lc,v4la}
\fmffixed{(0,whatever)}{v7,v4rc}
\fmffixed{(0,whatever)}{v4rc,v4ra}
\fmffixed{(whatever,0)}{v4la,v4ra}
\fmf{plain,left=0.25}{v5,v4la}
\fmf{plain}{v4la,v4lc}
\fmf{plain,tension=0,left=0.25}{v4lc,v1}
\fmf{plain,left=0}{v4lc,v2}
\fmf{plain,tension=0,right=0.25}{v4lc,v3}
\fmf{plain,tension=0}{v6,v4la}
\fmf{plain}{v7,v4ra}
\fmf{plain,tension=0,right=0.25}{v8,v4ra}
\fmf{plain,tension=0.66,left=0.5}{v4ra,v4rc}
\fmf{plain,tension=0.66,left=0.5}{v4rc,v4ra}
\fmf{plain,right=0.25}{v4rc,v4}
\fmffreeze
\fmfposition
\fmf{plain}{v4rc,v4la}
\fmf{plain,tension=1,left=0,width=1mm}{v5,v8}
\end{fmfchar*}}}
&\to
-\frac{(4\pi)^4}{k^4}\, M^3N\, I_{4\mathbf{bbb}}\,\chi(1)
=\frac{\lambda^3\hat\lambda}{16}\Big(-\frac{\pi^2}{2\varepsilon}\Big)\chi(1)
\col\\
V_{\text{r}41}=
\settoheight{\eqoff}{$\times$}%
\setlength{\eqoff}{0.5\eqoff}%
\addtolength{\eqoff}{-8.5\unitlength}%
\raisebox{\eqoff}{%
\fmfframe(0,1)(0,1){%
\begin{fmfchar*}(20,15)
\chionerangefourl
\fmfi{photon,left=0.25}{vmm{dir 30}..{dir -30}ve}
\fmfi{photon,right=0.25}{vmm{dir -30}..{dir 30}ve}
\fmf{plain,tension=1,left=0,width=1mm}{v4,ved1}
\end{fmfchar*}}}
&\to
\frac{(4\pi)^4}{2k^4}\,M^3N\, I_4\, \chi(1)
=\frac{\lambda^3\hat\lambda}{32}\Big(-\frac{1}{2\varepsilon^2}+\frac{2}{\varepsilon}\Big)\chi(1)
\col\\
V_{\text{r}42}=
\settoheight{\eqoff}{$\times$}%
\setlength{\eqoff}{0.5\eqoff}%
\addtolength{\eqoff}{-8.5\unitlength}%
\raisebox{\eqoff}{%
\fmfframe(0,1)(0,1){%
\begin{fmfchar*}(20,15)
\chionerangefourl
\fmfi{photon,left=0.25}{vm6{dir 75}..{dir 15}ve}
\fmfi{photon,right=0.25}{vm6{dir 15}..{dir 75}ve}
\fmf{plain,tension=1,left=0,width=1mm}{v4,ved1}
\end{fmfchar*}}}
&\to
\frac{(4\pi)^4}{2k^4}\, M^3N\,I_4\,\chi(1)
=\frac{\lambda^3\hat\lambda}{32}\Big(-\frac{1}{2\varepsilon^2}+\frac{2}{\varepsilon}\Big)\chi(1)
\col\\
V_{\text{r}43}=
\settoheight{\eqoff}{$\times$}%
\setlength{\eqoff}{0.5\eqoff}%
\addtolength{\eqoff}{-8.5\unitlength}%
\raisebox{\eqoff}{%
\fmfframe(0,1)(0,1){%
\begin{fmfchar*}(20,15)
\chionerangefourl
\fmfi{photon,left=0.25}{vi6--ve}
\fmfi{photon,right=0.25}{vo6--ve}
\fmf{plain,tension=1,left=0,width=1mm}{v4,ved1}
\end{fmfchar*}}}
&\to
\frac{(4\pi)^4}{k^4}\,M^3N\,I_{42\mathbf{b}bd}\,\chi(1)
=\frac{\lambda^3\hat\lambda}{16}\Big(\frac{1}{2\varepsilon^2}-\frac{1}{\varepsilon}\Big(2-\frac{\pi^2}{4}\Big)\Big)\chi(1)
\pnt
\end{aligned}
\end{equation}
Also in this case one has to consider the diagrams obtained by reflecting the previous ones.
The total contribution to the renormalization constant is then
\begin{equation}
\begin{aligned}\label{Zr4}
\mathcal{Z}_{\text{r}4,
\text{odd}}
=
\frac{\lambda\hat\lambda}{16}(\lambda^2+\hat\lambda^2)\frac{\pi^2}{4\varepsilon}\chi(1)
\pnt
\end{aligned}
\end{equation}

\subsubsection{Range three interactions}

The range three interactions arise from two-loop corrections to the propagators and vertices
involved in the two-loop diagram (\ref{two-loop}).
It is important to note that, due to the finiteness rules of section \ref{sec:finitenesscond},  
overall UV divergences
can arise only from corrections to the lower vertex or one of the three lower chiral propagators.
According to the analysis of \cite{Leoni:2010az}, the two-loop corrections to the 
chiral two-  and four-point functions  are plagued by IR divergences  even if free of UV poles.
This is due to the particular structure of the gauge superfield propagator 
and cubic vertices
in ${\cal N}=2$ superspace. 
We stress that IR divergences do not appear
in component fields \cite{Minahan:2009wg}, since in three dimensions IR dangerous cubic vertices contribute non-trivial momentum factors to the numerators of the loop integrals. In superspace,
the appearance of IR divergences in intermediate steps
can be cured by using a non-standard gauge fixing procedure 
first introduced in four dimensions in \cite{Abbott:1984pz}
 and adapted in \cite{Leoni:2010az} to the 
 three-dimensional
  case.
Since we are interested only in the overall UV divergences of the
diagrams, a computational strategy could be to ignore purely IR
divergent diagrams and to IR-regulate diagrams that involve both UV and IR 
divergences in such a way as to extract the purely UV poles.
For example, this is illustrated in 
appendix
 \ref{IRintegrals} 
where we can regulate the IR divergences by inserting external momenta in IR 
divergent diagrams.
However, in the main body of the paper we have decided to keep track of the IR divergences and 
check at the end their cancellation. Such 
a
 check is described in appendix \ref{IR_cancellation}.

The interested reader should look at 
appendix
 \ref{two-loop-sub} for a description of the 
two-loop corrections to the two- and four-point functions needed in the calculations of this section.

The contributions with only UV divergences are given by
\begin{equation}
\begin{aligned}
S_{\text{r}3}=
\settoheight{\eqoff}{$\times$}%
\setlength{\eqoff}{0.5\eqoff}%
\addtolength{\eqoff}{-8.5\unitlength}%
\raisebox{\eqoff}{%
\fmfframe(0,1)(-5,1){%
\begin{fmfchar*}(20,15)
\fmftop{v3}
\fmfbottom{v4}
\fmfforce{(0.125w,h)}{v3}
\fmfforce{(0.125w,0)}{v4}
\fmffixed{(0.25w,0)}{v2,v1}
\fmffixed{(0.25w,0)}{v3,v2}
\fmffixed{(0.25w,0)}{v4,v5}
\fmffixed{(0.25w,0)}{v5,v6}
\fmf{plain,tension=0,left=0.25}{v1,vc}
\fmf{plain,tension=1}{vc,v2}
\fmf{plain,tension=0,right=0.25}{v3,vc}
\fmf{phantom,tension=0,right=0.25}{va,v4}
\fmf{phantom,tension=1}{v5,va}
\fmf{plain,tension=0,left=0.25}{va,v6}
\fmf{plain,tension=1}{vc,va}
\fmffreeze
\fmffixed{(0,0.166h)}{vc2,vc3}
\fmffixed{(whatever,0.055h)}{vc3,va}
\fmf{plain,right=0.25}{vc2,v4}
\fmf{plain,right=0.25}{v5,vc2}
\fmf{plain,right=0.5}{vc2,vc3}
\fmf{plain,right=0.5}{vc3,vc2}
\fmf{plain,tension=0,right=0.5}{vc3,va}
\fmf{plain,tension=0,right=0.5}{va,vc3}
\fmf{plain,tension=1,left=0,width=1mm}{v4,v6}
\end{fmfchar*}}}
&\to -\frac{2(4\pi)^4}{k^4}\,M^3N\,I_{42\mathbf{bbb2}}\,\chi(1)
=\frac{\lambda^3\hat\lambda}{16}\Big(-\frac{\pi^2}{2\varepsilon}\Big)\chi(1)
\col\\
V_{\text{r}31a}=
\settoheight{\eqoff}{$\times$}%
\setlength{\eqoff}{0.5\eqoff}%
\addtolength{\eqoff}{-8.5\unitlength}%
\raisebox{\eqoff}{%
\fmfframe(0,1)(-5,1){%
\begin{fmfchar*}(20,15)
\fmftop{v3}
\fmfbottom{v4}
\fmfforce{(0.125w,h)}{v3}
\fmfforce{(0.125w,0)}{v4}
\fmffixed{(0.25w,0)}{v2,v1}
\fmffixed{(0.25w,0)}{v3,v2}
\fmffixed{(0.25w,0)}{v4,v5}
\fmffixed{(0.25w,0)}{v5,v6}
\chioneca{v1}{v2}{v3}{v4}{v5}{v6}
\fmfi{photon}{vmm{dir 30}..{dir -120}vm6}
\fmfi{photon}{vmm{dir -30}..{dir -60}vm6}
\fmf{plain,tension=1,left=0,width=1mm}{v4,v6}
\end{fmfchar*}}}
&\to\frac{(4\pi)^4}{2k^4}\,M^3N\,I_4\,\chi(1)
=\frac{\lambda^3\hat\lambda}{16}
\Big(-\frac{1}{4\varepsilon^2}+\frac{1}{\varepsilon}\Big)\chi(1)
\col
\\
V_{\text{r}31b}=
\settoheight{\eqoff}{$\times$}%
\setlength{\eqoff}{0.5\eqoff}%
\addtolength{\eqoff}{-8.5\unitlength}%
\raisebox{\eqoff}{%
\fmfframe(0,1)(-5,1){%
\begin{fmfchar*}(20,15)
\fmftop{v3}
\fmfbottom{v4}
\fmfforce{(0.125w,h)}{v3}
\fmfforce{(0.125w,0)}{v4}
\fmffixed{(0.25w,0)}{v2,v1}
\fmffixed{(0.25w,0)}{v3,v2}
\fmffixed{(0.25w,0)}{v4,v5}
\fmffixed{(0.25w,0)}{v5,v6}
\chioneca{v1}{v2}{v3}{v4}{v5}{v6}
\fmfi{photon}{vm4{dir 0}..{dir -45}vm5}
\fmfi{photon}{vm4{dir -90}..{dir 45}vm5}
\fmf{plain,tension=1,left=0,width=1mm}{v4,v6}
\end{fmfchar*}}}
&\to\frac{(4\pi)^4}{k^4}\,M^3N(I_4+I_{42\mathbf{b}bd})\chi(1)=\frac{\lambda^3\hat\lambda}{16}
\frac{\pi^2}{4\varepsilon}\chi(1)
\col
\\
V_{\text{r}32a}=
\settoheight{\eqoff}{$\times$}%
\setlength{\eqoff}{0.5\eqoff}%
\addtolength{\eqoff}{-8.5\unitlength}%
\raisebox{\eqoff}{%
\fmfframe(0,1)(-5,1){%
\begin{fmfchar*}(20,15)
\fmftop{v3}
\fmfbottom{v4}
\fmfforce{(0.125w,h)}{v3}
\fmfforce{(0.125w,0)}{v4}
\fmffixed{(0.25w,0)}{v2,v1}
\fmffixed{(0.25w,0)}{v3,v2}
\fmffixed{(0.25w,0)}{v4,v5}
\fmffixed{(0.25w,0)}{v5,v6}
\chioneca{v1}{v2}{v3}{v4}{v5}{v6}
\fmfi{photon}{vmm{dir 0}..{dir -90}vi6}
\fmfi{photon}{vmm{dir 30}..{dir -120}vo6}
\fmf{plain,tension=1,left=0,width=1mm}{v4,v6}
\end{fmfchar*}}}
&\to\frac{(4\pi)^4}{k^4}\,M^3N\,I_{42\mathbf{b}bd}\,\chi(1)
=\frac{\lambda^3\hat\lambda}{16}
\Big(\frac{1}{2\varepsilon^2}+\frac{1}{\varepsilon}\Big(-2+\frac{\pi^2}{4}\Big)\Big)\chi(1)
\col
\\
V_{\text{r}32b}=
\settoheight{\eqoff}{$\times$}%
\setlength{\eqoff}{0.5\eqoff}%
\addtolength{\eqoff}{-8.5\unitlength}%
\raisebox{\eqoff}{%
\fmfframe(0,1)(-5,1){%
\begin{fmfchar*}(20,15)
\fmftop{v3}
\fmfbottom{v4}
\fmfforce{(0.125w,h)}{v3}
\fmfforce{(0.125w,0)}{v4}
\fmffixed{(0.25w,0)}{v2,v1}
\fmffixed{(0.25w,0)}{v3,v2}
\fmffixed{(0.25w,0)}{v4,v5}
\fmffixed{(0.25w,0)}{v5,v6}
\chioneca{v1}{v2}{v3}{v4}{v5}{v6}
\fmfi{photon}{vm4{dir 0}..{dir 0}vi5}
\fmfi{photon}{vm4{dir -90}..{dir 0}vo5}
\fmf{plain,tension=1,left=0,width=1mm}{v4,v6}
\end{fmfchar*}}}
&\to
-\frac{(4\pi)^4}{2k^4}\,M^3N\,I_{422\mathbf{q}\text{tr}ABCD}\,\chi(1)
=\frac{\lambda^3\hat\lambda}{16}
\Big(-\frac{\pi^2}{6\varepsilon}\Big)\chi(1)
\col
\\
V_{\text{r}33a}=
\settoheight{\eqoff}{$\times$}%
\setlength{\eqoff}{0.5\eqoff}%
\addtolength{\eqoff}{-8.5\unitlength}%
\smash[b]{%
\raisebox{\eqoff}{%
\fmfframe(0,1)(-5,1){%
\begin{fmfchar*}(20,15)
\fmftop{v3}
\fmfbottom{v4}
\fmfforce{(0.125w,h)}{v3}
\fmfforce{(0.125w,0)}{v4}
\fmffixed{(0.25w,0)}{v2,v1}
\fmffixed{(0.25w,0)}{v3,v2}
\fmffixed{(0.25w,0)}{v4,v5}
\fmffixed{(0.25w,0)}{v5,v6}
\chioneca{v1}{v2}{v3}{v4}{v5}{v6}
\fmfi{photon}{vm4{dir 135}..{dir 0}vmm}
\fmfi{photon}{vm6{dir 45}..{dir 180}vmm}
\fmf{plain,tension=1,left=0,width=1mm}{v4,v6}
\end{fmfchar*}}}}
&\to\frac{(4\pi)^4}{k^4}\,(MN)^2 I_{422\mathbf{q}\text{tr}ABbd}\,\chi(1)\\
&=\frac{(\lambda\hat\lambda)^2}{16}
\Big(-\frac{1}{\varepsilon^2}+\frac{1}{\varepsilon}\Big(4-\frac{2\pi^2}{3}\Big)\Big)\chi(1)
\col
\\
V_{\text{r}33b}=
\settoheight{\eqoff}{$\times$}%
\setlength{\eqoff}{0.5\eqoff}%
\addtolength{\eqoff}{-8.5\unitlength}%
\raisebox{\eqoff}{%
\fmfframe(0,1)(-5,1){%
\begin{fmfchar*}(20,15)
\fmftop{v3}
\fmfbottom{v4}
\fmfforce{(0.125w,h)}{v3}
\fmfforce{(0.125w,0)}{v4}
\fmffixed{(0.25w,0)}{v2,v1}
\fmffixed{(0.25w,0)}{v3,v2}
\fmffixed{(0.25w,0)}{v4,v5}
\fmffixed{(0.25w,0)}{v5,v6}
\chioneca{v1}{v2}{v3}{v4}{v5}{v6}
\fmfi{photon}{vm4{dir -30}..{dir 0}vm5}
\fmfi{photon}{vm6{dir -150}..{dir 180}vm5}
\fmf{plain,tension=1,left=0,width=1mm}{v4,v6}
\end{fmfchar*}}}
&\to\frac{(4\pi)^4}{k^4}(MN)^2I_{422\mathbf{q}\text{tr}ABCD}\,\chi(1)
=\frac{(\lambda\hat\lambda)^2}{16}
\frac{\pi^2}{3\varepsilon}\chi(1)
\col
\\
V_{\text{r}34}=
\settoheight{\eqoff}{$\times$}%
\setlength{\eqoff}{0.5\eqoff}%
\addtolength{\eqoff}{-8.5\unitlength}%
\smash[b]{%
\raisebox{\eqoff}{%
\fmfframe(0,1)(-5,1){%
\begin{fmfchar*}(20,15)
\fmftop{v3}
\fmfbottom{v4}
\fmfforce{(0.125w,h)}{v3}
\fmfforce{(0.125w,0)}{v4}
\fmffixed{(0.25w,0)}{v2,v1}
\fmffixed{(0.25w,0)}{v3,v2}
\fmffixed{(0.25w,0)}{v4,v5}
\fmffixed{(0.25w,0)}{v5,v6}
\chioneca{v1}{v2}{v3}{v4}{v5}{v6}
\fmfi{photon}{vm4{dir -45}..{dir 0}vm5}
\fmfi{photon}{vi4{dir 120}..{dir -30}vo4}
\fmf{plain,tension=1,left=0,width=1mm}{v4,v6}
\end{fmfchar*}}}}
&\to
\frac{(4\pi)^4}{k^4}(MN)^2\big(\,
2I_{42\mathbf{b}be}
-I_{422\mathbf{q}\text{tr}ABbd}
\\
&\hphantom{{}={}\frac{(4\pi)^4}{k^4}(MN)^2\big(}
+2(2 I_{221be}-I_{221dc})G(2-2\lambda,1)G(2-3\lambda,1)
\\
&
\hphantom{{}={}\frac{(4\pi)^4}{k^4}(MN)^2\big(}
-2(I_{42\mathbf{b}bd}+I_{42\mathbf{b}be})\big)\chi(1)
\\
&=
\frac{(\lambda\hat\lambda)^2}{16}
\Big(-\frac{\pi^2}{3\varepsilon}\Big)\chi(1)
\pnt\\
\end{aligned}
\end{equation}

The contributions with both UV and IR divergences are given by
\begin{equation}
\begin{aligned}
V_{\text{r}35}=
\settoheight{\eqoff}{$\times$}%
\setlength{\eqoff}{0.5\eqoff}%
\addtolength{\eqoff}{-8.5\unitlength}%
\smash[b]{%
\raisebox{\eqoff}{%
\fmfframe(0,1)(-5,1){%
\begin{fmfchar*}(20,15)
\fmftop{v3}
\fmfbottom{v4}
\fmfforce{(0.125w,h)}{v3}
\fmfforce{(0.125w,0)}{v4}
\fmffixed{(0.25w,0)}{v2,v1}
\fmffixed{(0.25w,0)}{v3,v2}
\fmffixed{(0.25w,0)}{v4,v5}
\fmffixed{(0.25w,0)}{v5,v6}
\chioneca{v1}{v2}{v3}{v4}{v5}{v6}
\fmfi{photon}{vm4{dir -45}..{dir 0}vm5}
\vacpolp[0.33]{vm4{dir -45}..{dir 0}vm5}
\fmf{plain,tension=1,left=0,width=1mm}{v4,v6}
\end{fmfchar*}}}}
&\to-\frac{(4\pi)^4}{k^4}\big(MN(4MN-M^2)\big)
\big(I_4-I_{4\text{UVIR}}+I_{42\mathbf{b}bd}\big)\chi(1)\\
&=\frac{\lambda\hat\lambda}{16}(4\lambda\hat\lambda-\lambda^2)
\Big(-\frac{1}{2\varepsilon^2}+\frac{2}{\varepsilon}\Big(-2-\frac{\pi^2}{8}+\gamma-\ln 4\pi\Big)\Big)\chi(1)
\col\\
V_{\text{r}36}=
\settoheight{\eqoff}{$\times$}%
\setlength{\eqoff}{0.5\eqoff}%
\addtolength{\eqoff}{-8.5\unitlength}%
\smash[b]{%
\raisebox{\eqoff}{%
\fmfframe(0,1)(-5,1){%
\begin{fmfchar*}(20,15)
\fmftop{v3}
\fmfbottom{v4}
\fmfforce{(0.125w,h)}{v3}
\fmfforce{(0.125w,0)}{v4}
\fmffixed{(0.25w,0)}{v2,v1}
\fmffixed{(0.25w,0)}{v3,v2}
\fmffixed{(0.25w,0)}{v4,v5}
\fmffixed{(0.25w,0)}{v5,v6}
\chioneca{v1}{v2}{v3}{v4}{v5}{v6}
\vacpolp[0.33]{p4}
\fmf{plain,tension=1,left=0,width=1mm}{v4,v6}
\end{fmfchar*}}}}
&\to\frac{(4\pi)^4}{k^4}MN
\Big(2MNI_{4\mathbf{bbb}}
-\frac{1}{2}\big(8MN-(M^2+N^2)\big)I_{4\text{UVIR}}\Big)\chi(1)\\
&=\frac{\lambda\hat\lambda}{16}
\Big(\lambda\hat\lambda\frac{\pi^2}{\varepsilon}
+\big(8\lambda\hat\lambda-(\lambda^2+\hat\lambda^2)\big)
\Big(\frac{1}{4\varepsilon^2}+\frac{1}{\varepsilon}\big(2-\gamma+\ln 4\pi\big)\Big)\Big)\chi(1)
\pnt
\end{aligned}
\end{equation}
Note that the expressions for the integrals that appear in the results 
have their UV
subdivergences subtracted. The suffix UVIR appears on
integrals which due to different arrangements of their external 
momenta contribute both UV and IR divergences.
The UV poles can be extracted by adding external momentum to the cubic vertex which causes the IR divergence, i.e.\ one replaces $I_{4\text{UVIR}}\to I_4$.
This then yields
\begin{equation}
\begin{aligned}
V^{\text{UV}}_{\text{r}35}
&=\frac{\lambda\hat\lambda}{16}(4\lambda\hat\lambda-\lambda^2)\Big(-\frac{1}{2\varepsilon^2}+\frac{1}{\varepsilon}\Big(2-\frac{\pi^2}{4}\Big)\Big)\chi(1)
\col
\\
V^{\text{UV}}_{\text{r}36}
&=\frac{\lambda\hat\lambda}{16}
\Big(\lambda\hat\lambda\frac{\pi^2}{\varepsilon}
+(8\lambda\hat\lambda-(\lambda^2+\hat\lambda^2))\Big(\frac{1}{4\varepsilon^2}-\frac{1}{\varepsilon}\Big)\Big)\chi(1)
\col
\end{aligned}
\end{equation}
In appendix \ref{IR_cancellation} we explicitly demonstrate that this 
result is also obtained if instead of choosing an IR safe momentum 
configuration all relevant diagrams with IR divergence are considered, i.e.\
the IR divergences cancel out in the final result.

The contribution of the range three interactions to the renormalization 
constant ${\cal Z}$ is then given by
\begin{equation}
\begin{aligned}\label{Zr3}
\mathcal{Z}_{\text{r}{3},
\text{odd}}
&=-(1+{\cal{R}})
(S_{\text{r}3}+V_{\text{r}31a}+V_{\text{r}31b}+V_{\text{r}32a}+2V_{\text{r}32b}+2V_{\text{r}34}
+V^{\text{UV}}_{\text{r}35})
\\
&\phantom{{}={}}
-V_{\text{r}33a}-V_{\text{r}33b}-3V^{\text{UV}}_{\text{r}36}
\\
&=
\frac{\lambda\hat\lambda}{16}\Big(
\lambda\hat\lambda\Big(-\frac{1}{\varepsilon^2}
+\frac{1}{\varepsilon}\Big(4+\frac{2\pi^2}{3}\Big)\Big)
+(\lambda^2+\hat\lambda^2)\frac{\pi^2}{12\varepsilon}
\Big)\chi(1)
\pnt
\end{aligned}
\end{equation}

\subsection{Final result}

We are now ready to put together the parts of our calculations necessary to extract the four-loop
dilatation operator. 
As discussed before the dilatation operator for odd sites is obtained by extracting the 
$1/\varepsilon$ pole from the renormalization constant.
Summing up the contributions to the $1/\varepsilon$ pole 
from  \eqref{Zr5}, 
\eqref{Zr4} and \eqref{Zr3}, we obtain
\begin{equation}
\begin{aligned}\label{Z4}
\bar\lambda^4\mathcal{Z}_{4,\text{odd}}|_{\frac{1}{\varepsilon}}
&=
\big(\mathcal{Z}_{\text{r}5,\text{odd}}
+\mathcal{Z}_{\text{r}4,\text{odd}}
+\mathcal{Z}_{\text{r}3,\text{odd}}
\big)|_{\frac{1}{\varepsilon}}
\\
&=
\frac{\lambda\hat\lambda}{16\varepsilon}
\Big[
-2\lambda\hat\lambda
(\chi(1,3)+\chi(3,1))
+\Big(\lambda\hat\lambda
\Big(4+\frac{2\pi^2}{3}\Big)
+(\lambda^2+\hat\lambda^2)\frac{\pi^2}{3}
\Big)\chi(1)
\Big]
\col
\end{aligned}
\end{equation}
that, rewritten in terms of $\bar{\lambda}$ and $\sigma$ of \eqref{barlambdasigmadef},
gives
\begin{equation}
\begin{aligned}\label{lnZ4odd}
\bar\lambda^4\mathcal{Z}_{4,\text{odd}}|_{\frac{1}{\varepsilon}}
&=\frac{\bar\lambda^4}{16\varepsilon}
\Big[
-2(\chi(1,3)+\chi(3,1))
+\Big(4
\Big(1+\frac{\pi^2}{3}\Big)
+\sigma^2\frac{\pi^2}{3}\Big)\chi(1)
\Big]
\pnt
\end{aligned}
\end{equation}

As already observed, in the $\ln {\cal Z}$ the higher order poles must be absent.
This is
 a useful consistency check of our computation.
Additional diagrams that do not contribute to the dilatation 
operator but have non-vanishing double poles have to be taken into account.
Some of them consist of two separate two-loop interactions.
Furthermore, one has to consider the diagrams that lead to 
interactions between magnons at odd and even sites and contribute 
only to the double pole when summed up.
In appendix
 \ref{doublePoles}, we prove that when all these double poles 
are taken into account, their sum is indeed cancelled by the two-loop 
contribution in the expansion of $\ln\mathcal{Z}$.
The dilatation operator for odd sites is then obtained from \eqref{lnZ4odd} 
by multiplying the 
$1/\varepsilon$  pole by $8$.
With $\zeta(2)=\frac{\pi^2}{6}$, it reads
\begin{equation}
\begin{aligned}\label{D4odd}
\mathcal{D}_{4,\text{odd}}(\sigma)
&=
(2+(4+\sigma^2)\zeta(2))\chi(1)-\chi(1,3)-\chi(3,1)
\pnt
\end{aligned}
\end{equation}
By comparing the previous result with equation (\ref{D4})
we read off the four-loop
 coefficient of the function $h^2(\bar\lambda,\sigma)$
\begin{equation}
\begin{aligned}\label{h4sigma}
h_4(\sigma)&=-(4+\sigma^2)\zeta(2)
\pnt
\end{aligned}
\end{equation}
This result coincides with the one computed in \cite{Minahan:2009wg}.
It is interesting to note that, in contrast to the component calculation in \cite{Minahan:2009wg}, the integrals that contribute here to the dilatation operator show a
correlation between the quadratic and the rational simple pole in $\varepsilon$:
their relative coefficient is always $-4$ as for the simplest four-loop integral $I_4$ in \eqref{I4}. The rational term in \eqref{D4odd} and therefore 
its absence in \eqref{h4sigma} is hence correlated with the quadratic pole that
itself is determined by the two-loop result \eqref{Z2}.

\section{Possible scenarios for an all-loop function }
\label{sec:hallorder}

In this section we discuss our attempts to find   an all-loop function for $h^2(\bla,\s)$.

In the  ABJM case where $\s=0$, $h^2(\bla,0)=h^2(\la)$, there is a surprisingly simple function that matches the weak coupling behavior up to 
four-loop
order and also matches the leading strong coupling behavior.  To this end we define $t\equiv2\pi i\la$, which is a natural variable that also appears in expressions for supersymmetric 
ABJ(M)
Wilson loops \cite{Kapustin:2009kz,Marino:2009jd,Drukker:2010nc}.  We then consider a rescaled function $g(t)=(2\pi)^2\,h^2(\la)$.  In terms of $g(t)$ the magnon dispersion relation becomes
\begin{equation}\label{disp}
\veps(p)=\sqrt{\frac{1}{4}+\frac{g(t)}{\pi^2}\sin^2\frac{p}{2}}\,,
\ee
and so has a form more in line with the $\NN=4$ dispersion relation where in that case $g(t)$ in (\ref{disp}) is replaced with $\la$.

In terms of $g(t)$, the proposed all-loop function is
\be\label{gf}
g(t)=-(1-t)\log(1-t)-(1+t)\log(1+t)\,,
\ee
whose weak coupling expansion is
\be
\begin{aligned}
g(t)&=-\sum_{n=1}^\infty \frac{t^{2n}}{n(2n-1)}=-t^2-\frac{1}{6}\,t^4-\frac{1}{15}\,t^6+\mathcal{O}(t^8)\\
&=(2\pi)^2\left(\la^2-4\,\zeta(2)\,\la^4+6\,\zeta(4)\,\la^6+\mathcal{O}(\bla^8)\right)\,.
\end{aligned}
\ee
An obvious test is to compute $h^2(\la)$ to six-loop order, where the all-loop function in (\ref{gf}) 
predicts
the value $h_6=\frac{(2\pi)^4}{15}$.  A six-loop computation is admittedly very difficult, but we believe it is manageable using the $\NN=2$ superspace formulation.

At strong coupling the expansion is
\be\label{gstrong}
\begin{aligned}
g(t)&=-i\pi\,t-2\,\log t-2+\mathcal{O}(t^{-1})\\
&=(2\pi)^2\left(\frac{\la}{2}-\frac{1}{(2\pi)^2}\log(2\pi\la)-2+\mathcal{O}(\la^{-1})\right)\,.
\end{aligned}
\ee
The dominant term agrees with the leading strong coupling expansion from the string sigma-model.  But also observe that the first correction corresponds to a two-loop contribution; a one-loop correction is absent.  This disagrees with the prediction in \cite{McLoughlin:2008he} arising from the one-loop correction to the energy for a folded-string \cite{McLoughlin:2008ms,Alday:2008ut,Krishnan:2008zs,Gromov:2008fy,Mikhaylov:2010ib}.  In this language one would expect a $g(t)$ with leading asymptotic expansion 
\be
g(t)=-i\pi\,t-2\sqrt{-i\pi t}\ln(2)+\dots\,.
\ee
However,  if one chooses   a different prescription for summing over mode frequencies, where one essentially groups the modes into heavy and light  \cite{Gromov:2008fy}, then $g(t)$ no longer has the  $\sqrt{t}$ term, agreeing  with the large $t$ expansion \eqref{gstrong}.\footnote{See \cite{Abbott:2010yb} for a further discussion of this.  These authors also show that the same choices of prescriptions appear in finite size corrections for giant magnons \cite{Shenderovich:2008bs,Astolfi:2008ji} and lead to the same one-loop contributions to $h^2(\la)$.}

The function in \eqref{gf}  does not appear to have an easy generalization to the  ABJ case where $\s\ne0$.  Such a function would be expected to be invariant under the transformation \cite{Aharony:2008gk}
\begin{equation}\label{latrans}
\lambda\to\hat\lambda\col\qquad\hat\lambda\to2\hat\lambda-\lambda+1\pnt
\end{equation}
Under (\ref{latrans}) the perturbative regime is mapped into strong coupling, making its verification difficult.  Some evidence that $h^2(\bla,\s)$ is consistent with (\ref{latrans}) was presented in \cite{Minahan:2010nn}.  
One possible hint about the all-loop structure is that  the four-loop
contribution to 
$h^2(\bar\lambda,\sigma)$
can be rewritten as
\begin{equation}
\bar\lambda^4(4+\sigma^2)=\lambda\hat\lambda(\lambda+\hat\lambda)^2\pnt
\end{equation}
which is zero if $\la=-\hat\la$
It would be interesting to see if the higher order corrections remain zero under this condition.  However, it is not clear how this could square with the strong coupling behavior nor with an invariance under the transformation in \eqref{latrans}.

Another possibility is that  $h^2(\bar{\la},\s)$ is 
somehow
 related to recent results
concerning supersymmetric Wilson loops in the ABJ(M) models.  
In this latter case,
 it was found using localization \cite{Nekrasov:2002qd,Pestun:2007rz} that the Wilson loop expectation value could be reduced to a matrix model on a Lens space \cite{Kapustin:2009kz}.  This matrix model is solvable in the planar limit \cite{Marino:2002fk,Halmagyi:2003ze} and hence all-loop predictions can be extracted.  In particular, for ABJM the perturbative free energy of the matrix model is \cite{Marino:2009jd}
\be
F(t)=N^2\left(\log(t)+\frac{1}{36}t^2+\mathcal{O}(t^4)\right)\,.
\ee
It is tempting to look for a connection between $F(t)$ and $g(t)$.  One might try
\be\label{gFree}
(g(t))^{1/2}=-\frac{i}{N^2}t^2\frac{\partial F}{\partial t}=-i\,t-\frac{i}{18}\,t^3+\mathcal{O}(t^5)\,.
\ee
The full expansion also is maximally transcendental, 
but here one finds that the $t^3$ term is off by a factor of $2/3$.   At strong coupling the free energy is asymptotically \cite{Drukker:2010nc}
\be
F(t)\approx-N^2\frac{2\pi^{3/2}}{3}\,(-it)^{-1/2}\,.
\ee
Applying the same rule as in (\ref{gFree}) one finds
\be
(g(t))^{1/2}=-\frac{i}{N^2}t^2\frac{\partial F}{\partial t}\approx \frac{\pi}{3} (-i\pi t)^{1/2}\,,
\ee
which differs by an overall factor of $\pi/3$ from the square root of the leading term in (\ref{gstrong}).

\section{Wrapping interactions}
\label{sec:wrapping}

\enlargethispage{\baselineskip}
To obtain the complete four-loop spectrum of operators 
in the $SU(2)\times SU(2)$ subsector, we have to consider the wrapping 
interactions for the non-protected operators that consist of
up to four elementary fields.
The only non-trivial
 operator is in the $\mathbf{20}$ of $SU(4)$ and 
has $L=2$, i.e.\
exactly four elementary fields.

The only wrapping diagrams which according to the initially discussed 
finiteness theorems based on power counting can contribute to the dilatation 
operator are given by
\begin{equation}
\begin{aligned}
W_1=
\settoheight{\eqoff}{$\times$}%
\setlength{\eqoff}{0.5\eqoff}%
\addtolength{\eqoff}{-10\unitlength}%
\raisebox{\eqoff}{%
\fmfframe(2,1)(2,4){%
\begin{fmfchar*}(20,15)
\fmftop{v1}
\fmfbottom{v6}
\fmfforce{(-0.1250w,h)}{v1}
\fmfforce{(-0.125w,0)}{v6}
\fmffixed{(0.25w,0)}{v1,v2}
\fmffixed{(0.25w,0)}{v2,v3}
\fmffixed{(0.25w,0)}{v3,v4}
\fmffixed{(0.25w,0)}{v4,v5}
\fmffixed{(0.25w,0)}{v6,v7}
\fmffixed{(0.25w,0)}{v7,v8}
\fmffixed{(0.25w,0)}{v8,v9}
\fmffixed{(0.25w,0)}{v9,v10}
\fmffixed{(0,0)}{v10,vr}
\fmffixed{(0.25w,0)}{v6,vl}
\fmf{plain}{v7,v4la}
\fmf{plain}{v4la,v4lc}
\fmf{plain,tension=0,left=0.25}{v4lc,v1}
\fmf{plain,left=0}{v4lc,v2}
\fmf{plain,tension=0,right=0.25}{v4lc,v3}
\fmf{plain,tension=0,left=0.25}{v8,v4ra}
\fmf{plain}{v9,v4ra}
\fmf{plain,tension=0,right=0.25}{v10,v4ra}
\fmf{plain}{v4ra,v4rc}
\fmf{plain,left=0}{v4rc,v4}
\fmffreeze
\fmfposition
\fmf{plain,tension=1,left=0,width=1mm}{v7,v10}
\fmfi{plain}{vloc(__v4la){dir 0}..{dir 0}vloc(__v4rc)}
\plainwrap{v4la}{vl}{vr}{v4rc}
\end{fmfchar*}}}
&\to
-\frac{2(4\pi)^4}{k^4}(MN)^2
I_4\,\chi(1)
=\frac{(\lambda\hat\lambda)^4}{16}\Big(\frac{1}{\varepsilon^2}-\frac{4}{\varepsilon}\Big)\chi(1)
\col\\
W_2=
\settoheight{\eqoff}{$\times$}%
\setlength{\eqoff}{0.5\eqoff}%
\addtolength{\eqoff}{-10\unitlength}%
\raisebox{\eqoff}{%
\fmfframe(2,1)(2,4){%
\begin{fmfchar*}(20,15)
\chionetwog
\fmf{plain,tension=1,left=0,width=1mm}{v5,v8}
\fmfi{wiggly}{vgm1{dir -90}..{dir -90}vgu4}
\wigglywrap{vgm1}{v5}{v8}{vgd4}
\end{fmfchar*}}}
&\to
-\frac{2(4\pi)^4}{k^4}(MN)^2
I_{42\mathbf{bb0}cd}\,\chi(1)
=\frac{(\lambda\hat\lambda)^4}{16}\Big(-\frac{1}{2\varepsilon^2}+\frac{3}{\varepsilon}\Big)\chi(1)
\col\\
W_3=
\settoheight{\eqoff}{$\times$}%
\setlength{\eqoff}{0.5\eqoff}%
\addtolength{\eqoff}{-10\unitlength}%
\raisebox{\eqoff}{%
\fmfframe(2,1)(2,4){%
\begin{fmfchar*}(20,15)
\chionetwog
\fmf{plain,tension=1,left=0,width=1mm}{v5,v8}
\fmfi{wiggly}{vgm1{dir -90}..{dir -90}vgm3}
\wigglywrap{vgm1}{v5}{v8}{vgm4}
\end{fmfchar*}}}
&\to
\frac{(4\pi)^4}{k^4}(MN)^2
I_{422\mathbf{b}\text{tr}ABcd}\,\chi(1)
=\frac{(\lambda\hat\lambda)^4}{16}\Big(\frac{1}{\varepsilon^2}-\frac{2}{\varepsilon}\Big)\chi(1)
\col\\
W_4=
\settoheight{\eqoff}{$\times$}%
\setlength{\eqoff}{0.5\eqoff}%
\addtolength{\eqoff}{-10\unitlength}%
\raisebox{\eqoff}{%
\fmfframe(2,1)(2,4){%
\begin{fmfchar*}(20,15)
\chionerangefourl
\fmfi{photon}{vmm--vemm}
\fmf{plain,tension=1,left=0,width=1mm}{v4,ved1}
\wigglywrap{vmm}{v4}{ved1}{vemm}
\end{fmfchar*}}}
&\to
-\frac{2(4\pi)^4}{k^4}(MN)^2I_4\,\chi(1)
=\frac{(\lambda\hat\lambda)^4}{16}\Big(\frac{1}{\varepsilon^2}-\frac{4}{\varepsilon}\Big)\chi(1)
\col\\
W_5=
\settoheight{\eqoff}{$\times$}%
\setlength{\eqoff}{0.5\eqoff}%
\addtolength{\eqoff}{-10\unitlength}%
\raisebox{\eqoff}{%
\fmfframe(2,1)(2,4){%
\begin{fmfchar*}(20,15)
\chionerangefourl
\fmfi{photon}{vm6--vem6}
\fmf{plain,tension=1,left=0,width=1mm}{v4,ved1}
\wigglywrap{vm4}{v4}{ved1}{vem6}
\end{fmfchar*}}}
&\to
\frac{(4\pi)^4}{k^4}(MN)^2I_{422\mathbf{q}\text{tr}ABbd}\chi(1)
=
\frac{(\lambda\hat\lambda)^4}{16}\Big(-\frac{1}{\varepsilon^2}+\frac{1}{\varepsilon}\Big(
4-\frac{2}{3}\pi^2\Big)\Big)\chi(1)
\pnt
\end{aligned}
\end{equation}
There are four distinct diagrams of type $W_2$ and two of type 
$W_3$.
The sum of the wrapping diagrams is therefore given by
\begin{equation}
\begin{aligned}
W&=W_1+4W_2+2W_3+W_4+W_5
=
\frac{(\lambda\hat\lambda)^4}{16}\Big[\frac{1}{\varepsilon^2}
+\frac{2}{\varepsilon}\Big(2-\frac{\pi^2}{3}\Big)\Big]\chi(1)
\pnt
\end{aligned}
\end{equation}

Multiplying the $1/\varepsilon$ pole of $W$ by $-8$, we 
obtain the wrapping contribution to the dilatation operator. It reads
\begin{equation}
\begin{aligned}
{\cal D}_{4,\text{odd}}^{\text{w}}
=-(2-2\zeta(2))\chi(1)
\pnt
\label{Dwrapped}
\end{aligned}
\end{equation}
Now, by subtracting  from \eqref{D4odd}  the range five contribution and inserting 
$h_4(\sigma)=-(4+\sigma^2)\zeta(2)$, the
subtracted dilatation operator becomes
\begin{equation}
\begin{aligned}
{\cal D}_{4,\text{odd}}^{\text{sub}}(\sigma)
&=(2-h_4(\sigma))\chi(1)
=\big(2+(4+\sigma^2)\zeta(2)\big)\chi(1)
\label{Dsub}
\pnt
\end{aligned}
\end{equation}
The dilatation operator for length four states then reads
\begin{equation}
\begin{aligned}
{\cal D}_{4,\text{odd}}^{\text{range } 4}(\sigma)
={\cal D}_{4,\text{odd}}^{\text{sub}}(\sigma)
+{\cal D}_{4,\text{odd}}^{\text{w}}
=(6+\sigma^2)\zeta(2)\chi(1)
\col
\end{aligned}
\end{equation}
and it coincides with the results obtained in terms of component 
fields \cite{Minahan:2009aq,Minahan:2009wg}.

Note that the separation of the dilatation operator into wrapping and subtracted parts differs in the superfield calculation from
the one obtained in component fields in \cite{Minahan:2009aq,Minahan:2009wg}.
The sum of the two terms is, however, the same in the two calculations, and 
hence the resulting anomalous dimensions for operators with
length $2L=4$ agree.

\section{Conclusions}
\label{sec:concl}

In this paper we have computed $h_4(\s)$ using the $\NN=2$ superspace formalism.  The computation is greatly simplified from the component version 
\cite{Minahan:2009aq,Minahan:2009wg}
because 
the manifest supersymmetry in combination with finiteness conditions
leads to a large reduction in the number of Feynman diagrams.

With this reduction in diagrams, it should be possible to tackle more challenging computations, including the six-loop
 term $h_6(\s)$.  Six loops would give one more data point and might provide further insights into an all-loop function.

Alternatively, one could also apply the superspace formalism to four loops but beyond the $SU(2)\times SU(2)$ sector.  This would not give us further information on $h^2(\bla,\s)$, but it would provide a check of 
higher-loop
integrability in both ABJM and ABJ models.
 One reason that integrability  in the ABJ case
  is not assured is because at strong coupling a nonzero $\s$ would correspond to a nonzero $\theta$-angle for the world-sheet, which is normally thought to destroy integrability.  However, at the lowest order in perturbation theory, the spin-chain is integrable in all sectors, even when $\s\ne0$ \cite{Bak:2008vd,Minahan:2009te}.   It would be interesting to see how this plays out at higher loops.

\section*{Acknowledgements}

The work of M.\ L.\,, A.\ M.\ and A.\ S.\ has been supported in part by the Italian MIUR-PRIN 
contract 20075ATT78.
The research of J.\ A.\ M.\ is supported in part by the
Swedish research council and the STINT foundation.  J.\ A.\ M.\  thanks the
CTP at MIT  and Nordita during the workshop ``Integrability in String and Gauge Theories; AdS/CFT Duality and its applications" for kind
hospitality  during the course of this work. 
O.~O.~S.\@ thanks the Centre for Mathematical Science at City University in London for kind hospitality during course of this work.
C.\ S.\ thanks the department of Physics in Milan for kind
hospitality  during the course of this work. 
During the first stages of this work, G.\ T.-M. was supported by the J. S. Toll Professorship, 
the University of Maryland Center for String \& Particle Theory, and National Science 
Foundation Grant PHY-0354401. After March 2010, G.\ T.-M. was supported by the European 
Commission Marie Curie Intra-European Fellowships under the contract 
PIEF-GA-2009-236454.  
G.\ T.-M.
thanks the department of Physics in Milan for support and hospitality 
at different stages of this work.



\appendix

\section{Conventions and identities}
\label{app:conventions}

We use 
three-dimensional
 spinor and superspace notations adapted from \cite{Gates:1983nr}.
We directly work in the Wick
 rotated Euclidean space-time with metric 
$g_{\mu\nu}=g^{\mu\nu}=\diag(1,1,1)$. For a given 
three-dimensional
 spinor field $\psi_{\alpha}$, we raise
and lower spinor indices as
\begin{equation}
\begin{aligned}
\psi^\alpha=C^{\alpha\beta}\psi_\beta\col\qquad \psi_\alpha=\psi^\beta C_{\beta\alpha}
\pnt
\end{aligned}
\end{equation}
where we use the spinor metric $C^{\alpha\beta}$ defined by
\begin{equation}
\begin{aligned}
C^{\alpha\beta}&=\begin{pmatrix}0 & i \\ -i & 0\end{pmatrix}\col\qquad
C_{\alpha\beta}=\begin{pmatrix}0 & -i \\ i & 0\end{pmatrix}\col\qquad\\
\end{aligned}
\end{equation}
For the contraction of spinor indices we use the notation
\begin{equation}
\begin{aligned}
\psi\chi=\psi^\alpha\chi_\alpha=\chi^\alpha\psi_\alpha=\chi\psi
\col\qquad
\psi^2=\frac{1}{2}\psi^\alpha\psi_\alpha\pnt
\end{aligned}
\end{equation}

The $\gamma$-matrices 
obey the relation
\begin{equation}
(\gamma^\mu)^\alpha{}_\gamma(\gamma^\nu)^\gamma{}_\beta
=-g^{\mu\nu}\delta^\alpha{}_\beta-\epsilon^{\mu\nu\rho}(\gamma_\rho)^\alpha{}_\beta
\pnt
\end{equation}
where the Levi-Civita tensor 
is such that $\epsilon^{012}=1$.
When one spinor index is lowered or raised the $\gamma$-matrices are symmetric 
\begin{equation}
(\gamma^\mu)_{\alpha\beta}=(\gamma^\mu)_{\alpha}{}^{\delta}C_{\delta\beta}
=(\gamma^\mu)_{\beta\alpha}
~,~~~
(\gamma^\mu)^{\alpha\beta}=C^{\alpha\delta}(\gamma^\mu)_{\delta}{}^{\beta}
=(\gamma^\mu)^{\beta\alpha}
~
\pnt
\end{equation}
The trace of product of $\gamma$-matrices satisfies 
\begin{equation}
\begin{aligned}
\tr(\gamma^\mu\gamma^\nu)
&
=(\gamma^\mu)^\alpha{}_\beta(\gamma^\nu)^\beta{}_\alpha
=-2g^{\mu\nu}
\col\\
\tr(\gamma^\mu\gamma^\nu\gamma^\rho)
&
=-(\gamma^\mu)^\alpha{}_\beta(\gamma^\nu)^\beta{}_\gamma(\gamma^\rho)^\gamma{}_\alpha
=-2\epsilon^{\mu\nu\rho}
\col\\
\tr(\gamma^\mu\gamma^\nu\gamma^\rho\gamma^\sigma)
&
=(\gamma^\mu)^\alpha{}_\beta(\gamma^\nu)^\beta{}_\gamma(\gamma^\rho)^\gamma{}_\delta
(\gamma^\sigma)^\delta{}_\alpha
=2(g^{\mu\nu}g^{\rho\sigma}-g^{\mu\rho}g^{\nu\sigma}+g^{\mu\sigma}g^{\nu\rho})
\pnt
\end{aligned}
\end{equation}
We use the convention that the first of two 
contracted indices is always an upper index; this 
is used in the previous formulas in the definition of the trace of products of gamma 
matrices
and it is very useful for $\D$-algebra manipulations \cite{Gates:1983nr}.

Using the $\gamma$-matrices we can move from vector to bi-spinor indices 
thanks to the following definitions
\begin{equation}
\begin{aligned}
x^{\alpha\beta}&=\frac{1}{2}(\gamma_\mu)^{\alpha\beta}x^\mu\col\qquad
x^\mu=(\gamma^\mu)_{\alpha\beta}x^{\alpha\beta}\col\\
p_{\alpha\beta}&=(\gamma^\mu)_{\alpha\beta}p_\mu\col\qquad
p_\mu=\frac{1}{2}(\gamma_\mu)^{\alpha\beta}p_{\alpha\beta}\col\\
A_{\alpha\beta}&=\frac{1}{\sqrt{2}}(\gamma^\mu)_{\alpha\beta}A_\mu\col\qquad
A_\mu=\frac{1}{\sqrt{2}}(\gamma_\mu)^{\alpha\beta}A_{\alpha\beta}\col\\
\end{aligned}
\end{equation}
respectively for coordinates, momenta and fields.
As usual, 
here the momentum $p_\mu$ is related to the vector derivative 
$\partial_\mu=\frac{\partial}{\partial x^\mu}$
by Fourier transform and $p_\mu=i \partial_\mu$.

The three-dimensional,
 ${\cal N}=2$ superspace spinor covariant derivatives 
$\D_\alpha,{\barD}_{\alpha}$ satisfy the algebra
\begin{equation}
\begin{aligned}
\acomm{\D_\alpha}{\D_\beta}&=\acomm{\barD_\alpha}{\barD_\beta}=0\col\quad
\acomm{\D_\alpha}{\barD_\beta}=p_{\alpha\beta}
\pnt
\end{aligned}
\end{equation}

The metric $\epsilon_{AB}$ for the SU(2) flavour indices is given by
\begin{equation}
\begin{aligned}
\epsilon_{12}=1\col\qquad
\epsilon^{12}
=1\col\qquad
\epsilon^{AB}\epsilon_{CD}=\delta^A_C\delta^B_D-\delta^A_D\delta^B_C
\pnt
\end{aligned}
\end{equation}
The flavour indices are raised and lowered as
\begin{equation}
\psi^{A}=\epsilon^{AB}\psi_{B}~,~~~
\psi_A=\psi^{B}\epsilon_{BA}~.
\end{equation}
For the integration over the superspace our conventions are
$\int\de^2\theta=\frac{1}{2}\partial^\alpha\partial_\alpha$,
$\int\de^2{\bar\theta}=\frac{1}{2}{\bar\partial}^\alpha{\bar\partial}_\alpha$
and
$\int\de^4{\theta}=\int\de^2{\theta}\de^2{\bar\theta}$,
such that
\begin{equation}
\begin{aligned}
&\int\de^3x\de^2\theta=\int\de^3x\,\D^2|_{\theta={\bar\theta}=0}
~,~~~
\int\de^3x\de^2{\bar\theta}=\int\de^3x\,{\barD}^2|_{\theta={\bar\theta}=0}~,~~~
\\
&\int\de^3x\de^4\theta=\int\de^3x\,\D^2{\barD}^2|_{\theta={\bar\theta}=0}
~.
\end{aligned}
\end{equation}
The $\theta$-space $\delta$-function is given by
\begin{equation}
\delta^4(\theta-\theta')=(\theta-\theta')^2({\bar\theta}-{\bar\theta}')^2
~.
\end{equation}

\section{Feynman rules in superspace}
\label{app:Feynmanrules}

We use the Wick rotated Feynman rules, i.e.\ we have $\e^{-iS}\to\e^S$
in the path integral.
The propagators are given by
\begin{equation}\label{propagators}
\begin{aligned}
\settoheight{\eqoff}{$\times$}%
\setlength{\eqoff}{0.5\eqoff}%
\addtolength{\eqoff}{-3.75\unitlength}%
\raisebox{\eqoff}{%
\fmfframe(2,2)(2,2){%
\begin{fmfchar*}(15,7.5)
\fmfleft{v1}
\fmfright{v2}
\fmfforce{0.0625w,0.5h}{v1}
\fmfforce{0.9375w,0.5h}{v2}
\fmf{photon}{v1,v2}
\fmffreeze
\fmfposition
\fmfipath{pm[]}
\fmfiset{pm1}{vpath(__v1,__v2)}
\nvml{1}{$\scriptstyle p$}
\end{fmfchar*}}}
&{}={}
\langle V(p) V(-p)\rangle
=-\langle \hat V(p) \hat V(-p)\rangle
=\frac{1}{p^2}\D\barD
\delta^4(\theta_1-\theta_2)\col\\
\settoheight{\eqoff}{$\times$}%
\setlength{\eqoff}{0.5\eqoff}%
\addtolength{\eqoff}{-3.75\unitlength}%
\raisebox{\eqoff}{%
\fmfframe(2,2)(2,2){%
\begin{fmfchar*}(15,7.5)
\fmfleft{v1}
\fmfright{v2}
\fmfforce{0.0625w,0.5h}{v1}
\fmfforce{0.9375w,0.5h}{v2}
\fmf{plain}{v1,v2}
\fmffreeze
\fmfposition
\fmfipath{pm[]}
\fmfiset{pm1}{vpath(__v1,__v2)}
\nvml{1}{$\scriptstyle p$}
\fmfiv{label=$\scriptstyle A$,l.dist=2}{vloc(__v1)}
\fmfiv{label=$\scriptstyle B$,l.dist=2}{vloc(__v2)}
\end{fmfchar*}}}
&{}={}
\langle  Z^B(p)\bar Z_A(-p)\rangle
=\langle \bar W^B(p)W_A(-p)\rangle
=\frac{\delta_A^B}{p^2}\delta^4(\theta_1-\theta_2)\col\\
\settoheight{\eqoff}{$\times$}%
\setlength{\eqoff}{0.5\eqoff}%
\addtolength{\eqoff}{-3.75\unitlength}%
\raisebox{\eqoff}{%
\fmfframe(2,2)(2,2){%
\begin{fmfchar*}(15,7.5)
\fmfleft{v1}
\fmfright{v2}
\fmfforce{0.0625w,0.5h}{v1}
\fmfforce{0.9375w,0.5h}{v2}
\fmf{dots}{v1,v2}
\fmffreeze
\fmfposition
\fmfipath{pm[]}
\fmfiset{pm1}{vpath(__v1,__v2)}
\nvml{1}{$\scriptstyle p$}
\end{fmfchar*}}}
&{}={}
\langle \bar c'(p)c(-p)\rangle
=-\langle c'(p)\bar c(-p)\rangle
\\
&{}={}
-\langle \hat{\bar c}'(p)\hat c(-p)\rangle
=\langle \hat c'(p)\hat{\bar c}(-p)\rangle
=\frac{1}{p^2}\delta^4(\theta_1-\theta_2)\col\\
\end{aligned}
\end{equation}
where diagonality in the gauge group indices and a factor $\frac{4\pi}{k}$ for each propagator have been suppressed.

The vertices are obtained by taking the functional derivatives of the Wick rotated action (no factors of $i$) w.r.t.\ the corresponding superfields; we will give only the vertices involved in the computations of our paper.
When a functional derivatives w.r.t.\ the (anti)-chiral superfields is taken, factors of $(\D^2)$
$\barD^2$ are generated in the vertices.
Omitting
factors $\frac{k}{4\pi}$, for the three point vertices we obtain
\begin{equation}\label{cvertices}
\begin{gathered}
\begin{aligned}
V_{V^3}
&=
\left(\cvert{photon}{photon}{photon}{}{$\scriptstyle\D^\alpha$}{$\scriptstyle\barD_{\!\alpha}$}
-\cvert{photon}{photon}{photon}{}{$\scriptstyle\barD^{\!\alpha}$}{$\scriptstyle\D_\alpha$}
\right)
\frac{1}{2}\tr\big(T^a\comm{T^b}{T^c}\big)
\col\\
\end{aligned}\\
\begin{aligned}
V_{VZ^B\bar Z_C}
&=
\cvert{photon}{plain}{plain}{}{$\scriptstyle\barD^2$}{$\scriptstyle\D^2$}
\delta_B^C\tr\big(T^aB^{\underline{b}}B_{\underline{c}}\big)
\col\qquad
V_{\hat VW_B\bar W^C}
=
\cvert{photon}{plain}{plain}{}{$\scriptstyle\barD^2$}{$\scriptstyle\D^2$}
\delta_C^B\tr\big(T^{\hat{a}}B_{\underline{b}}B^{\underline{c}}\big)
\col\\
V_{\hat V\bar Z_BZ^C}
&=
\cvert{photon}{plain}{plain}{}{$\scriptstyle\D^2$}{$\scriptstyle\barD^2$}
(-1)\delta_C^B\tr\big(T^{\hat{a}}B_{\underline{b}}B^{\underline{c}}\big)
\col\qquad
V_{V\bar W^BW_C}
=
\cvert{photon}{plain}{plain}{}{$\scriptstyle\D^2$}{$\scriptstyle\barD^2$}
(-1)\delta_B^C\tr\big(T^{a}B^{\underline{b}}B_{\underline{c}}\big)
\col\\
\end{aligned}\\
\begin{aligned}
V_{Vcc'}
&=
\cvert{photon}{dots}{dots}{}{$\scriptstyle\barD^2$}{$\scriptstyle\barD^2$}\frac{1}{2}\tr\big(T^a\comm{T^b}{T^c}\big)
\col\qquad
V_{Vc\bar c'}
=
\cvert{photon}{dots}{dots}{}{$\scriptstyle\barD^2$}{$\scriptstyle\D^2$}\frac{1}{2}\tr\big(T^a\comm{T^b}{T^c}\big)
\col\\
V_{V\bar cc'}
&=
\cvert{photon}{dots}{dots}{}{$\scriptstyle\D^2$}{$\scriptstyle\barD^2$}\frac{1}{2}\tr\big(T^a\comm{T^b}{T^c}\big)
\col\qquad
V_{V\bar c\bar c'}
=
\cvert{photon}{dots}{dots}{}{$\scriptstyle\D^2$}{$\scriptstyle\D^2$}\frac{1}{2}\tr\big(T^a\comm{T^b}{T^c}\big)
\col\\
\end{aligned}
\end{gathered}
\end{equation}
where the colour indices are labeled $(a,b,c)$ counter clockwise
starting with the leg to the left. 
Besides the matrices $T^a$ and $T^{\hat a}$ transforming in the adjoint of the respective gauge groups $U(M)$ and $U(N)$, we have introduced matrices 
$B^{\underline{a}}$ and $B_{\underline{a}}$,
with underlined $\underline{a}=1,\cdots,MN$ indices 
that transform in the $({\bf M}, {\bf \bar{N}})$ and
$({\bf N}, {\bf \bar{M}})$ of the gauge group $U(M)\times U(N)$.
The previous notations are useful because one can 
effectively consider all the matrices to be the same for $M=N$ and then only at the end one can 
easily recover 
the different factors of $M$ and $N$ coming 
from the colour contractions.

The quartic vertices used in the paper are
\begin{equation}\label{qvertices}
\begin{gathered}
\begin{aligned}
V_{V^2Z^C\bar Z_D}
&=
\qvert{photon}{photon}{plain}{plain}{}{}{$\scriptstyle\barD^2$}{$\scriptstyle\D^2$}\frac{1}{2}\delta_C^D\big[\tr\big(\{T^a,T^b\}B^{\underline{c}}B_{\underline{d}}\big)
\big]
\col\\
V_{\hat V^2\bar Z_CZ^D}
&=
\qvert{photon}{photon}{plain}{plain}{}{}{$\scriptstyle\D^2$}{$\scriptstyle\barD^2$}\frac{1}{2}\delta_D^C\big[\tr\big(\{T^{\hat{a}},T^{\hat{b}}\}B_{\underline{c}}B^{\underline{d}}\big)
\big]
\col\\
V_{VZ^B\hat V\bar Z_D}
&=
\qvert{photon}{plain}{photon}{plain}{}{$\scriptstyle\barD^2$}{}{$\scriptstyle\D^2$}(-1)\delta_B^D\tr\big(T^aB^{\underline{b}}T^{\hat{c}}B_{\underline{d}}\big)
\col
\end{aligned}
\end{gathered}
\end{equation}
where the colour indices are labeled $(a,b,c,d)$ counter clockwise
starting with the leg in the upper left corner.
The vertices $V_{\hat V^2W_C\bar W^D}$, $V_{V^2\bar W^CW_D}$, $V_{\hat VW_BV\bar W^D}$ 
involving the $W_A$ and $\bar{W}^{A}$ superfields 
are respectively identical to the previous three vertices up to trivial modifications in the flavour
and colour structures.

The quartic superpotential vertices are
\begin{equation}
\begin{gathered}
\begin{aligned}
V_{Z^AW_BZ^CW_D}
&=
\qvert{plain}{plain}{plain}{plain}{$\scriptstyle\barD^2$}{$\scriptstyle\barD^2$}{$\scriptstyle\barD^2$}{}i\epsilon^{AC}\epsilon_{BD}\big[
\tr\big(B^{\underline{a}}B_{\underline{b}}B^{\underline{c}}B_{\underline{d}}\big)
-\tr\big(B^{\underline{c}}B_{\underline{b}}B^{\underline{a}}B_{\underline{d}}\big)
\big]
\col\\
V_{\bar Z_A\bar W^B\bar Z_C\bar W^D}
&=
\qvert{plain}{plain}{plain}{plain}{$\scriptstyle\D^2$}{$\scriptstyle\D^2$}{$\scriptstyle\D^2$}{}i\epsilon_{AC}\epsilon^{BD}\big[
\tr\big(B_{\underline{a}}B^{\underline{b}}B_{\underline{c}}B^{\underline{d}}\big)
-\tr\big(B_{\underline{c}}B^{\underline{b}}B_{\underline{a}}B^{\underline{d}}\big)
\big]
\col
\end{aligned}
\end{gathered}
\end{equation}
where again  the colour indices are labeled $(a,b,c,d)$ counter clockwise
starting with the leg in the upper left corner.
Note also that, in a standard way, one of the ($\D^2$) $\barD^2$ factors has been absorbed 
into the (anti)chiral integration such that the integration
measure of the (anti)chiral vertex is promoted to the full superspace measure.

\section{Integrals}
\label{Integrals}

In this section we collect the integrals required for our paper.
The results are based on the Appendices H, I, J of \cite{Minahan:2009wg}
where the reader should look to have a complete description of the notations 
and results that we are using.

The integrals are computed by using dimensional regularization in Euclidean space 
with $D$ dimensions and
\begin{equation}
D=2(\lambda+1)=3-2\varepsilon\col
\qquad
\lambda=\frac{1}{2}-\varepsilon
\pnt
\end{equation}
As usual we will expand the integrals in the limit $\varepsilon\to0$ up to the order needed for 
our computations.
The parameter $\lambda$ in this appendix
 should not be confused
with the 't Hooft coupling that appears in the main body of the paper.
The integrals have a simple dependence on the external momentum $p_\mu$ which we will
omit. 
Relations between four-loop expressions are understood to hold for 
the pole parts up to disregarded finite contributions.

\subsection{Integrals with only UV divergences}

We need the following two-loop integral
\begin{equation}\label{I2}
\begin{aligned}
I_2=
\settoheight{\eqoff}{$\times$}%
\setlength{\eqoff}{0.5\eqoff}%
\addtolength{\eqoff}{-5\unitlength}%
\raisebox{\eqoff}{%
\begin{fmfchar*}(15,10)
  \fmfleft{in}
  \fmfright{out1}
\fmf{plain}{in,v1}
\fmf{plain}{out,v2}
\fmfforce{(0,0.5h)}{in}
\fmfforce{(w,0.5h)}{out}
\fmffixed{(0.75w,0)}{v1,v2}
  \fmf{plain,left=0.5
  ,l.side=left,l.dist=2}{v1,v2}
  \fmf{plain,right=0.5}{v1,v2}
  \fmf{plain}{v1,v2}
\end{fmfchar*}}
&=G(1,1)G(1-\lambda,1)
\pnt
\end{aligned}
\end{equation}
The reader can look at  the 
appendix H of \cite{Minahan:2009wg} for our notations in using
the $G$-functions.
Furthermore, we need 
the following two-loop integrals with two contracted momenta in their numerators
\begin{equation}
\begin{aligned}
\settoheight{\eqoff}{$\times$}%
\setlength{\eqoff}{0.5\eqoff}%
\addtolength{\eqoff}{-10\unitlength}%
\settoheight{\eqoff}{$\times$}%
\setlength{\eqoff}{0.5\eqoff}%
\addtolength{\eqoff}{-10\unitlength}%
I_{221be}=
\raisebox{\eqoff}{%
\fmfframe(1,0)(1,0){%
\begin{fmfchar*}(9,18)
\fmftop{vt}
\fmfbottom{vb}
\fmffixed{(0,0.1h)}{vo,vt1}
\fmffixed{(0,0.1h)}{vb1,vi}
\fmffixed{(0,0.75h)}{vi,vo}
\fmffixed{(0.8w,0)}{v1,v2}
\fmf{phantom}{vt1,vt}
\fmf{phantom}{vb,vb1}
\fmf{derplains}{vi,vb1}
\fmf{plain}{vt1,vo}
\fmf{plain,right=0.25}{v1,vi}
\fmf{derplains,right=0.25}{vi,v2}
\fmf{derplain,right=0.25}{vo,v1}
\fmf{plain,right=0.25}{v2,vo}
\fmf{derplain}{v1,v2}
\end{fmfchar*}}}
&=\frac{1}{2}(-G_1(1,1)G(1,1)-G(1,1)G_1(2-\lambda,1)
+G_1(1,1)G_1(2-\lambda,1))
\col\\
\settoheight{\eqoff}{$\times$}%
\setlength{\eqoff}{0.5\eqoff}%
\addtolength{\eqoff}{-10\unitlength}%
I_{221dc}=
\raisebox{\eqoff}{%
\fmfframe(1,0)(1,0){%
\begin{fmfchar*}(9,18)
\fmftop{vt}
\fmfbottom{vb}
\fmffixed{(0,0.1h)}{vo,vt1}
\fmffixed{(0,0.1h)}{vb1,vi}
\fmffixed{(0,0.75h)}{vi,vo}
\fmffixed{(0.8w,0)}{v1,v2}
\fmf{phantom}{vt1,vt}
\fmf{phantom}{vb,vb1}
\fmf{derplains}{vi,vb1}
\fmf{plain}{vt1,vo}
\fmf{plain,right=0.25}{v1,vi}
\fmf{derplain,right=0.25}{vi,v2}
\fmf{derplain,right=0.25}{vo,v1}
\fmf{plain,right=0.25}{v2,vo}
\fmf{derplains}{v1,v2}
\end{fmfchar*}}}
&
=-G_1(1,1)G_1(2-\lambda,1)
\pnt
\end{aligned}
\end{equation}

At four loops there are many integrals involved in the computations.
Here we list the results for the pole parts of the UV logarithmically divergent
integrals where the subdivergences
have already been subtracted. 
Four-loop integrals with no momenta in their numerators are
\begin{equation}\label{I4}
\begin{aligned}
I_4=
\settoheight{\eqoff}{$\times$}%
\setlength{\eqoff}{0.5\eqoff}%
\addtolength{\eqoff}{-5.5\unitlength}%
\raisebox{\eqoff}{%
\fmfframe(0,3)(0,-7){%
\begin{fmfchar*}(15,15)
  \fmfleft{in}
  \fmfright{out}
  \fmftop{top}
\fmf{plain}{in,v1}
\fmf{plain}{out,v2}
  \fmf{phantom}{top,v3}
\fmfpoly{phantom}{v2,v3,v1}
\fmffixed{(whatever,0)}{in,v1}
\fmffixed{(whatever,0)}{out,v2}
\fmffixed{(0.75w,0)}{v1,v2}
  \fmf{plain,left=0.25}{v1,v3}
  \fmf{plain,right=0.25}{v1,v3}
  \fmf{plain}{v1,v3}
  \fmf{plain,left=0.25}{v3,v2}
  \fmf{plain,right=0.25}{v3,v2}
  \fmf{plain}{v1,v2}
\end{fmfchar*}}}
&=\frac{1}{(8\pi)^4}\Big(-\frac{1}{2\varepsilon^2}+\frac{2}{\varepsilon}\Big)
\col\\
I_{4\mathbf{bbb}}=
\settoheight{\eqoff}{$\times$}%
\setlength{\eqoff}{0.5\eqoff}%
\addtolength{\eqoff}{-5.5\unitlength}%
\raisebox{\eqoff}{%
\fmfframe(0,3)(0,-7){%
\begin{fmfchar*}(15,15)
  \fmfleft{in}
  \fmfright{out}
  \fmftop{top}
\fmf{plain}{in,v1}
\fmf{plain}{out,v2}
  \fmf{phantom}{top,v3}
\fmfpoly{phantom}{v2,v3,v1}
\fmffixed{(whatever,0)}{in,v1}
\fmffixed{(whatever,0)}{out,v2}
\fmffixed{(0.75w,0)}{v1,v2}
  \fmf{plain,left=0.25}{v1,v3}
  \fmf{plain,right=0.25}{v1,v3}
  \fmf{plain,left=0.25}{v3,v2}
  \fmf{plain,right=0.25}{v3,v2}
  \fmf{plain,left=0.25}{v1,v2}
  \fmf{plain,right=0.25}{v1,v2}
\end{fmfchar*}}}
&
=\frac{1}{(8\pi)^4}\frac{\pi^2}{2\varepsilon}
\pnt
\end{aligned}
\end{equation}
Four-loop integrals with two contracted momenta in their numerators
are
\begin{equation}
\begin{aligned}
I_{42\mathbf{bbb2}}=
\Ifourtwobbbtwo{plain}{plain}{plain}{derplain}{derplain}{plain}
&=\frac{1}{(8\pi)^4}\frac{\pi^2}{4\varepsilon}
\col\\
I_{42\mathbf{bb0}cd}=
\Ifourtwobbbthree{plain}{plain}{derplain}{derplain}{plain}{plain}
&=\frac{1}{(8\pi)^4}\Big(\frac{1}{4\varepsilon^2}-\frac{3}{2\varepsilon}\Big)
\col\\
I_{42\mathbf{b}bd}
=
\Ifourtwoe{plain}{derplain}{plain}{derplain}{plain}{plain}{plain}
&=\frac{1}{(8\pi)^4}\Big(\frac{1}{2\varepsilon^2}-\frac{1}{\varepsilon}\Big(2-\frac{\pi^2}{4}\Big)\Big)
\col\\
I_{42\mathbf{b}be}
\Ifourtwoe{plain}{derplain}{plain}{plain}{derplain}{plain}{plain}
&=\frac{1}{(8\pi)^4}\Big(-\frac{1}{4\varepsilon^2}\Big)
\pnt
\end{aligned}
\end{equation}
Let us consider now four-loop integrals with four 
pairwise contracted momenta in their numerators.
The following ones
\begin{equation}
\begin{aligned}
I_{422\mathbf{b}ABcd}
=
\smash[b]{\Ifourtwotwob{derplain}{derplain}{plain}{plain}{plain}{plain}{derplains}{derplains}}
\col\qquad
I_{422\mathbf{b}AcBd}
=
\smash[b]{\Ifourtwotwob{derplain}{derplains}{plain}{plain}{plain}{plain}{derplain}{derplains}}
\col\qquad
I_{422\mathbf{b}AdBc}
=
\Ifourtwotwob{derplain}{derplains}{plain}{plain}{plain}{plain}{derplains}{derplain}
\col\qquad
\end{aligned}
\end{equation}
appear in a fixed combination which 
can be recast into the form
\begin{equation}
\begin{aligned}
I_{422\mathbf{b}\text{tr}ABcd}
&=
-\tr\Ifourtwotwob{derplain}{derplain}{plain}{plain}{plain}{plain}{derplain}{derplain}
=
-2(I_{422\mathbf{b}ABcd}-I_{422\mathbf{b}AcBd}+I_{422\mathbf{b}AdBc})\\
&=
2
\settoheight{\eqoff}{$\times$}%
\setlength{\eqoff}{0.5\eqoff}%
\addtolength{\eqoff}{-10\unitlength}%
\raisebox{\eqoff}{%
\fmfframe(0,0)(0,0){%
\begin{fmfchar*}(20,20)
 \fmfleft{vl}
 \fmfright{vr}
 \fmftop{vt}
 \fmfbottom{vb}
 \fmf{phantom,tension=1}{vl,v1}
 \fmf{phantom,tension=1}{vr,v3}
 \fmf{phantom,tension=1}{vt,v4}
 \fmf{plain,tension=1}{vb,v2}
\fmffixed{(0.9w,0)}{v1,v3}
\fmfpoly{phantom}{v1,v2,v3,v4}
 \fmf{plain,right=0.25}{v1,v2}
 \fmf{plain,right=0.25}{v2,v3}
 \fmf{derplain,right=0.25}{v3,v4}
 \fmf{derplain,right=0.25}{v4,v1}
\fmffreeze
 \fmf{plain}{v0,v1}
 \fmf{phantom}{v0,v2}
 \fmf{plain}{v0,v3}
 \fmf{phantom}{v0,v4}
\fmffreeze
 \fmf{plain,right=0.25}{v0,v4}
 \fmf{plain,left=0.25}{v0,v4}
\end{fmfchar*}}}
+
2
\settoheight{\eqoff}{$\times$}%
\setlength{\eqoff}{0.5\eqoff}%
\addtolength{\eqoff}{-10\unitlength}%
\raisebox{\eqoff}{%
\fmfframe(-1,0)(-1,0){%
\begin{fmfchar*}(20,20)
 \fmfleft{vl}
 \fmfright{vr}
 \fmftop{vt}
 \fmfbottom{vb}
 \fmf{phantom,tension=1}{vl,v1}
 \fmf{phantom,tension=1}{vr,v3}
 \fmf{phantom,tension=1}{vt,v4}
 \fmf{plain,tension=1}{vb,v2}
\fmffixed{(0.9w,0)}{v1,v3}
\fmfpoly{phantom}{v1,v2,v3,v4}
 \fmf{phantom,right=0.25}{v1,v2}
 \fmf{phantom,right=0.25}{v2,v3}
 \fmf{phantom,right=0.25}{v3,v4}
 \fmf{phantom,right=0.25}{v4,v1}
\fmffreeze
 \fmf{phantom}{v0,v1}
 \fmf{phantom}{v0,v2}
 \fmf{phantom}{v0,v3}
 \fmf{phantom}{v0,v4}
\fmffreeze
 \fmf{plain,right=0.25}{v0,v4}
 \fmf{plain,left=0.25}{v0,v4}
 \fmf{plain,right=0.75}{v0,v4}
 \fmf{plain,left=0.75}{v0,v4}
 \fmf{plain,right=0.75}{v4,v2}
 \fmf{plain,left=0.75}{v4,v2}
\end{fmfchar*}}}
-
4
\settoheight{\eqoff}{$\times$}%
\setlength{\eqoff}{0.5\eqoff}%
\addtolength{\eqoff}{-10\unitlength}%
\raisebox{\eqoff}{%
\fmfframe(-1,0)(-1,0){%
\begin{fmfchar*}(20,20)
 \fmfleft{vl}
 \fmfright{vr}
 \fmftop{vt}
 \fmfbottom{vb}
 \fmf{phantom,tension=1}{vl,v1}
 \fmf{phantom,tension=1}{vr,v3}
 \fmf{phantom,tension=1}{vt,v4}
 \fmf{plain,tension=1}{vb,v2}
\fmffixed{(0.9w,0)}{v1,v3}
\fmfpoly{phantom}{v1,v2,v3,v4}
 \fmf{phantom,right=0.25}{v1,v2}
 \fmf{derplain,right=0.25}{v2,v3}
 \fmf{derplain,right=0.25}{v3,v4}
 \fmf{phantom,right=0.25}{v4,v1}
\fmffreeze
 \fmf{phantom}{v0,v1}
 \fmf{plain}{v0,v2}
 \fmf{plain}{v0,v3}
 \fmf{plain}{v0,v4}
\fmffreeze
 \fmf{plain,right=0.5}{v0,v4}
 \fmf{plain,left=0.5}{v0,v4}
 \fmf{plain,right=0.75}{v4,v2}
\end{fmfchar*}}}
=
\frac{1}{(8\pi)^4}\Big(\frac{1}{\varepsilon^2}-\frac{2}{\varepsilon}\Big)
\col
\end{aligned}
\end{equation}
Here we have taken the trace of $\gamma$-matrices contracted with the momenta in the integral.
We thereby read off the momenta in a cycle, but keep their direction as indicated by the arrows.

We also need  the integrals
\begin{equation}
\begin{aligned}
I_{422\mathbf{q}ABbd}
=
\smash[b]{\Ifourtwotwoq{derplain}{derplain}{plain}{plain}{plain}{derplains}{plain}{derplains}}
&=
\frac{1}{(8\pi)^4}\Big(
\frac{1}{4\varepsilon^2}+\frac{1}{4\varepsilon}\Big)
\col\\
~\\
I_{422\mathbf{q}AdBb}
=
\smash[b]{
\Ifourtwotwoq{derplain}{derplains}{plain}{plain}{plain}{derplains}{plain}{derplain}}
&=
\frac{1}{(8\pi)^4}\Big(\frac{1}{2\varepsilon^2}-\frac{1}{\varepsilon}\Big(
1-\frac{\pi^2}{4}\Big)\Big)
\col\\
~\\
I_{422\mathbf{q}AbBd}
=
\smash[b]{\Ifourtwotwoq{derplain}{derplains}{plain}{plain}{plain}{derplain}{plain}{derplains}}
&=
\frac{1}{(8\pi)^4}\Big(\frac{1}{4\varepsilon^2}+\frac{1}{\varepsilon}\Big(
\frac{5}{4}-\frac{\pi^2}{12}\Big)\Big)
\pnt\\
~\\
\end{aligned}
\end{equation}
The linear combinations of integrals originating from the traces of 
$\gamma$-matrices read
\begin{equation}
\begin{aligned}
I_{422\mathbf{q}\text{tr}ABbd}
=-
\smash[b]{\tr\Ifourtwotwoq{derplain}{derplain}{plain}{plain}{plain}{derplain}{plain}{derplain}}
&=
-2(I_{422\mathbf{q}ABbd}-I_{422\mathbf{q}AbBd}+I_{422\mathbf{q}AdBb})\\
&
=\frac{1}{(8\pi)^4}\Big(-\frac{1}{\varepsilon^2}
+\frac{1}{\varepsilon}\Big(4-\frac{2}{3}\pi^2\Big)\Big)
\col\\
I_{422\mathbf{q}\text{tr}ABCD}
=
\tr\Ifourtwotwoq{derplain}{derplain}{derplain}{derplain}{plain}{plain}{plain}{plain}
&=\frac{1}{(8\pi)^4}\frac{\pi^2}{3\varepsilon}
\pnt
\end{aligned}
\end{equation}

There is an interesting relation involving the traces. It reads
\begin{equation}
\begin{aligned}
I_{422\mathbf{q}\text{tr}ABCD}
&=I_{422\mathbf{q}\text{tr}ABbd}
+2I_4+4I_{42\mathbf{b}bd}
&=\frac{1}{(8\pi)^4}\frac{\pi^2}{3\varepsilon}
\pnt
\end{aligned}
\end{equation}

\subsection{Integrals with IR divergences}
\label{IRintegrals}

In this subsection we collect the integrals having poles in $\varepsilon$ 
which are due to IR divergences. By suffixes 
IR and UVIR we thereby label 
integrals which have one or both IR and UV  divergences.

The simplest two-loop integral with both an IR and an UV divergence
is the logarithmically divergent tadpole
\begin{equation}\label{I2tp}
\begin{aligned}
I_{2\text{tp}}
=I_{2\text{UVIR}}
=
\settoheight{\eqoff}{$\times$}%
\setlength{\eqoff}{0.5\eqoff}%
\addtolength{\eqoff}{-5.5\unitlength}%
\raisebox{\eqoff}{%
\fmfframe(0,3)(0,-7){%
\begin{fmfchar*}(15,15)
  \fmfleft{in}
  \fmfright{out}
  \fmftop{top}
\fmf{plain}{in,v1}
\fmf{plain}{out,v2}
  \fmf{phantom}{top,v3}
\fmfpoly{phantom}{v2,v3,v1}
\fmffixed{(whatever,0)}{in,v1}
\fmffixed{(whatever,0)}{out,v2}
\fmffixed{(0.75w,0)}{v1,v2}
  \fmf{plain}{v1,vc}
  \fmf{plain}{vc,v2}
\fmffreeze
  \fmf{plain,left=0.5}{vc,v3}
  \fmf{plain,right=0.5}{vc,v3}
  \fmf{plain}{vc,v3}
\end{fmfchar*}}}
&=0
\pnt
\end{aligned}
\end{equation}
It is zero in dimensional regularization, i.e.\ the IR and the UV divergence 
cancel against each other. The UV divergence can be extracted by reshuffling 
the external momentum. 
In  particular, the UV divergence of $I_{2\text{tp}}$ is $I_2$ defined in 
\eqref{I2};
then the IR divergence  of $I_{2\text{tp}}$ is  $-I_2$.

The simplest two-loop integral with only an IR divergence 
 is given by
\begin{equation}\label{I2IR}
\begin{aligned}
I_{2\text{IR}}
=
\settoheight{\eqoff}{$\times$}%
\setlength{\eqoff}{0.5\eqoff}%
\addtolength{\eqoff}{-5.5\unitlength}%
\raisebox{\eqoff}{%
\fmfframe(0,3)(0,-7){%
\begin{fmfchar*}(15,15)
  \fmfleft{in}
  \fmfright{out}
  \fmftop{top}
\fmf{plain}{in,v1}
\fmf{plain}{out,v2}
  \fmf{phantom}{top,v3}
\fmfpoly{phantom}{v2,v3,v1}
\fmffixed{(whatever,0)}{in,v1}
\fmffixed{(whatever,0)}{out,v2}
\fmffixed{(0.75w,0)}{v1,v2}
  \fmf{plain,left=0.25}{v1,v3}
  \fmf{plain,right=0.25}{v1,v3}
  \fmf{plain}{v3,v2}
  \fmf{plain}{v1,v2}
\end{fmfchar*}}}
=G(1,1)G(2-\lambda,1)
=\frac{1}{(8\pi)^2}\Big(-\frac{1}{\varepsilon}+2(1+\gamma-\ln 4\pi)
+O(\varepsilon)\Big)
\pnt
\end{aligned}
\end{equation}
One four-loop integral with both, an IR and a UV divergence is
given by\footnote{Note that, according to \cite{Minahan:2009wg}, with $\Kop()$ we mean
the extraction of the pole parts of a function of $\varepsilon$.}
\begin{equation}\label{I4IR}
\begin{aligned}
I_{4\text{UVIR}}
=
\settoheight{\eqoff}{$\times$}%
\setlength{\eqoff}{0.5\eqoff}%
\addtolength{\eqoff}{-5.5\unitlength}%
\raisebox{\eqoff}{%
\fmfframe(0,3)(0,-7){%
\begin{fmfchar*}(15,15)
  \fmfleft{in}
  \fmfright{out}
  \fmftop{top}
\fmf{plain}{in,v1}
\fmf{plain}{out,v2}
  \fmf{phantom}{top,v3}
\fmfpoly{phantom}{v2,v3,v1}
\fmffixed{(whatever,0)}{in,v1}
\fmffixed{(whatever,0)}{out,v2}
\fmffixed{(0.75w,0)}{v1,v2}
  \fmf{plain,left=0.25}{v1,v3}
  \fmf{plain,right=0.25}{v1,v3}
  \fmf{plain,left=0.25}{v1,v2}
  \fmf{plain,right=0.25}{v1,v2}
  \fmf{plain}{v1,v2}
  \fmf{plain}{v3,v2}
\end{fmfchar*}}}
&=\Kop(G(1,1)^2G(1-\lambda,1)G(1-2\lambda,2-\lambda))\\
&=\frac{1}{(8\pi)^4}\Big(-\frac{1}{2\varepsilon^2}+\frac{2}{\varepsilon}(-2+\gamma-\ln 4\pi)\Big)
\pnt
\end{aligned}
\end{equation}
Its IR divergence is extracted as
\begin{equation}
\begin{aligned}
I_{4\text{UVIR}}
-I_{4}
=\frac{1}{(8\pi)^4}\Big(\frac{2}{\varepsilon}(-3+\gamma-\ln 4\pi)\Big)
\col
\end{aligned}
\end{equation}
where $I_4$ removes the overall UV divergence, since $I_{\text{4UVIR}}$ 
does not have a UV subdivergence.

The simplest four-loop integral with only an IR divergence 
as overall divergence
is given by
\begin{equation}
\begin{aligned}
I_{4\text{IR}}
=
\settoheight{\eqoff}{$\times$}%
\setlength{\eqoff}{0.5\eqoff}%
\addtolength{\eqoff}{-10\unitlength}%
\smash[b]{%
\raisebox{\eqoff}{%
\fmfframe(1,0)(1,0){%
\begin{fmfchar*}(15,20)
  \fmfleft{vl}
  \fmfright{vr}
  \fmftop{vt}
  \fmfbottom{vb}
\fmfforce{(0,0.5h)}{vl}
\fmfforce{(w,0.5h)}{vr}
  \fmf{phantom,tension=1}{vt,v1}
  \fmf{plain,tension=1}{vl,v2}
  \fmf{phantom,tension=1}{vb,v3}
  \fmf{plain,tension=1}{vr,v4}
\fmffixed{(0,0.9h)}{v3,v1}
\fmfpoly{phantom}{v1,v2,v4}
\fmfpoly{phantom}{v3,v4,v2}
  \fmf{plain,left=0.25}{v2,v1}
  \fmf{plain}{v2,v1}
  \fmf{plain,right=0.25}{v2,v1}
  \fmf{plain,right=0.25}{v4,v1}
  \fmf{plain,right=0.25}{v2,v3}
  \fmf{plain,left=0.25}{v2,v3}
  \fmf{plain,right=0.25}{v3,v4}
\end{fmfchar*}}}}
&=\Kop(G(1,1)^2G(1-\lambda,1)G(2-2\lambda,2-\lambda))-\Kop(I_2)I_{2\text{IR}}\\
&=\frac{1}{(8\pi)^4}\Big(\frac{2}{\varepsilon}(-3
+
\gamma
-\ln 4\pi)\Big)
\pnt
\end{aligned}
\end{equation}
Here we have subtracted the UV subdivergence.

\section{Relevant one- and two-loop subdiagrams}
\label{two-loop-sub}

In this 
appendix
 we collect the results for the planar contributions to the
one-loop vector 
superfield
 two-point function 
and the chiral 
superfield
two-loop contributions to the two and four-point 
functions. 
The two-point functions have been first computed in 
\cite{Akerblom:2009gx,Bianchi:2009ja,Bianchi:2009rf} in the Landau gauge. 
Such analysis has been extended to general gauges in \cite{Leoni:2010az} where
the four point functions have also first been given for the ABJM case. Here we give the results
extended to the $U(M)\times U(N)$ ABJ case in the Landau gauge.
The 
two-loop corrections to the chiral propagator and superpotential enter
as subdiagrams 
in the evaluation of the dilatation operator given in section \ref{Dilatation-OP}.

\subsection{One-loop vector two-point function}

For the $U(M)$ vector superfield $V$ 
the one-loop two-point function gets contributions from three kind of diagrams respectively 
having matter, ghosts and vector superfields propagating in the one-loop bubble.

The contribution coming from the chiral matter superfields is
\begin{equation}
\begin{aligned}
\Sigma_{V,{\rm matter}}
=\vacpolD{photon}{plain}{photon}{}{}{}{}{}{}{}
\to 2N\delta^{ab}G(1,1)
\D^\alpha\barD^2\D_\alpha
\pnt
\end{aligned}
\end{equation}

The ghosts correction is
\begin{equation}
\begin{aligned}
\Sigma_{V,{\rm ghosts}}
=\vacpolD{photon}{dots}{photon}{}{}{}{}{}{}{}
\to \frac{1}{2}M\delta^{ab}G(1,1)
\big(-\D^\alpha\barD^2\D_\alpha+\acomm{\D^2}{\barD^2}\big)
\pnt
\end{aligned}
\end{equation}

The diagrams involving a loop of vectors sum up to the following contribution
\begin{equation}
\begin{aligned}
\Sigma_{V,{\rm vectors}}
=\vacpolD{photon}{photon}{photon}{}{}{}{}{}{}{}
\to\frac{1}{2}M\delta^{ab}G(1,1)\big(-\acomm{\D^2}{\barD^2}\big)
\pnt
\end{aligned}
\end{equation}

The total contribution to the two-point function for the $V$ superfield is then
\begin{equation}
\begin{aligned}
\Sigma_{V}
=
\settoheight{\eqoff}{$\times$}%
\setlength{\eqoff}{0.5\eqoff}%
\addtolength{\eqoff}{-7.5\unitlength}%
\raisebox{\eqoff}{%
\fmfframe(1,0)(1,0){%
\begin{fmfchar*}(20,15)
\fmfleft{vc1}
\fmfright{vc2}
\fmfforce{(0.0625w,0.5h)}{vc1}
\fmfforce{(0.9375w,0.5h)}{vc2}
\fmf{photon,tension=0.5,left=#1}{vc1,vc2}
\fmffreeze
\fmfposition
\vacpol{vc1}{vc2}
\end{fmfchar*}}}
&\to
\frac{1}{2}\delta^{ab}G(1,1)(4N-M)\D^\alpha\barD^2\D_\alpha
\pnt
\end{aligned}
\end{equation}
The corrections to the $U(N)$ gauge vector $\hat{V}$ two point function are clearly the same with 
the only difference that one has to exchange $M$ with $N$ in the results.

\subsection{Two-loop chiral two-point function}
\label{app:chiralse}

The non-vanishing contributions to the two-point function of chiral superfields can be seen to arise 
from the following diagrams
\begin{equation}
\begin{aligned}
\swftwoone{plain}{1}{-1}
&\to 2MNI_2\col
\\
\swftwoone{photon}{1}{-1}
&\to 2MNI_2
\col\\
\swftwoone{photon}{1}{0.5}
&\to-\frac{1}{2}M^2I_2
\col\\
\smash[b]{\swftwofive{1}}
&\to-MN
(G(1,1))^2
\col\\
~\\
\smash[b]{\swftwoseven{1}}
&\to(4N-M)MG(1,1)G_1(1,2-\lambda)
\\
&=\frac{1}{2}
(4N-M)M(I_{2\text{tp}}-I_2+I_{2\text{IR}})\\
\swftwoeight{1}
&\to
-\frac{1}{2}(4N-M)MI_{2\text{tp}}
\col\\
\end{aligned}
\end{equation}
where, in each contribution, 
we have omitted a factor $\D^2\barD^2$ together with the colour and flavour structures.
As discussed in section \ref{IRintegrals}, the tadpole integral $I_{2\text{tp}}$
is zero in dimensional 
regularization. However, we keep track of it by splitting its UV and IR divergent parts. 
This is necessary for the check of the cancellation of the IR divergences performed in 
appendix \ref{IR_cancellation}.

Taking into account reflections of the diagrams at the vertical and horizontal 
axes where necessary, and summing up the contributions, the result reads
\begin{equation}
\begin{aligned}
\Sigma_C&=
\settoheight{\eqoff}{$\times$}%
\setlength{\eqoff}{0.5\eqoff}%
\addtolength{\eqoff}{-7.5\unitlength}%
\raisebox{\eqoff}{%
\fmfframe(1,0)(1,0){%
\begin{fmfchar*}(20,15)
\fmfleft{v1}
\fmfright{v2}
\fmfforce{(0.0625w,0.5h)}{v1}
\fmfforce{(0.9375w,0.5h)}{v2}
\fmf{plain}{v1,v2}
\fmffreeze
\fmfposition
\vacpol{v1}{v2}
\end{fmfchar*}}}
\to-2MN
(G(1,1))^2
+\frac{1}{2}(8MN-(M^2+N^2))I_{2\text{IR}}
\pnt
\end{aligned}
\end{equation}
Note that the result is UV finite and it includes an IR divergent term which turns out to be gauge 
dependent \cite {Leoni:2010az} and, according to the discussion in 
appendix \ref{IR_cancellation}, 
does
 not contribute to the dilatation operator.

\subsection{Two-loop chiral four-point function}

The two-loop renormalization of the superpotential has been studied in \cite{Leoni:2010az}.
Here we summarize the results and extend them to the ABJ $U(M)\times U(N)$ case.
It holds
\begin{equation}
\begin{aligned}
\settoheight{\eqoff}{$\times$}%
\setlength{\eqoff}{0.5\eqoff}%
\addtolength{\eqoff}{-11\unitlength}%
\raisebox{\eqoff}{%
\fmfframe(1,1)(1,1){%
\begin{fmfchar*}(20,20)
\fmftop{v1}
\fmfbottom{v2}
\fmfforce{(0w,h)}{v1}
\fmfforce{(0w,0)}{v2}
\fmffixed{(w,0)}{v1,v4}
\fmffixed{(w,0)}{v2,v3}
\fmf{plain}{v1,vc1}
\fmf{plain}{v2,vc1}
\fmf{plain}{v3,vc2}
\fmf{plain}{v4,vc2}
\fmf{plain,left=0.5}{vc1,vc}
\fmf{plain,right=0.5}{vc1,vc}
\fmf{plain,left=0.5}{vc,vc2}
\fmf{plain,right=0.5}{vc,vc2}
\end{fmfchar*}}}
& \to  -(4\pi)^2\lambda^2
(p_1+p_2)^2
\settoheight{\eqoff}{$\times$}%
\setlength{\eqoff}{0.5\eqoff}%
\addtolength{\eqoff}{-11\unitlength}%
\raisebox{\eqoff}{%
\fmfframe(1,1)(1,1){%
\begin{fmfchar*}(20,20)
\fmftop{v1}
\fmfbottom{v2}
\fmfforce{(0w,h)}{v1}
\fmfforce{(0w,0)}{v2}
\fmffixed{(w,0)}{v1,v4}
\fmffixed{(w,0)}{v2,v3}
\fmf{plain}{v1,vc1}
\fmf{plain}{v2,vc1}
\fmf{plain}{v3,vc2}
\fmf{plain}{v4,vc2}
\fmf{plain,left=0.5}{vc1,vc}
\fmf{plain,right=0.5}{vc1,vc}
\fmf{plain,left=0.5}{vc,vc2}
\fmf{plain,right=0.5}{vc,vc2}
\end{fmfchar*}}}
\col\\
\settoheight{\eqoff}{$\times$}%
\setlength{\eqoff}{0.5\eqoff}%
\addtolength{\eqoff}{-11\unitlength}%
\raisebox{\eqoff}{%
\fmfframe(1,1)(1,1){%
\begin{fmfchar*}(20,20)
\fmftop{v1}
\fmfbottom{v2}
\fmfforce{(0w,h)}{v1}
\fmfforce{(0w,0)}{v2}
\fmffixed{(w,0)}{v1,v4}
\fmffixed{(w,0)}{v2,v3}
\superpot{v1}{v2}{v3}{v4}
\fmfi{photon}{vm1{dir -90}..{dir -90}vm2}
\fmfi{photon}{vm1{dir -150}..{dir -30}vm2}
\end{fmfchar*}}}
& \to 
\frac{(4\pi)^2\lambda^2}{2}
(p_1+p_2)^2
\settoheight{\eqoff}{$\times$}%
\setlength{\eqoff}{0.5\eqoff}%
\addtolength{\eqoff}{-11\unitlength}%
\raisebox{\eqoff}{%
\fmfframe(1,1)(1,1){%
\begin{fmfchar*}(20,20)
\fmftop{v1}
\fmfbottom{v2}
\fmfforce{(0w,h)}{v1}
\fmfforce{(0w,0)}{v2}
\fmffixed{(w,0)}{v1,v4}
\fmffixed{(w,0)}{v2,v3}
\superpot{v1}{v2}{v3}{v4}
\fmfi{plain}{vm1{dir -90}..{dir -90}vm2}
\fmfi{plain}{vm1{dir -150}..{dir -30}vm2}
\end{fmfchar*}}}
\col\\
\settoheight{\eqoff}{$\times$}%
\setlength{\eqoff}{0.5\eqoff}%
\addtolength{\eqoff}{-11\unitlength}%
\raisebox{\eqoff}{%
\fmfframe(1,1)(1,1){%
\begin{fmfchar*}(20,20)
\fmftop{v1}
\fmfbottom{v2}
\fmfforce{(0w,h)}{v1}
\fmfforce{(0w,0)}{v2}
\fmffixed{(w,0)}{v1,v4}
\fmffixed{(w,0)}{v2,v3}
\superpot{v1}{v2}{v3}{v4}
\fmfi{photon}{vm1{dir -135}..{dir -45}vo2}
\fmfi{photon}{vm1{dir -90}..{dir -45}vi2}
\end{fmfchar*}}}
&\to
-\frac{(4\pi)^2\lambda^2}{2}
\left(
\tr(\gamma^\mu\gamma^\nu\gamma^\alpha\gamma^\beta)
\settoheight{\eqoff}{$\times$}%
\setlength{\eqoff}{0.5\eqoff}%
\addtolength{\eqoff}{-11\unitlength}%
\raisebox{\eqoff}{%
\fmfframe(1,1)(1,1){%
\begin{fmfchar*}(20,20)
\fmftop{v1}
\fmfbottom{v2}
\fmfforce{(0w,h)}{v1}
\fmfforce{(0w,0)}{v2}
\fmffixed{(w,0)}{v1,v4}
\fmffixed{(w,0)}{v2,v3}
\superpot[phantom]{v1}{v2}{v3}{v4}
\fmfi{derplain,label=$\scriptstyle\mu$,l.dist=2}{vm1--vloc(__v1)}
\fmfi{plain}{vm1--vloc(__vc)}
\fmfi{derplain,label=$\scriptstyle\alpha$,l.dist=2}{vo2--vi2}
\fmfi{derplain,label=$\scriptstyle\nu$,l.side=left,l.dist=2}{vo2--vloc(__v2)}
\fmfi{derplain,label=$\scriptstyle\beta$,l.dist=2}{vi2--vloc(__vc)}
\fmfi{plain}{vloc(__vc)--vloc(__v3)}
\fmfi{plain}{vloc(__vc)--vloc(__v4)}
\fmfi{plain}{vm1{dir -135}..{dir -45}vo2}
\fmfi{plain}{vm1{dir -90}..{dir -45}vi2}
\end{fmfchar*}}}
+
2p_2^2
\settoheight{\eqoff}{$\times$}%
\setlength{\eqoff}{0.5\eqoff}%
\addtolength{\eqoff}{-11\unitlength}%
\raisebox{\eqoff}{%
\fmfframe(1,1)(1,1){%
\begin{fmfchar*}(20,20)
\fmftop{v1}
\fmfbottom{v2}
\fmfforce{(0w,h)}{v1}
\fmfforce{(0w,0)}{v2}
\fmffixed{(w,0)}{v1,v4}
\fmffixed{(w,0)}{v2,v3}
\superpot[phantom]{v1}{v2}{v3}{v4}
\fmfi{plain}{vm1--vloc(__vc)}
\fmfi{plain}{vm1--vloc(__v1)}
\fmfi{derplain}{vo2--vi2}
\fmfi{plain}{vo2--vloc(__v2)}
\fmfi{derplain}{vi2--vloc(__vc)}
\fmfi{plain}{vloc(__vc)--vloc(__v3)}
\fmfi{plain}{vloc(__vc)--vloc(__v4)}
\fmfi{plain}{vm1{dir -135}..{dir -45}vo2}
\fmfi{plain}{vm1{dir -90}..{dir -45}vi2}
\end{fmfchar*}}}
\right)
\col\\
\settoheight{\eqoff}{$\times$}%
\setlength{\eqoff}{0.5\eqoff}%
\addtolength{\eqoff}{-11\unitlength}%
\raisebox{\eqoff}{%
\fmfframe(1,1)(1,1){%
\begin{fmfchar*}(20,20)
\fmftop{v1}
\fmfbottom{v2}
\fmfforce{(0w,h)}{v1}
\fmfforce{(0w,0)}{v2}
\fmffixed{(w,0)}{v1,v4}
\fmffixed{(w,0)}{v2,v3}
\superpot{v1}{v2}{v3}{v4}
\fmfi{photon}{vm1{dir -135}..{dir -45}vm2}
\fmfi{photon}{vm2{dir -45}..{dir 45}vm3}
\end{fmfchar*}}}
& \to
(4\pi)^2\lambda\hat\lambda
\tr(\gamma^\mu\gamma^\nu\gamma^\alpha\gamma^\beta)
\settoheight{\eqoff}{$\times$}%
\setlength{\eqoff}{0.5\eqoff}%
\addtolength{\eqoff}{-11\unitlength}%
\raisebox{\eqoff}{%
\fmfframe(1,1)(1,1){%
\begin{fmfchar*}(20,20)
\fmftop{v1}
\fmfbottom{v2}
\fmfforce{(0w,h)}{v1}
\fmfforce{(0w,0)}{v2}
\fmffixed{(w,0)}{v1,v4}
\fmffixed{(w,0)}{v2,v3}
\superpot[phantom]{v1}{v2}{v3}{v4}
\fmfi{derplain,label=$\scriptstyle\mu$,l.dist=2}{vm1--vloc(__v1)}
\fmfi{plain}{vm1--vloc(__vc)}
\fmfi{derplain,label=$\scriptstyle\beta$,l.side=left,l.dist=2}{vm1--vloc(__vc)}
\fmfi{derplain,label=$\scriptstyle\nu$,l.side=left,l.dist=2}{vm3--vloc(__v3)}
\fmfi{derplain,label=$\scriptstyle\alpha$,l.dist=2}{vm3--vloc(__vc)}
\fmfi{plain}{vloc(__vc)--vloc(__v2)}
\fmfi{plain}{vloc(__vc)--vloc(__v4)}
\fmfi{plain}{vm1{dir -135}..{dir -45}vm2}
\fmfi{plain}{vm2{dir -45}..{dir 45}vm3}
\end{fmfchar*}}}
\col\\
\settoheight{\eqoff}{$\times$}%
\setlength{\eqoff}{0.5\eqoff}%
\addtolength{\eqoff}{-11\unitlength}%
\raisebox{\eqoff}{%
\fmfframe(1,1)(1,1){%
\begin{fmfchar*}(20,20)
\fmftop{v1}
\fmfbottom{v2}
\fmfforce{(0w,h)}{v1}
\fmfforce{(0w,0)}{v2}
\fmffixed{(w,0)}{v1,v4}
\fmffixed{(w,0)}{v2,v3}
\superpot{v1}{v2}{v3}{v4}
\fmfi{photon}{vm1{dir -135}..{dir -45}vm2}
\fmfi{photon}{vo1{dir 45}..{dir -135}vi1}
\end{fmfchar*}}}
&\to
(4\pi)^2\lambda\hat\lambda
\tr(\gamma^\mu\gamma^\nu\gamma^\rho\gamma^\alpha\gamma^\beta\gamma^\gamma)
\settoheight{\eqoff}{$\times$}%
\setlength{\eqoff}{0.5\eqoff}%
\addtolength{\eqoff}{-11\unitlength}%
\raisebox{\eqoff}{%
\fmfframe(1,1)(1,1){%
\begin{fmfchar*}(20,20)
\fmftop{v1}
\fmfbottom{v2}
\fmfforce{(0w,h)}{v1}
\fmfforce{(0w,0)}{v2}
\fmffixed{(w,0)}{v1,v4}
\fmffixed{(w,0)}{v2,v3}
\superpot[phantom]{v1}{v2}{v3}{v4}
\fmfi{derplain,label=$\scriptstyle\mu$,l.dist=2}{vo1--vloc(__v1)}
\fmfi{derplain,label=$\scriptstyle\alpha$,l.dist=2}{vo1--vm1}
\fmfi{derplain,label=$\scriptstyle\beta$,l.dist=2}{vm1--vi1}
\fmfi{derplain,label=$\scriptstyle\gamma$,l.side=left,l.dist=2}{vi1--vloc(__vc)}
\fmfi{derplain,label=$\scriptstyle\nu$,l.side=left,l.dist=2}{vm2--vloc(__v2)}
\fmfi{derplain,label=$\scriptstyle\rho$,l.dist=2}{vm2--vloc(__vc)}
\fmfi{plain}{vloc(__vc)--vloc(__v3)}
\fmfi{plain}{vloc(__vc)--vloc(__v4)}
\fmfi{plain}{vm1{dir -135}..{dir -45}vm2}
\fmfi{plain}{vo1{dir 45}..{dir -135}vi1}
\end{fmfchar*}}}
\pnt
\end{aligned}
\end{equation}
Here the external momenta $(p_1,\cdots,p_4)$ are ordered counterclockwise with $p_1$ the 
momentum of the upper-left leg.

The last contribution is rather 
complicated.
 However, it can be simplified
by using momentum conservation to eliminate $p_2^\nu$ in the trace and
the symmetrization inside the trace as
\begin{equation}
\begin{aligned}
&\frac{1}{2}(
\tr(\gamma^\mu\gamma^\nu\gamma^\rho\gamma^\alpha\gamma^\beta\gamma^\gamma)
+\tr(\gamma^\rho\gamma^\nu\gamma^\mu\gamma^\alpha\gamma^\beta\gamma^\gamma))\\
&=-g^{\mu\nu}\tr(\gamma^\rho\gamma^\alpha\gamma^\beta\gamma^\gamma)
+g^{\mu\rho}\tr(\gamma^\nu\gamma^\alpha\gamma^\beta\gamma^\gamma)
-g^{\nu\rho}\tr(\gamma^\mu\gamma^\alpha\gamma^\beta\gamma^\gamma)
\pnt
\end{aligned}
\end{equation}
One then obtains
\begin{equation}
\begin{aligned}
&\tr(\gamma^\mu\gamma^\nu\gamma^\rho\gamma^\alpha\gamma^\beta\gamma^\gamma)
\settoheight{\eqoff}{$\times$}%
\setlength{\eqoff}{0.5\eqoff}%
\addtolength{\eqoff}{-11\unitlength}%
\raisebox{\eqoff}{%
\fmfframe(1,1)(1,1){%
\begin{fmfchar*}(20,20)
\fmftop{v1}
\fmfbottom{v2}
\fmfforce{(0w,h)}{v1}
\fmfforce{(0w,0)}{v2}
\fmffixed{(w,0)}{v1,v4}
\fmffixed{(w,0)}{v2,v3}
\superpot[phantom]{v1}{v2}{v3}{v4}
\fmfi{derplain,label=$\scriptstyle\mu$,l.dist=2}{vo1--vloc(__v1)}
\fmfi{derplain,label=$\scriptstyle\alpha$,l.dist=2}{vo1--vm1}
\fmfi{derplain,label=$\scriptstyle\beta$,l.dist=2}{vm1--vi1}
\fmfi{derplain,label=$\scriptstyle\gamma$,l.side=left,l.dist=2}{vi1--vloc(__vc)}
\fmfi{derplain,label=$\scriptstyle\nu$,l.side=left,l.dist=2}{vm2--vloc(__v2)}
\fmfi{derplain,label=$\scriptstyle\rho$,l.dist=2}{vm2--vloc(__vc)}
\fmfi{plain}{vloc(__vc)--vloc(__v3)}
\fmfi{plain}{vloc(__vc)--vloc(__v4)}
\fmfi{plain}{vm1{dir -135}..{dir -45}vm2}
\fmfi{plain}{vo1{dir 45}..{dir -135}vi1}
\end{fmfchar*}}}
\\
&=\tr(\gamma^\rho\gamma^\alpha\gamma^\beta\gamma^\gamma)
\left(
p_1^2
\settoheight{\eqoff}{$\times$}%
\setlength{\eqoff}{0.5\eqoff}%
\addtolength{\eqoff}{-11\unitlength}%
\raisebox{\eqoff}{%
\fmfframe(1,1)(1,1){%
\begin{fmfchar*}(20,20)
\fmftop{v1}
\fmfbottom{v2}
\fmfforce{(0w,h)}{v1}
\fmfforce{(0w,0)}{v2}
\fmffixed{(w,0)}{v1,v4}
\fmffixed{(w,0)}{v2,v3}
\superpot[phantom]{v1}{v2}{v3}{v4}
\fmfi{plain}{vo1--vloc(__v1)}
\fmfi{derplain,label=$\scriptstyle\alpha$,l.dist=2}{vo1--vm1}
\fmfi{derplain,label=$\scriptstyle\beta$,l.dist=2}{vm1--vi1}
\fmfi{derplain,label=$\scriptstyle\gamma$,l.side=left,l.dist=2}{vi1--vloc(__vc)}
\fmfi{plain}{vm2--vloc(__v2)}
\fmfi{derplain,label=$\scriptstyle\rho$,l.dist=2}{vm2--vloc(__vc)}
\fmfi{plain}{vloc(__vc)--vloc(__v3)}
\fmfi{plain}{vloc(__vc)--vloc(__v4)}
\fmfi{plain}{vm1{dir -135}..{dir -45}vm2}
\fmfi{plain}{vo1{dir 45}..{dir -135}vi1}
\end{fmfchar*}}}
+
\settoheight{\eqoff}{$\times$}%
\setlength{\eqoff}{0.5\eqoff}%
\addtolength{\eqoff}{-11\unitlength}%
\raisebox{\eqoff}{%
\fmfframe(1,1)(1,1){%
\begin{fmfchar*}(20,20)
\fmftop{v1}
\fmfbottom{v2}
\fmfforce{(0w,h)}{v1}
\fmfforce{(0w,0)}{v2}
\fmffixed{(w,0)}{v1,v4}
\fmffixed{(w,0)}{v2,v3}
\superpot[phantom]{v1}{v2}{v3}{v4}
\fmfi{derplain,label=$\scriptstyle\rho$,l.dist=2}{vo1--vloc(__v1)}
\fmfi{derplain,label=$\scriptstyle\alpha$,l.dist=2}{vo1--vm1}
\fmfi{derplain,label=$\scriptstyle\beta$,l.dist=2}{vm1--vi1}
\fmfi{derplain,label=$\scriptstyle\gamma$,l.side=left,l.dist=2}{vi1--vloc(__vc)}
\fmfi{plain}{vm2--vloc(__v2)}
\fmfi{plain}{vm2--vloc(__vc)}
\fmfi{plain}{vloc(__vc)--vloc(__v3)}
\fmfi{plain}{vloc(__vc)--vloc(__v4)}
\fmfi{plain}{vm1{dir -135}..{dir 45}vloc(__vc)}
\fmfi{plain}{vo1{dir 45}..{dir -135}vi1}
\end{fmfchar*}}}
+2
\settoheight{\eqoff}{$\times$}%
\setlength{\eqoff}{0.5\eqoff}%
\addtolength{\eqoff}{-11\unitlength}%
\raisebox{\eqoff}{%
\fmfframe(1,1)(1,1){%
\begin{fmfchar*}(20,20)
\fmftop{v1}
\fmfbottom{v2}
\fmfforce{(0w,h)}{v1}
\fmfforce{(0w,0)}{v2}
\fmffixed{(w,0)}{v1,v4}
\fmffixed{(w,0)}{v2,v3}
\superpot[phantom]{v1}{v2}{v3}{v4}
\fmfi{derplains}{vo1--vloc(__v1)}
\fmfi{derplain,label=$\scriptstyle\alpha$,l.dist=2}{vo1--vm1}
\fmfi{derplain,label=$\scriptstyle\beta$,l.dist=2}{vm1--vi1}
\fmfi{dblderplains,label=$\scriptstyle\gamma$,l.side=left,l.dist=2}{vi1--vloc(__vc)}
\fmfi{plain}{vm2--vloc(__v2)}
\fmfi{derplain,label=$\scriptstyle\rho$,l.dist=2}{vm2--vloc(__vc)}
\fmfi{plain}{vloc(__vc)--vloc(__v3)}
\fmfi{plain}{vloc(__vc)--vloc(__v4)}
\fmfi{plain}{vm1{dir -135}..{dir -45}vm2}
\fmfi{plain}{vo1{dir 45}..{dir -135}vi1}
\end{fmfchar*}}}
-
\settoheight{\eqoff}{$\times$}%
\setlength{\eqoff}{0.5\eqoff}%
\addtolength{\eqoff}{-11\unitlength}%
\raisebox{\eqoff}{%
\fmfframe(1,1)(1,1){%
\begin{fmfchar*}(20,20)
\fmftop{v1}
\fmfbottom{v2}
\fmfforce{(0w,h)}{v1}
\fmfforce{(0w,0)}{v2}
\fmffixed{(w,0)}{v1,v4}
\fmffixed{(w,0)}{v2,v3}
\superpot[phantom]{v1}{v2}{v3}{v4}
\fmfi{derplain,label=$\scriptstyle\gamma$,l.dist=2}{vo1--vloc(__v1)}
\fmfi{derplain,label=$\scriptstyle\alpha$,l.dist=2}{vo1--vm1}
\fmfi{derplain,label=$\scriptstyle\beta$,l.dist=2}{vm1--vloc(__vc)}
\fmfi{plain}{vm2--vloc(__v2)}
\fmfi{derplain,label=$\scriptstyle\rho$,l.dist=2}{vm2--vloc(__vc)}
\fmfi{plain}{vloc(__vc)--vloc(__v3)}
\fmfi{plain}{vloc(__vc)--vloc(__v4)}
\fmfi{plain}{vm1{dir -135}..{dir -45}vm2}
\fmfi{plain}{vo1{dir 45}..{dir -135}vloc(__vc)}
\end{fmfchar*}}}
\right)\\
&=
\tr(\gamma^\rho\gamma^\alpha\gamma^\beta\gamma^\gamma)
\left(
\settoheight{\eqoff}{$\times$}%
\setlength{\eqoff}{0.5\eqoff}%
\addtolength{\eqoff}{-11\unitlength}%
\raisebox{\eqoff}{%
\fmfframe(1,1)(1,1){%
\begin{fmfchar*}(20,20)
\fmftop{v1}
\fmfbottom{v2}
\fmfforce{(0w,h)}{v1}
\fmfforce{(0w,0)}{v2}
\fmffixed{(w,0)}{v1,v4}
\fmffixed{(w,0)}{v2,v3}
\fmffixed{(whatever,0.5h)}{vc1,v1}
\fmffixed{(whatever,0)}{vc1,vc2}
\fmf{plain,tension=2}{v1,vm1}
\fmf{derplain,label=$\scriptstyle\alpha$,l.dist=2,tension=2}{vm1,vc1}
\fmf{plain}{v2,vc1}
\fmf{plain}{v3,vc2}
\fmf{plain}{v4,vc2}
\fmf{derplain,right=0.5,label=$\scriptstyle\rho$,l.dist=2}{vc1,vc2}
\fmffreeze
\fmffixed{(whatever,0)}{vm1,vc}
\fmf{derplain,label=$\scriptstyle\beta$,l.dist=2}{vc1,vc}
\fmf{derplain,label=$\scriptstyle\gamma$,l.side=left,l.dist=2}{vc,vc2}
\fmffreeze
\fmf{plain,left=0.5}{vm1,vc}
\end{fmfchar*}}}
+
\settoheight{\eqoff}{$\times$}%
\setlength{\eqoff}{0.5\eqoff}%
\addtolength{\eqoff}{-11\unitlength}%
\raisebox{\eqoff}{%
\fmfframe(1,1)(1,1){%
\begin{fmfchar*}(20,20)
\fmftop{v1}
\fmfbottom{v2}
\fmfforce{(0w,h)}{v1}
\fmfforce{(0w,0)}{v2}
\fmffixed{(w,0)}{v1,v4}
\fmffixed{(w,0)}{v2,v3}
\superpot[phantom]{v1}{v2}{v3}{v4}
\fmfi{derplain,label=$\scriptstyle\rho$,l.dist=2}{vo1--vloc(__v1)}
\fmfi{derplain,label=$\scriptstyle\alpha$,l.dist=2}{vo1--vm1}
\fmfi{derplain,label=$\scriptstyle\beta$,l.dist=2}{vm1--vi1}
\fmfi{derplain,label=$\scriptstyle\gamma$,l.side=left,l.dist=2}{vi1--vloc(__vc)}
\fmfi{plain}{vm2--vloc(__v2)}
\fmfi{plain}{vm2--vloc(__vc)}
\fmfi{plain}{vloc(__vc)--vloc(__v3)}
\fmfi{plain}{vloc(__vc)--vloc(__v4)}
\fmfi{plain}{vm1{dir -135}..{dir 45}vloc(__vc)}
\fmfi{plain}{vo1{dir 45}..{dir -135}vi1}
\end{fmfchar*}}}
-
\settoheight{\eqoff}{$\times$}%
\setlength{\eqoff}{0.5\eqoff}%
\addtolength{\eqoff}{-11\unitlength}%
\raisebox{\eqoff}{%
\fmfframe(1,1)(1,1){%
\begin{fmfchar*}(20,20)
\fmftop{v1}
\fmfbottom{v2}
\fmfforce{(0w,h)}{v1}
\fmfforce{(0w,0)}{v2}
\fmffixed{(w,0)}{v1,v4}
\fmffixed{(w,0)}{v2,v3}
\superpot[phantom]{v1}{v2}{v3}{v4}
\fmfi{plain}{vo1--vloc(__v1)}
\fmfi{derplain,label=$\scriptstyle\alpha$,l.dist=2}{vo1--vm1}
\fmfi{derplain,label=$\scriptstyle\beta$,l.dist=2}{vm1--vloc(__vc)}
\fmfi{plain}{vm2--vloc(__v2)}
\fmfi{derplain,label=$\scriptstyle\rho$,l.dist=2}{vm2--vloc(__vc)}
\fmfi{plain}{vloc(__vc)--vloc(__v3)}
\fmfi{plain}{vloc(__vc)--vloc(__v4)}
\fmfi{derplain,label=$\scriptstyle\gamma$,l.dist=2}{vm2{dir 135}..{dir 45}vm1}
\fmfi{plain}{vo1{dir 45}..{dir -135}vloc(__vc)}
\end{fmfchar*}}}
\right)
\col
\end{aligned}
\end{equation}
where  we have used
\begin{equation}
p_1^2(k-p_1)^\gamma+2p_1\cdot(k-p_1)(k-p_1)^\gamma-(k-p_1)^2p_1^\gamma
=k^2(k-p_1)^\gamma-(k-p_1)^2k^\gamma
\,,
\end{equation}
with $k$ being one of the loop momenta.

The contribution involving the one-loop vacuum polarization reads
\begin{equation}
\begin{aligned}
\settoheight{\eqoff}{$\times$}%
\setlength{\eqoff}{0.5\eqoff}%
\addtolength{\eqoff}{-11\unitlength}%
\raisebox{\eqoff}{%
\fmfframe(1,1)(1,1){%
\begin{fmfchar*}(20,20)
\fmftop{v1}
\fmfbottom{v2}
\fmfforce{(0w,h)}{v1}
\fmfforce{(0w,0)}{v2}
\fmffixed{(w,0)}{v1,v4}
\fmffixed{(w,0)}{v2,v3}
\superpot{v1}{v2}{v3}{v4}
\fmfi{photon}{vm1{dir -135}..{dir -45}vm2}
\vacpolp[0.25]{vm1{dir -135}..{dir -45}vm2}
\end{fmfchar*}}}
&\to
(4\pi)^2\frac{1}{2}
(4\lambda\hat\lambda-\lambda^2)
\\
&\phantom{{}={}}
\left(
-(p_1+p_2)^2
\settoheight{\eqoff}{$\times$}%
\setlength{\eqoff}{0.5\eqoff}%
\addtolength{\eqoff}{-11\unitlength}%
\raisebox{\eqoff}{%
\fmfframe(1,1)(1,1){%
\begin{fmfchar*}(20,20)
\fmftop{v1}
\fmfbottom{v2}
\fmfforce{(0w,h)}{v1}
\fmfforce{(0w,0)}{v2}
\fmffixed{(w,0)}{v1,v4}
\fmffixed{(w,0)}{v2,v3}
\superpot[plain]{v1}{v2}{v3}{v4}
\fmfi{plain}{vm1{dir -112.5}..{dir -67.5}vm2}
\fmfi{plain}{vm1{dir -67.5}..{dir -112.5}vm2}
\end{fmfchar*}}}
+
p_1^2
\settoheight{\eqoff}{$\times$}%
\setlength{\eqoff}{0.5\eqoff}%
\addtolength{\eqoff}{-11\unitlength}%
\raisebox{\eqoff}{%
\fmfframe(1,1)(1,1){%
\begin{fmfchar*}(20,20)
\fmftop{v1}
\fmfbottom{v2}
\fmfforce{(0w,h)}{v1}
\fmfforce{(0w,0)}{v2}
\fmffixed{(w,0)}{v1,v4}
\fmffixed{(w,0)}{v2,v3}
\superpot{v1}{v2}{v3}{v4}
\fmfiset{p5}{vm1{dir -135}..{dir -45}vm2}
\svertex{vm5}{p5}
\fmfi{plain}{vm1{dir -150}..{dir -75}vm5}
\fmfi{plain}{vm1{dir -90}..{dir -135}vm5}
\fmfi{plain}{vm5{dir 0}..{dir 0}vloc(__vc)}
\end{fmfchar*}}}
+
p_2^2
\settoheight{\eqoff}{$\times$}%
\setlength{\eqoff}{0.5\eqoff}%
\addtolength{\eqoff}{-11\unitlength}%
\raisebox{\eqoff}{%
\fmfframe(1,1)(1,1){%
\begin{fmfchar*}(20,20)
\fmftop{v1}
\fmfbottom{v2}
\fmfforce{(0w,h)}{v1}
\fmfforce{(0w,0)}{v2}
\fmffixed{(w,0)}{v1,v4}
\fmffixed{(w,0)}{v2,v3}
\superpot{v1}{v2}{v3}{v4}
\fmfiset{p5}{vm1{dir -135}..{dir -45}vm2}
\svertex{vm5}{p5}
\fmfi{plain}{vm2{dir 150}..{dir 75}vm5}
\fmfi{plain}{vm2{dir 90}..{dir 135}vm5}
\fmfi{plain}{vm5{dir 0}..{dir 0}vloc(__vc)}
\end{fmfchar*}}}
\right)
\pnt
\end{aligned}
\end{equation}
Considering a factor $-1$ from the cancellation of the propagator 
connecting the
chiral vertex to the two-loop self energy, we obtain from the $\D$-algebra
manipulations
\begin{equation}
\begin{aligned}
\settoheight{\eqoff}{$\times$}%
\setlength{\eqoff}{0.5\eqoff}%
\addtolength{\eqoff}{-11\unitlength}%
\raisebox{\eqoff}{%
\fmfframe(1,1)(1,1){%
\begin{fmfchar*}(20,20)
\fmftop{v1}
\fmfbottom{v2}
\fmfforce{(0w,h)}{v1}
\fmfforce{(0w,0)}{v2}
\fmffixed{(w,0)}{v1,v4}
\fmffixed{(w,0)}{v2,v3}
\superpot{v1}{v2}{v3}{v4}
\vacpol{v1}{vc}
\end{fmfchar*}}}
&\to
(4\pi)^2
\left(
2\lambda\hat\lambda p_1^2
\settoheight{\eqoff}{$\times$}%
\setlength{\eqoff}{0.5\eqoff}%
\addtolength{\eqoff}{-11\unitlength}%
\raisebox{\eqoff}{%
\fmfframe(1,1)(1,1){%
\begin{fmfchar*}(20,20)
\fmftop{v1}
\fmfbottom{v2}
\fmfforce{(0w,h)}{v1}
\fmfforce{(0w,0)}{v2}
\fmffixed{(w,0)}{v1,v4}
\fmffixed{(w,0)}{v2,v3}
\superpot[phantom]{v1}{v2}{v3}{v4}
\fmfi{plain}{vo1--vloc(__v1)}
\fmfi{plain}{vloc(__vc)--vloc(__v2)}
\fmfi{plain}{vloc(__vc)--vloc(__v3)}
\fmfi{plain}{vloc(__vc)--vloc(__v4)}
\fmfcmd{
vm5=1/2[vloc(__vc),vo1];}
\fmfi{plain}{vo1{dir 30}..{dir -120}vm5}
\fmfi{plain}{vo1{dir -120}..{dir 30}vm5}
\fmfi{plain}{vm5{dir 30}..{dir -120}vloc(__vc)}
\fmfi{plain}{vm5{dir -120}..{dir 30}vloc(__vc)}
\end{fmfchar*}}}
-\frac{1}{2}(8\lambda\hat\lambda-(\lambda^2+\hat\lambda^2))
p_1^2
\settoheight{\eqoff}{$\times$}%
\setlength{\eqoff}{0.5\eqoff}%
\addtolength{\eqoff}{-11\unitlength}%
\raisebox{\eqoff}{%
\fmfframe(1,1)(1,1){%
\begin{fmfchar*}(20,20)
\fmftop{v1}
\fmfbottom{v2}
\fmfforce{(0w,h)}{v1}
\fmfforce{(0w,0)}{v2}
\fmffixed{(w,0)}{v1,v4}
\fmffixed{(w,0)}{v2,v3}
\superpot{v1}{v2}{v3}{v4}
\fmfiset{p5}{vm1{dir -135}..{dir -45}vm2}
\svertex{vm5}{p5}
\fmfi{plain}{vm1{dir -150}..{dir -75}vm5}
\fmfi{plain}{vm1{dir -90}..{dir -135}vm5}
\fmfi{plain}{vm5{dir 0}..{dir 0}vloc(__vc)}
\end{fmfchar*}}}
\right)
\pnt
\end{aligned}
\end{equation}

Let us conclude by mentioning a useful property that was used in 
appendix \ref{IR_cancellation}.
In the combination
\begin{equation}\label{twoloopIRcancel}
\begin{aligned}
\settoheight{\eqoff}{$\times$}%
\setlength{\eqoff}{0.5\eqoff}%
\addtolength{\eqoff}{-11\unitlength}%
\raisebox{\eqoff}{%
\fmfframe(1,1)(1,1){%
\begin{fmfchar*}(20,20)
\fmftop{v1}
\fmfbottom{v2}
\fmfforce{(0w,h)}{v1}
\fmfforce{(0w,0)}{v2}
\fmffixed{(w,0)}{v1,v4}
\fmffixed{(w,0)}{v2,v3}
\superpot{v1}{v2}{v3}{v4}
\vacpol{v1}{vc}
\end{fmfchar*}}}
+\frac{1}{2}
\left(
\settoheight{\eqoff}{$\times$}%
\setlength{\eqoff}{0.5\eqoff}%
\addtolength{\eqoff}{-11\unitlength}%
\raisebox{\eqoff}{%
\fmfframe(1,1)(1,1){%
\begin{fmfchar*}(20,20)
\fmftop{v1}
\fmfbottom{v2}
\fmfforce{(0w,h)}{v1}
\fmfforce{(0w,0)}{v2}
\fmffixed{(w,0)}{v1,v4}
\fmffixed{(w,0)}{v2,v3}
\superpot{v1}{v2}{v3}{v4}
\fmfi{photon}{vm1{dir -135}..{dir -45}vm2}
\vacpolp[0.25]{vm1{dir -135}..{dir -45}vm2}
\end{fmfchar*}}}
+
\settoheight{\eqoff}{$\times$}%
\setlength{\eqoff}{0.5\eqoff}%
\addtolength{\eqoff}{-11\unitlength}%
\raisebox{\eqoff}{%
\fmfframe(1,1)(1,1){%
\begin{fmfchar*}(20,20)
\fmftop{v1}
\fmfbottom{v2}
\fmfforce{(0w,h)}{v1}
\fmfforce{(0w,0)}{v2}
\fmffixed{(w,0)}{v1,v4}
\fmffixed{(w,0)}{v2,v3}
\superpot{v1}{v2}{v3}{v4}
\fmfi{photon}{vm1{dir 45}..{dir -45}vm4}
\vacpolp[0.25]{vm1{dir 45}..{dir -45}vm4}
\end{fmfchar*}}}
\right)
\end{aligned}
\end{equation}
the infrared divergence from the integrals involving the first leg is 
cancelled out.

There are two other diagrams with non-trivial $\D$-algebra and colour structure
\begin{equation}
\begin{aligned}
\settoheight{\eqoff}{$\times$}%
\setlength{\eqoff}{0.5\eqoff}%
\addtolength{\eqoff}{-11\unitlength}%
\raisebox{\eqoff}{%
\fmfframe(1,1)(1,1){%
\begin{fmfchar*}(20,20)
\fmftop{v1}
\fmfbottom{v2}
\fmfforce{(0w,h)}{v1}
\fmfforce{(0w,0)}{v2}
\fmffixed{(w,0)}{v1,v4}
\fmffixed{(w,0)}{v2,v3}
\superpot{v1}{v2}{v3}{v4}
\fmfi{photon}{vi1{dir -135}..{dir -45}vi2}
\fmfi{photon}{vo1{dir -135}..{dir -45}vo2}
\end{fmfchar*}}}
~,~~~
\settoheight{\eqoff}{$\times$}%
\setlength{\eqoff}{0.5\eqoff}%
\addtolength{\eqoff}{-11\unitlength}%
\raisebox{\eqoff}{%
\fmfframe(1,1)(1,1){%
\begin{fmfchar*}(20,20)
\fmftop{v1}
\fmfbottom{v2}
\fmfforce{(0w,h)}{v1}
\fmfforce{(0w,0)}{v2}
\fmffixed{(w,0)}{v1,v4}
\fmffixed{(w,0)}{v2,v3}
\superpot{v1}{v2}{v3}{v4}
\fmfi{photon}{vi1{dir -135}..{dir -45}vi2}
\fmfi{photon}{vo2{dir -45}..{dir 45}vo3}
\end{fmfchar*}}}
\pnt
\end{aligned}
\end{equation}
Interestingly, these can be seen to be
 proportional to the very same integrals which appear in components \cite{Minahan:2009wg}.
 The two diagrams are zero due to the vanishing of the one-loop triangle subdiagrams.

\section{Cancellation of IR divergences}
\label{IR_cancellation}

In order to check the cancellation of the IR divergences, together with the contributions having both UV and IR divergences given in section \ref{FComputation},
we 
have to include diagrams that have pure IR poles
and would have been excluded by
the UV 
finiteness conditions
 of 
subsection \ref{sec:finitenesscond}.
The cancellation of IR divergences in the combination 
\eqref{twoloopIRcancel} means that chiral and anti-chiral vertices 
with any number of legs and with external propagators 
are free of IR divergences from perturbative corrections.
In the following we will hence attach propagators to the external fields of
the diagrams 
that appear at four loops as quantum corrections of a 
chiral composite operator.
This does not affect the UV poles, since the chiral self-energy is UV finite
as demonstrated in appendix \ref{app:chiralse}.

The contributions to the $\chi(1)$ structure with only an IR divergence are given by
\begin{equation}
\begin{aligned}
V_{\text{r}44}=
\settoheight{\eqoff}{$\times$}%
\setlength{\eqoff}{0.5\eqoff}%
\addtolength{\eqoff}{-8.5\unitlength}%
\smash[b]{%
\raisebox{\eqoff}{%
\fmfframe(0,1)(0,1){%
\begin{fmfchar*}(20,15)
\chionerangefourl
\fmfi{photon,left=0.25}{vm1--ve}
\vacpolp[0.25]{vm1--ve}
\fmf{plain,tension=1,left=0,width=1mm}{v4,ved1}
\end{fmfchar*}}}}
&\to
\frac{(4\pi)^4}{k^4}MN(4MN-M^2)
I_{4\text{IR}}\,\chi(1)\\
&=
\frac{\lambda\hat\lambda}{16}(4\lambda\hat\lambda-\lambda^2)
\Big(\,\frac{2}{\varepsilon}(-3+\gamma-\ln 4\pi)\chi(1)\Big)
\col\\
V_{\text{r}45}=
\settoheight{\eqoff}{$\times$}%
\setlength{\eqoff}{0.5\eqoff}%
\addtolength{\eqoff}{-8.5\unitlength}%
\smash[b]{%
\raisebox{\eqoff}{%
\fmfframe(0,1)(0,1){%
\begin{fmfchar*}(20,15)
\chionerangefourl
\fmfi{photon,left=0.25}{vmm--ve}
\vacpolp[0.25]{vmm--ve}
\fmf{plain,tension=1,left=0,width=1mm}{v4,ved1}
\end{fmfchar*}}}}
&\to
-\frac{(4\pi)^4}{k^4}\frac{MN}{2}(4MN-M^2)
\big(I_{4\text{IR}}+I_{4\text{UVIR}}-I_4\big)\chi(1)\\
&=
-\frac{\lambda\hat\lambda}{16}(4\lambda\hat\lambda-\lambda^2)
\Big(\,\frac{2}{\varepsilon}(-3+\gamma-\ln 4\pi)\chi(1)\Big)
\col\\
V_{\text{r}46}=
\settoheight{\eqoff}{$\times$}%
\setlength{\eqoff}{0.5\eqoff}%
\addtolength{\eqoff}{-8.5\unitlength}%
\smash[b]{%
\raisebox{\eqoff}{%
\fmfframe(0,1)(0,1){%
\begin{fmfchar*}(20,15)
\chionerangefourl
\fmfi{photon,left=0.25}{vm6--ve}
\vacpolp[0.25]{vm6--ve}
\fmf{plain,tension=1,left=0,width=1mm}{v4,ved1}
\end{fmfchar*}}}}
&\to
\frac{(4\pi)^4}{k^4}\frac{MN}{2}(4MN-M^2)
\big(I_{4\text{UVIR}}-I_4+I_2I_{2\text{IR}}-\Kop(I_2)I_{2\text{IR}}\big)\chi(1)\\
&=
\frac{\lambda\hat\lambda}{16}(4\lambda\hat\lambda-\lambda^2)
\Big(\,\frac{2}{\varepsilon}(-3+\gamma-\ln 4\pi)\chi(1)\Big)
\col\\
V_{\text{r}37}=
\settoheight{\eqoff}{$\times$}%
\setlength{\eqoff}{0.5\eqoff}%
\addtolength{\eqoff}{-8.5\unitlength}%
\smash[b]{%
\raisebox{\eqoff}{%
\fmfframe(0,1)(-5,1){%
\begin{fmfchar*}(20,15)
\fmftop{v3}
\fmfbottom{v4}
\fmfforce{(0.125w,h)}{v3}
\fmfforce{(0.125w,0)}{v4}
\fmffixed{(0.25w,0)}{v2,v1}
\fmffixed{(0.25w,0)}{v3,v2}
\fmffixed{(0.25w,0)}{v4,v5}
\fmffixed{(0.25w,0)}{v5,v6}
\chioneca{v1}{v2}{v3}{v4}{v5}{v6}
\fmfi{photon}{vm1{dir -45}..{dir -135}vm6}
\vacpolp[0.33]{vm1{dir -45}..{dir -135}vm6}
\fmf{plain,tension=1,left=0,width=1mm}{v4,v6}
\end{fmfchar*}}}}
&\to
-\frac{(4\pi)^4}{k^4}\frac{MN}{2}(4MN-M^2)
(I_{4\text{IR}}+I_2I_{2\text{IR}}
-\Kop(I_2)I_{2\text{IR}})\chi(1)\\
&=
-\frac{\lambda\hat\lambda}{16}(4\lambda\hat\lambda-\lambda^2)
\Big(\,\frac{2}{\varepsilon}(-3+\gamma-\ln 4\pi)\chi(1)\Big)
\col
\\
V_{\text{r}38}=
\settoheight{\eqoff}{$\times$}%
\setlength{\eqoff}{0.5\eqoff}%
\addtolength{\eqoff}{-8.5\unitlength}%
\smash[b]{%
\raisebox{\eqoff}{%
\fmfframe(0,1)(-5,1){%
\begin{fmfchar*}(20,15)
\fmftop{v3}
\fmfbottom{v4}
\fmfforce{(0.125w,h)}{v3}
\fmfforce{(0.125w,0)}{v4}
\fmffixed{(0.25w,0)}{v2,v1}
\fmffixed{(0.25w,0)}{v3,v2}
\fmffixed{(0.25w,0)}{v4,v5}
\fmffixed{(0.25w,0)}{v5,v6}
\chioneca{v1}{v2}{v3}{v4}{v5}{v6}
\fmfi{photon}{vmm{dir 0}..{dir -120}vm6}
\vacpolp[0.33]{vmm{dir 0}..{dir -120}vm6}
\fmf{plain,tension=1,left=0,width=1mm}{v4,v6}
\end{fmfchar*}}}}
&\to
-\frac{(4\pi)^4}{k^4}\frac{MN}{2}(4MN-M^2)\big(
I_4-I_{4\text{UVIR}}-I_2I_{2\text{IR}}
+\Kop(I_2)I_{2\text{IR}}\big)\chi(1)\\
&=
\frac{\lambda\hat\lambda}{16}(4\lambda\hat\lambda-\lambda^2)
\Big(\,\frac{2}{\varepsilon}(-3+\gamma-\ln 4\pi)\chi(1)\Big)
\pnt
\end{aligned}
\end{equation}
where we have given only the IR pole terms, and the UV subdivergences  have been subtracted.

We also have to consider the 
correction of the chiral propagator that is a neighbour of the fields
interacting via $\chi(1)$
\begin{equation}
\begin{aligned}
V_{\text{r}3\text{s}}=
\settoheight{\eqoff}{$\times$}%
\setlength{\eqoff}{0.5\eqoff}%
\addtolength{\eqoff}{-8.5\unitlength}%
\smash[b]{%
\raisebox{\eqoff}{%
\fmfframe(0,1)(0,1){%
\begin{fmfchar*}(20,15)
\chionerangefourl
\vacpolp[0.25]{pe}
\fmf{plain,tension=1,left=0,width=1mm}{v4,ved1}
\end{fmfchar*}}}}
&\to
-\frac{(4\pi)^4}{k^4}\frac{MN}{2}\big(8MN-(M^2+N^2)\big)
\big(I_2-\Kop(I_2)\big)I_{2\text{IR}}\,
\chi(1)\\
&=
-\frac{\lambda\hat\lambda}{16}\big(8\lambda\hat\lambda-(\lambda^2+\hat\lambda^2)\big)
\Big(\frac{1}{\varepsilon}(-3+\gamma-\ln 4\pi)\chi(1)\Big)
\pnt
\end{aligned}
\end{equation}
According to \eqref{twoloopIRcancel}, one half of this
contribution has to be taken into account, since the other half should cancel
part of the IR divergence from an interaction of the isolated leg via
a one-loop corrected gauge propagator with its
neighbour to the right. Similar considerations hold also for the reflected 
diagram of $V_{\text{r}3\text{s}}$, such that the total contribution 
of these diagrams to the IR divergence is 
$\frac{1}{2}(1+{\cal R})V_{\text{r}3\text{s}}$.

Further IR divergent contributions from self energy corrections of the
three external and one internal line at the upper chiral vertex that
forms $\chi(1)$ cancel among respective diagrams in which two of these 
lines are interacting via one-loop corrected gauge propagator. 
This is guaranteed by \eqref{twoloopIRcancel} since in the considered 
propagators are attached to their external lines.

At this point a simple way 
to check the cancellation of the IR divergences is to sum up
all the contribution containing them and check that the result 
is the same 
as if from the very beginning we had 
omitted all IR divergent diagrams, and had only considered 
$V^{\text{UV}}_{\text{r}35}$ and $V^{\text{UV}}_{\text{r}36}$.
In fact, the sum
\begin{equation}
\begin{aligned}\label{ZUVIR}
&
-(1+{\cal{R}})(
V_{\text{r}35}
+V_{\text{r}44}
+V_{\text{r}45}
+V_{\text{r}46}
+V_{\text{r}37}
+V_{\text{r}38}
)
-3V_{\text{r}36}
-\frac{1}{2}(1+{\cal R})V_{\text{r}3\text{s}}
\\
&=
\frac{(4\pi)^4}{k^4}MN\Big(
-6MNI_{4\mathbf{bbb}}
+\frac{1}{2}(8MN-(M^2+N^2))(3I_{4}+2I_{42\mathbf{b}bd})\Big)\chi(1)
\\
&=\frac{\lambda\hat\lambda}{16}
\Big(
-\lambda\hat\lambda\frac{3\pi^2}{\varepsilon}
+\big(8\lambda\hat\lambda-(\lambda^2+\hat\lambda^2)\big)\Big(
-\frac{1}{4\varepsilon^2}
+\frac{1}{\varepsilon}\Big(1+\frac{\pi^2}{4}\Big)
\Big)\Big)\chi(1)
\end{aligned}
\end{equation}
turns out to be equal to
\begin{equation}
-(1+{\cal{R}})V^{\text{UV}}_{\text{r}35} 
-3V^{\text{UV}}_{\text{r}36}
\col
\end{equation}
which is the respective contribution of only the overall UV 
divergences from the diagrams with also an IR divergence to \eqref{Zr3}.

It is important to note that, besides the previously described check of the cancellation of the IR 
divergences, we have also performed the full computation of the range three contribution 
in the IR-safe $\eta$-gauge described in \cite{Leoni:2010az}. The result turns out to be the same.


\section{Double poles}
\label{doublePoles}

In this appendix we check explicitly the cancellation of the double poles in $\ln {\cal Z}$.
For that we need to consider diagrams which 
are responsible for interactions between magnons at odd and 
even sites which are proportional to chiral functions 
$\chi(1,2)$ and $\chi(2,3)$.
We start by computing those contributions, and then we prove the 
complete cancellation of the double poles.

\subsection{Odd- and even-site magnon interactions}
\label{app:Z4mixed}
The relevant diagrams that 
couple the odd and even site
magnons with each other are the following ones
\begin{equation}
\begin{aligned}
S_{\text{mixed}}=
\settoheight{\eqoff}{$\times$}%
\setlength{\eqoff}{0.5\eqoff}%
\addtolength{\eqoff}{-8.5\unitlength}%
\raisebox{\eqoff}{%
\fmfframe(0,1)(0,1){%
\begin{fmfchar*}(20,15)
\fmftop{v1}
\fmfbottom{v5}
\fmfforce{(0.125w,h)}{v1}
\fmfforce{(0.125w,0)}{v5}
\fmffixed{(0.25w,0)}{v1,v2}
\fmffixed{(0.25w,0)}{v2,v3}
\fmffixed{(0.25w,0)}{v3,v4}
\fmffixed{(0.25w,0)}{v5,v6}
\fmffixed{(0.25w,0)}{v6,v7}
\fmffixed{(0.25w,0)}{v7,v8}
\fmffixed{(0,whatever)}{v2,v4lc}
\fmffixed{(0,whatever)}{v4lc,v4la}
\fmffixed{(0,whatever)}{v7,v4rc}
\fmffixed{(0,whatever)}{v4rc,v4ra}
\fmffixed{(whatever,0)}{v4la,v4rc}
\fmf{plain,left=0.25}{v5,v4la}
\fmf{plain}{v4la,v4lc}
\fmf{plain,tension=0,left=0.25}{v4lc,v1}
\fmf{plain,left=0}{v4lc,v2}
\fmf{plain,tension=0,right=0.25}{v4lc,v3}
\fmf{plain,tension=0,left=0.25}{v6,v4ra}
\fmf{plain}{v7,v4ra}
\fmf{plain,tension=0,right=0.25}{v8,v4ra}
\fmf{plain}{v4ra,v4rc}
\fmf{plain,right=0.25}{v4rc,v4}
\fmffreeze
\fmfposition
\fmf{plain,left=0.5}{v4la,v4rc}
\fmf{plain,left=0.5}{v4rc,v4la}
\fmf{plain,tension=1,left=0,width=1mm}{v5,v8}
\end{fmfchar*}}}
&\to
\frac{(4\pi)^4}{k^4}(MN)^2
I_4\,\chi(1,2)
=\frac{(\lambda\hat\lambda)^2}{16}\Big(-\frac{1}{2\varepsilon^2}+\frac{2}{\varepsilon}\Big)\chi(1,2)
\col\\
V_{\text{mixed}1}
=
\settoheight{\eqoff}{$\times$}%
\setlength{\eqoff}{0.5\eqoff}%
\addtolength{\eqoff}{-8.5\unitlength}%
\raisebox{\eqoff}{%
\fmfframe(0,1)(0,1){%
\begin{fmfchar*}(20,15)
\chionetwog
\fmf{plain,tension=1,left=0,width=1mm}{v5,v8}
\fmfi{wiggly}{vgm1{dir -120}..{dir -60}vgu3}
\fmfi{wiggly}{vgm1{dir -120}..{dir -60}vgd3}
\end{fmfchar*}}}
&\to
\frac{(4\pi)^4}{k^4}(MN)^2
I_{42\mathbf{bb0}cd}\,\chi(1,2)
=\frac{(\lambda\hat\lambda)^2}{16}\Big(\frac{1}{4\varepsilon^2}-\frac{3}{2\varepsilon}\Big)\chi(1,2)
\col\\
V_{\text{mixed}2}
=
\settoheight{\eqoff}{$\times$}%
\setlength{\eqoff}{0.5\eqoff}%
\addtolength{\eqoff}{-8.5\unitlength}%
\smash[b]{%
\raisebox{\eqoff}{%
\fmfframe(0,1)(0,1){%
\begin{fmfchar*}(20,15)
\chionetwog
\fmf{plain,tension=1,left=0,width=1mm}{v5,v8}
\fmfi{wiggly}{vgm1{dir -120}..{dir -60}vgm3}
\fmfi{wiggly}{vgm1{dir -90}..{dir -90}vgm4}
\end{fmfchar*}}}}
&\to
-\frac{(4\pi)^4}{k^4}\frac{(MN)^2}{2}
I_{422\mathbf{b}\text{tr}ABcd}\,\chi(1,2)
=\frac{(\lambda\hat\lambda)^2}{16}\Big(-\frac{1}{2\varepsilon^2}+\frac{1}{\varepsilon}\Big)\chi(1,2)
\pnt\\
\end{aligned}
\end{equation}
In the sum of all contributions  one has to consider the reflected 
diagrams. The second contribution acquires an additional factor of 
two due to two distinct positions for the vector vertices which are not mapped to each other under 
reflection.
The result for the mixed renormalization constant 
reads\footnote{There is another contribution with identical prefactor
that involves the chiral function $\chi(2,3)$ that we associate to the 
even site sector.}
\begin{equation}
\begin{aligned}\label{Zmixed}
\mathcal{Z}_{4,\text{mixed}}&=-(1+{\cal R})(S_{\text{mixed}}+2V_{\text{mixed}1}+V_{\text{mixed}2})
=\frac{(\lambda\hat\lambda)^2}{16}\frac{1}{\varepsilon^2}\chi(1,2)
\pnt
\end{aligned}
\end{equation}
As expected \cite{Bak:2009mq}, the $1/\varepsilon$ pole is cancelled out such that
at four loops there is no contribution to the dilatation operator 
that couples the magnons at odd and even sites.

\subsection{Double pole cancellation}
\label{DoublePolesCancellation}

Summing up the contributions to the $1/\varepsilon^2$ poles 
of the odd-site sector to the four-loop
${\cal Z}$ from  \eqref{Zr5}, \eqref{Zr4}, \eqref{Zr3} and \eqref{Zmixed}, 
we obtain
\begin{equation}
\begin{aligned}\label{Z4e^2}
\bar\lambda^4(\mathcal{Z}_{4,\text{odd}}+\mathcal{Z}_{4,\text{mixed}})|_{\frac{1}{\varepsilon^2}}
&=
\big(\mathcal{Z}_{\text{r}5,\text{odd}}
+\mathcal{Z}_{4,\text{mixed}}
+\mathcal{Z}_{\text{r}4,\text{odd}}
+\mathcal{Z}_{\text{r}3,\text{odd}}\big)|_{\frac{1}{\varepsilon^2}}
\\
&=
\frac{{\bar{\lambda}}^4}{16\varepsilon^2}
\Big[
\,
\frac{1}{2}\big(\chi(1,3)+\chi(3,1)\big)
+\chi(1,2)
-\chi(1)
\Big]
\pnt
\end{aligned}
\end{equation}
In the definition of the dilatation operator,
the logarithm guarantees that all higher order poles in $\varepsilon$ 
cancel out, such that $\ln\mathcal{Z}$ only contains simple 
$\frac{1}{\varepsilon}$ poles. Inserting \eqref{opren}, the expansion reads
\begin{equation}\label{lnZ}
\ln\mathcal{Z}=\bar\lambda^2\mathcal{Z}_2+\bar\lambda^4\Big(\mathcal{Z}_4-\frac{1}{2}\mathcal{Z}^2_2\Big)+\mathcal{O}(\bar\lambda^6)\pnt
\end{equation}
 Let us now check
 the double pole cancellations in the $\bar{\lambda}^4$ term.
The two-loop contribution to the renormalization constant for operators of 
length $L$ can be written as
\begin{equation}\label{Z2}
\begin{aligned}
\bar\lambda^2\mathcal{Z}_2
&=-
\sum_{i=1}^{2L}
\settoheight{\eqoff}{$\times$}%
\setlength{\eqoff}{0.5\eqoff}%
\addtolength{\eqoff}{-4\unitlength}%
\raisebox{\eqoff}{%
\fmfframe(0,0)(-5,0){%
\begin{fmfchar*}(16,8)
\fmftop{vu3}
\fmfbottom{vd3}
\fmfforce{(0.125w,h)}{vu3}
\fmfforce{(0.125w,0)}{vd3}
\fmffixed{(0.25w,0)}{vu2,vu1}
\fmffixed{(0.25w,0)}{vu3,vu2}
\fmffixed{(0.25w,0)}{vd3,vd2}
\fmffixed{(0.25w,0)}{vd2,vd1}
\chioneca{vd3}{vd2}{vd1}{vu1}{vu2}{vu3}
\fmf{plain,tension=1,left=0,width=1mm}{vd1,vd3}
\fmfv{l=$\scriptscriptstyle i$,l.a=-90,l.dist=2}{vd3}
\end{fmfchar*}}}
=
-\frac{\lambda\hat\lambda}{4}\frac{1}{\varepsilon}(\chi(1)+\chi(2))
\col
\end{aligned}
\end{equation}
where we have indicated the sum over the sites explicitly. It
has an obvious decomposition into two parts acting exclusively
on even and on odd sites, respectively.
The square of the above result can be decomposed as follows
\begin{equation}\label{Z2inZ22}
\begin{aligned}
\frac{1}{2}\mathcal{Z}_2^2
&=\mathcal{Z}_{22,\text{dc}}
+\mathcal{Z}_{22,S}
\pnt
\end{aligned}
\end{equation}
The individual terms are given by
\begin{equation}
\begin{aligned}
\bar\lambda^4\mathcal{Z}_{22,\text{dc}}
&=
\sum_{j\ge i+3}^{2L}
\Big(
\settoheight{\eqoff}{$\times$}%
\setlength{\eqoff}{0.5\eqoff}%
\addtolength{\eqoff}{-4\unitlength}%
\raisebox{\eqoff}{%
\fmfframe(0,0)(-5,0){%
\begin{fmfchar*}(16,8)
\fmftop{vu3}
\fmfbottom{vd3}
\fmfforce{(0.125w,h)}{vu3}
\fmfforce{(0.125w,0)}{vd3}
\fmffixed{(0.25w,0)}{vu2,vu1}
\fmffixed{(0.25w,0)}{vu3,vu2}
\fmffixed{(0.25w,0)}{vd3,vd2}
\fmffixed{(0.25w,0)}{vd2,vd1}
\chioneca{vd3}{vd2}{vd1}{vu1}{vu2}{vu3}
\fmf{plain,tension=1,left=0,width=1mm}{vd1,vd3}
\fmfv{l=$\scriptscriptstyle i$,l.a=-90,l.dist=2}{vd3}
\end{fmfchar*}}}
\settoheight{\eqoff}{$\times$}%
\setlength{\eqoff}{0.5\eqoff}%
\addtolength{\eqoff}{-4\unitlength}%
\raisebox{\eqoff}{%
\fmfframe(0,0)(-5,0){%
\begin{fmfchar*}(16,8)
\fmftop{vu3}
\fmfbottom{vd3}
\fmfforce{(0.125w,h)}{vu3}
\fmfforce{(0.125w,0)}{vd3}
\fmffixed{(0.25w,0)}{vu2,vu1}
\fmffixed{(0.25w,0)}{vu3,vu2}
\fmffixed{(0.25w,0)}{vd3,vd2}
\fmffixed{(0.25w,0)}{vd2,vd1}
\chioneca{vd3}{vd2}{vd1}{vu1}{vu2}{vu3}
\fmf{plain,tension=1,left=0,width=1mm}{vd1,vd3}
\fmfv{l=$\scriptscriptstyle j$,l.a=-90,l.dist=2}{vd3}
\end{fmfchar*}}}
\Big)
\\
%
%
\bar\lambda^4\mathcal{Z}_{22,S}
&=
\frac{1}{2}\sum_{i=1}^{2L}\left(
\settoheight{\eqoff}{$\times$}%
\setlength{\eqoff}{0.5\eqoff}%
\addtolength{\eqoff}{-9\unitlength}%
\raisebox{\eqoff}{%
\fmfframe(0,1)(3.5,1){%
\begin{fmfchar*}(16,16)
\fmf{plain,tension=1,left=0,width=1mm,fore=(0.5,,0.5,,0.5)}{vm1,vm5}
\fmftop{vu5}
\fmfbottom{vd5}
\fmfforce{(0.125w,h)}{vu5}
\fmfforce{(0.125w,0)}{vd5}
\fmffixed{(0.25w,0)}{vu2,vu1}
\fmffixed{(0.25w,0)}{vu3,vu2}
\fmffixed{(0.25w,0)}{vu4,vu3}
\fmffixed{(0.25w,0)}{vu5,vu4}
\fmffixed{(0.25w,0)}{vd5,vd4}
\fmffixed{(0.25w,0)}{vd4,vd3}
\fmffixed{(0.25w,0)}{vd3,vd2}
\fmffixed{(0.25w,0)}{vd2,vd1}
\fmffixed{(0,0.5h)}{vd1,vm1}
\fmffixed{(0,0.5h)}{vd2,vm2}
\fmffixed{(0,0.5h)}{vd3,vm3}
\fmffixed{(0,0.5h)}{vd4,vm4}
\fmffixed{(0,0.5h)}{vd5,vm5}
\fmf{plain}{vd4,vm4}
\fmf{plain}{vd5,vm5}
\fmf{plain}{vm1,vu1}
\fmf{plain}{vm2,vu2}
%
\fmf{plain,tension=0,left=0.25}{vd3,vc1}
\fmf{plain,tension=1}{vc1,vd2}
\fmf{plain,tension=0,right=0.25}{vd1,vc1}
\fmf{plain,tension=0,right=0.25}{va1,vm1}
\fmf{plain,tension=1}{vm2,va1}
\fmf{plain,tension=0,left=0.25}{va1,vm3}
\fmf{plain,tension=1}{vc1,va1}
\fmf{plain,tension=0,left=0.25}{vm5,vc2}
\fmf{plain,tension=1}{vc2,vm4}
\fmf{plain,tension=0,right=0.25}{vm3,vc2}
\fmf{plain,tension=0,right=0.25}{va2,vu3}
\fmf{plain,tension=1}{vu4,va2}
\fmf{plain,tension=0,left=0.25}{va2,vu5}
\fmf{plain,tension=1}{vc2,va2}
\fmffreeze
\fmf{plain,tension=1,left=0,width=1mm}{vd1,vd5}
\fmfv{l=$\scriptscriptstyle i$,l.a=-90,l.dist=2}{vd3}
\end{fmfchar*}}}
+
\settoheight{\eqoff}{$\times$}%
\setlength{\eqoff}{0.5\eqoff}%
\addtolength{\eqoff}{-9\unitlength}%
\raisebox{\eqoff}{%
\fmfframe(0,1)(3.5,1){%
\begin{fmfchar*}(16,16)
\fmf{plain,tension=1,left=0,width=1mm,fore=(0.5,,0.5,,0.5)}{vm1,vm5}
\fmftop{vu5}
\fmfbottom{vd5}
\fmfforce{(0.125w,h)}{vu5}
\fmfforce{(0.125w,0)}{vd5}
\fmffixed{(0.25w,0)}{vu2,vu1}
\fmffixed{(0.25w,0)}{vu3,vu2}
\fmffixed{(0.25w,0)}{vu4,vu3}
\fmffixed{(0.25w,0)}{vu5,vu4}
\fmffixed{(0.25w,0)}{vd5,vd4}
\fmffixed{(0.25w,0)}{vd4,vd3}
\fmffixed{(0.25w,0)}{vd3,vd2}
\fmffixed{(0.25w,0)}{vd2,vd1}
\fmffixed{(0,0.5h)}{vd1,vm1}
\fmffixed{(0,0.5h)}{vd2,vm2}
\fmffixed{(0,0.5h)}{vd3,vm3}
\fmffixed{(0,0.5h)}{vd4,vm4}
\fmffixed{(0,0.5h)}{vd5,vm5}
\fmf{plain}{vm4,vu4}
\fmf{plain}{vm5,vu5}
\fmf{plain}{vd1,vm1}
\fmf{plain}{vd2,vm2}
%
\fmf{plain,tension=0,left=0.25}{vm3,vc1}
\fmf{plain,tension=1}{vc1,vm2}
\fmf{plain,tension=0,right=0.25}{vm1,vc1}
\fmf{plain,tension=0,right=0.25}{va1,vu1}
\fmf{plain,tension=1}{vu2,va1}
\fmf{plain,tension=0,left=0.25}{va1,vu3}
\fmf{plain,tension=1}{vc1,va1}
\fmf{plain,tension=0,left=0.25}{vd5,vc2}
\fmf{plain,tension=1}{vc2,vd4}
\fmf{plain,tension=0,right=0.25}{vd3,vc2}
\fmf{plain,tension=0,right=0.25}{va2,vm3}
\fmf{plain,tension=1}{vm4,va2}
\fmf{plain,tension=0,left=0.25}{va2,vm5}
\fmf{plain,tension=1}{vc2,va2}
\fmffreeze
\fmf{plain,tension=1,left=0,width=1mm}{vd1,vd5}
\fmfv{l=$\scriptscriptstyle i$,l.a=-90,l.dist=2}{vd5}
\end{fmfchar*}}}
%
%
+
\settoheight{\eqoff}{$\times$}%
\setlength{\eqoff}{0.5\eqoff}%
\addtolength{\eqoff}{-9\unitlength}%
\raisebox{\eqoff}{%
\fmfframe(0,1)(-0.5,1){%
\begin{fmfchar*}(16,16)
\fmf{plain,tension=1,left=0,width=1mm,fore=(0.5,,0.5,,0.5)}{vm1,vm4}
\fmftop{vu4}
\fmfbottom{vd4}
\fmfforce{(0.125w,h)}{vu4}
\fmfforce{(0.125w,0)}{vd4}
\fmffixed{(0.25w,0)}{vu2,vu1}
\fmffixed{(0.25w,0)}{vu3,vu2}
\fmffixed{(0.25w,0)}{vu4,vu3}
\fmffixed{(0.25w,0)}{vd4,vd3}
\fmffixed{(0.25w,0)}{vd3,vd2}
\fmffixed{(0.25w,0)}{vd2,vd1}
\fmffixed{(0,0.5h)}{vd1,vm1}
\fmffixed{(0,0.5h)}{vd2,vm2}
\fmffixed{(0,0.5h)}{vd3,vm3}
\fmffixed{(0,0.5h)}{vd4,vm4}
\fmf{plain}{vd4,vm4}
\fmf{plain}{vm1,vu1}
%
\fmf{plain,tension=0,left=0.25}{vd3,vc1}
\fmf{plain,tension=1}{vc1,vd2}
\fmf{plain,tension=0,right=0.25}{vd1,vc1}
\fmf{plain,tension=0,right=0.25}{va1,vm1}
\fmf{plain,tension=1}{vm2,va1}
\fmf{plain,tension=0,left=0.25}{va1,vm3}
\fmf{plain,tension=1}{vc1,va1}
\fmf{plain,tension=0,left=0.25}{vm4,vc2}
\fmf{plain,tension=1}{vc2,vm3}
\fmf{plain,tension=0,right=0.25}{vm2,vc2}
\fmf{plain,tension=0,right=0.25}{va2,vu2}
\fmf{plain,tension=1}{vu3,va2}
\fmf{plain,tension=0,left=0.25}{va2,vu4}
\fmf{plain,tension=1}{vc2,va2}
\fmffreeze
\fmf{plain,tension=1,left=0,width=1mm}{vd1,vd4}
\fmfv{l=$\scriptscriptstyle i$,l.a=-90,l.dist=2}{vd3}
\end{fmfchar*}}}
+
\settoheight{\eqoff}{$\times$}%
\setlength{\eqoff}{0.5\eqoff}%
\addtolength{\eqoff}{-9\unitlength}%
\raisebox{\eqoff}{%
\fmfframe(0,1)(-0.5,1){%
\begin{fmfchar*}(16,16)
\fmf{plain,tension=1,left=0,width=1mm,fore=(0.5,,0.5,,0.5)}{vm1,vm4}
\fmftop{vu4}
\fmfbottom{vd4}
\fmfforce{(0.125w,h)}{vu4}
\fmfforce{(0.125w,0)}{vd4}
\fmffixed{(0.25w,0)}{vu2,vu1}
\fmffixed{(0.25w,0)}{vu3,vu2}
\fmffixed{(0.25w,0)}{vu4,vu3}
\fmffixed{(0.25w,0)}{vd4,vd3}
\fmffixed{(0.25w,0)}{vd3,vd2}
\fmffixed{(0.25w,0)}{vd2,vd1}
\fmffixed{(0,0.5h)}{vd1,vm1}
\fmffixed{(0,0.5h)}{vd2,vm2}
\fmffixed{(0,0.5h)}{vd3,vm3}
\fmffixed{(0,0.5h)}{vd4,vm4}
\fmf{plain}{vm4,vu4}
\fmf{plain}{vd1,vm1}
%
\fmf{plain,tension=0,left=0.25}{vm3,vc1}
\fmf{plain,tension=1}{vc1,vm2}
\fmf{plain,tension=0,right=0.25}{vm1,vc1}
\fmf{plain,tension=0,right=0.25}{va1,vu1}
\fmf{plain,tension=1}{vu2,va1}
\fmf{plain,tension=0,left=0.25}{va1,vu3}
\fmf{plain,tension=1}{vc1,va1}
\fmf{plain,tension=0,left=0.25}{vd4,vc2}
\fmf{plain,tension=1}{vc2,vd3}
\fmf{plain,tension=0,right=0.25}{vd2,vc2}
\fmf{plain,tension=0,right=0.25}{va2,vm2}
\fmf{plain,tension=1}{vm3,va2}
\fmf{plain,tension=0,left=0.25}{va2,vm4}
\fmf{plain,tension=1}{vc2,va2}
\fmffreeze
\fmf{plain,tension=1,left=0,width=1mm}{vd1,vd4}
\fmfv{l=$\scriptscriptstyle i$,l.a=-90,l.dist=2}{vd4}
\end{fmfchar*}}}
%
%
+
\settoheight{\eqoff}{$\times$}%
\setlength{\eqoff}{0.5\eqoff}%
\addtolength{\eqoff}{-9\unitlength}%
\raisebox{\eqoff}{%
\fmfframe(0,1)(-4.5,1){%
\begin{fmfchar*}(16,16)
\fmf{plain,tension=1,left=0,width=1mm,fore=(0.5,,0.5,,0.5)}{vm1,vm3}
\fmftop{vu3}
\fmfbottom{vd3}
\fmfforce{(0.125w,h)}{vu3}
\fmfforce{(0.125w,0)}{vd3}
\fmffixed{(0.25w,0)}{vu2,vu1}
\fmffixed{(0.25w,0)}{vu3,vu2}
\fmffixed{(0.25w,0)}{vd3,vd2}
\fmffixed{(0.25w,0)}{vd2,vd1}
\fmffixed{(0,0.5h)}{vd1,vm1}
\fmffixed{(0,0.5h)}{vd2,vm2}
\fmffixed{(0,0.5h)}{vd3,vm3}
%
\fmf{plain,tension=0,left=0.25}{vm3,vc1}
\fmf{plain,tension=1}{vc1,vm2}
\fmf{plain,tension=0,right=0.25}{vm1,vc1}
\fmf{plain,tension=0,right=0.25}{va1,vu1}
\fmf{plain,tension=1}{vu2,va1}
\fmf{plain,tension=0,left=0.25}{va1,vu3}
\fmf{plain,tension=1}{vc1,va1}
\fmf{plain,tension=0,left=0.25}{vd3,vc2}
\fmf{plain,tension=1}{vc2,vd2}
\fmf{plain,tension=0,right=0.25}{vd1,vc2}
\fmf{plain,tension=0,right=0.25}{va2,vm1}
\fmf{plain,tension=1}{vm2,va2}
\fmf{plain,tension=0,left=0.25}{va2,vm3}
\fmf{plain,tension=1}{vc2,va2}
\fmffreeze
\fmf{plain,tension=1,left=0,width=1mm}{vd1,vd3}
\fmfv{l=$\scriptscriptstyle i$,l.a=-90,l.dist=2}{vd3}
\end{fmfchar*}}}
\right)
\\
&\to
\frac{1}{2}\frac{(4\pi)^4}{k^4}M^2N^2\Kop(I_2)^2
(\chi(1,3)+\chi(3,1)+2\chi(1,2)-2\chi(1))\\
&=\frac{(\lambda\hat\lambda)^2}{16}\frac{1}{2\varepsilon^2}
(\chi(1,3)+\chi(3,1)+2\chi(1,2)-2\chi(1))
\col\\
\end{aligned}
\end{equation}
where the arrow denotes that in the final result 
we have considered the chiral functions with 
odd indices only 
and $\chi(1,2)$ and neglected the ones with 
only even indices and $\chi(2,3)$.

According to \eqref{Z2inZ22}, the square of the two-loop contribution expands as
\begin{equation}
\frac{1}{2}(\bar\lambda\mathcal{Z}_2)^2=\frac{(\lambda\hat\lambda)^2}{16}\frac{1}{2\varepsilon^2}(\chi(1,3)+\chi(3,1)+2\chi(1,2)-2\chi(1))+\dots
\col
\label{Z^2}
\end{equation}
where we have neglected the chiral functions with only even arguments and 
$\chi(2,3)$.
We have also disregarded the terms $\mathcal{Z}_{22,\text{dc}}$
which trivially
cancel against four-loop diagrams that only 
contain double poles and hence become
disconnected when the composite operator is removed. 
We have omitted
to present these diagrams in the paper.

Comparing equations (\ref{Z4e^2}) and (\ref{Z^2}) we finally find 
our desired result
\begin{equation}
\Big(\mathcal{Z}_4-\frac{1}{2}\mathcal{Z}_2^2\Big)|_{\frac{1}{\varepsilon^2}}=0
\col
\end{equation}
where we have considered that the discussion is identical for the neglected 
contributions with chiral functions with even arguments and $\chi(2,3)$.

\end{fmffile}

\footnotesize
\bibliographystyle{JHEP}
\bibliography{references}

\end{document}
